\documentclass[final, 12pt,oneside]{class_diss}

\usepackage[numbers,square,sort&compress]{natbib}
\usepackage{amsmath,amssymb}
\usepackage{graphicx}
\usepackage{epstopdf}
\usepackage{amsthm}
\usepackage{picture}
\usepackage[usenames]{color}
\usepackage{verbatim}
\usepackage{fancyhdr}
\usepackage{xstring}
\usepackage{framed}
\usepackage[pdftex,plainpages=false,pdfpagelabels]{hyperref}
\usepackage{paralist}
\usepackage{latexsym}
\usepackage{mathtools}
\usepackage{multirow}
\usepackage{rotating}

\theoremstyle{definition}
\newtheorem{defin}{Definition}

\definecolor{Pink}{rgb}{1.0, 0.5, 0.5}
\definecolor{Maroon}{rgb}{0.8, 0.0, 0.0}
\definecolor{shadecolor}{gray}{0.9}

\hypersetup{
    linktocpage=true,
    colorlinks=true,
    bookmarks=true,
    citecolor=blue,
    urlcolor=red,
    linkcolor=Maroon,
    citebordercolor={1 0 0},
    urlbordercolor={1 0 0},
    linkbordercolor={.7 .8 .8},
    breaklinks=true,
    pdfpagelabels=true,
    }

\begin{document}
\title{Continuous-time Infinite Dynamic Topic Models}
\author{Wesam Elshamy}
\date{\today}



\thispagestyle{empty}


\pdfbookmark[0]{Title Page}{PDFTitlePage}

\begin{center}

   \vspace{1cm}


   \Large CONTINUOUS-TIME INFINITE DYNAMIC TOPIC MODELS\\

   \vspace{0.5cm}

   by\\

   \vspace{0.5cm}


   WESAM SAMY ELSHAMY\\

   \vspace{0.5cm}


   B.S., Ain Shams University, Egypt, 2004\\
   M.Sc., Cairo University, Egypt, 2007\\

   \vspace{0.65cm}
   \rule{2in}{0.5pt}\\
   \vspace{0.85cm}

   {\Large AN ABSTRACT OF A DISSERTATION}\\

   \vspace{0.5cm}
   submitted in partial fulfillment of the\\
   requirements for the degree\\

   \vspace{0.5cm}


   {\Large DOCTOR OF PHILOSOPHY}\\
   \vspace{0.5cm}


   Department of Computing and Information Sciences\\
   College of Engineering\\

   \vspace{0.5cm}
   {\Large KANSAS STATE UNIVERSITY}\\
   Manhattan, Kansas\\


   2013\\
   \vspace{1cm}

\end{center}

\begin{abstract}
   \setcounter{page}{-1}
   \pdfbookmark[0]{Abstract}{PDFAbstractPage}
   \pagestyle{empty}
\vspace{1cm}
\setlength{\baselineskip}{0.8cm}

Topic models are probabilistic models for discovering topical themes in collections of documents.  In real world applications, these models provide us with the means of organizing what would otherwise be unstructured collections.  They can help us cluster a huge collection into different topics or find a subset of the collection that resembles the topical theme found in an article at hand.

  The first wave of topic models developed were able to discover the prevailing topics in a big collection of documents spanning a period of time.  It was later realized that these time-invariant models were not capable of modeling 1) the time varying number of topics they discover and 2) the time changing structure of these topics.  Few models were developed to address this two deficiencies.  The online-hierarchical Dirichlet process models the documents with a time varying number of topics.  It varies the structure of the topics over time as well.  However, it relies on document order, not timestamps to evolve the model over time.  The continuous-time dynamic topic model evolves topic structure in continuous-time.  However, it uses a fixed number of topics over time.

In this dissertation, I present a model, the continuous-time infinite dynamic topic model, that combines the advantages of these two models 1) the online-hierarchical Dirichlet process, and 2) the continuous-time dynamic topic model.  More specifically, the model I present is a probabilistic topic model that does the following: 1) it changes the number of topics over continuous time, and 2) it changes the topic structure over continuous-time.

I compared the model I developed with the two other models with different setting values.  The results obtained were favorable to my model and showed the need for having a model that has a continuous-time varying number of topics and topic structure.

   \vfill
\end{abstract}


\newpage


\thispagestyle{empty}


\begin{center}

   \vspace{1cm}


   \Large CONTINUOUS-TIME INFINITE DYNAMIC TOPIC MODELS\\
   \vspace{0.5cm}

   by \\

   \vspace{0.5cm}


   WESAM SAMY ELSHAMY\\

   \vspace{0.5cm}


   B.S., Ain Shams University, Egypt, 2004\\
   M.Sc., Cairo University, Egypt, 2007\\

   \vspace{0.65cm}
   \rule{2in}{0.5pt}\\
   \vspace{0.85cm}

   {\Large A DISSERTATION}\\

   \vspace{0.5cm}
   submitted in partial fulfillment of the\\
   requirements for the degree\\

   \vspace{0.5cm}


   {\Large DOCTOR OF PHILOSOPHY}\\
   \vspace{0.5cm}


   Department of Computing and Information Sciences\\
   College of Engineering\\

   \vspace{0.5cm}
   {\Large KANSAS STATE UNIVERSITY}\\
   Manhattan, Kansas\\


   2013\\
   \vspace{1cm}

\end{center}

\begin{flushleft}
   \hspace{10cm}Approved by:\\
   \vspace{ 1cm}
   \hspace{10cm}Major Professor\\


   \hspace{10cm}William Henry Hsu\\
\end{flushleft}




\newpage

\thispagestyle{empty}

\begin{center}

{\bf \Huge Copyright}

\vspace{1cm}


   \Large Wesam Samy Elshamy\\

   \vspace{0.5cm}


   2013\\

   \vspace{0.5cm}

\end{center}

\begin{abstract}
  \pagestyle{empty}
\vspace{1cm}
\setlength{\baselineskip}{0.8cm}

Topic models are probabilistic models for discovering topical themes in collections of documents.  In real world applications, these models provide us with the means of organizing what would otherwise be unstructured collections.  They can help us cluster a huge collection into different topics or find a subset of the collection that resembles the topical theme found in an article at hand.

  The first wave of topic models developed were able to discover the prevailing topics in a big collection of documents spanning a period of time.  It was later realized that these time-invariant models were not capable of modeling 1) the time varying number of topics they discover and 2) the time changing structure of these topics.  Few models were developed to address this two deficiencies.  The online-hierarchical Dirichlet process models the documents with a time varying number of topics.  It varies the structure of the topics over time as well.  However, it relies on document order, not timestamps to evolve the model over time.  The continuous-time dynamic topic model evolves topic structure in continuous-time.  However, it uses a fixed number of topics over time.

In this dissertation, I present a model, the continuous-time infinite dynamic topic model, that combines the advantages of these two models 1) the online-hierarchical Dirichlet process, and 2) the continuous-time dynamic topic model.  More specifically, the model I present is a probabilistic topic model that does the following: 1) it changes the number of topics over continuous time, and 2) it changes the topic structure over continuous-time.

I compared the model I developed with the two other models with different setting values.  The results obtained were favorable to my model and showed the need for having a model that has a continuous-time varying number of topics and topic structure.

  \vfill
\end{abstract}

\newpage
\pagenumbering{roman}


\setcounter{page}{6}

\phantomsection
\addcontentsline{toc}{chapter}{Table of Contents}

\tableofcontents
\listoftables
\listoffigures


\newpage
\begin{center}
{\bf \Huge Acknowledgments}
\end{center}
\vspace{1cm}
\setlength{\baselineskip}{0.8cm}


I am greatly thankful to all those who helped me or game me advice while working on this dissertation and the research that led to it.

Specifically, I would like to thank my advisor William Hsu for guiding and supporting me since I came to Kansas State University over five years ago.  He gave me the right amount of freedom to explore different problems and different approaches to solve them.  His guidance and feedback motivated and inspired me.  I learned a lot from Bill throughout those years.

I wish to thank Tracey Hsu (Bill's wife) for proof reading and editing my dissertation.  She is very attentive to details and my dissertation is in a better shape now thanks to her.

Dr Doina Cragea's advice helped me refine the work I presented in this dissertation.  I would like to thank her for that and for the advice and guidance she gave me while working on an independent study class with her.

I was fortunate to have the help of undergraduate programmers in our group.  Xinghuang Leon Hsu developed the crawler and parser used for crawling the BBC News corpus.  He spent precious hours and days working on it making sure the corpus has what I need.

I enjoyed having good discussions with Surya Teja Kallumadi over the past few years.  We discussed research and non-research problems.  I want to thank him for hosting me in the summer of 2011 in Boston.

Special thanks to my beautiful wife Rachel who believed I will finish my PhD one day.  She did not give up even though that day was a moving target.  Always moving forward.

Finally, I am greatly indebted to my parents.  Without their love and support throughout my life I could not have accomplished this.

\phantomsection
\addcontentsline{toc}{chapter}{Acknowledgements}

\newpage
\pagenumbering{arabic}
\setcounter{page}{1}

\cleardoublepage
\chapter{Introduction}
\label{cha:introduction}

\section{Goal}
\subsection{Problem statement}
In this thesis, I develop a continuous-time dynamic topic model.  This model is an extension of the latent Dirichlet allocation (LDA) which is atemporal.  I add two temporal components to: it i) the word distribution per topic, and ii) the number of topics.  Both evolve in continuous time.

There exists similar temporal dynamic topic models.  \citet{ahmed10} presented a model where the word distribution per topic and the number of topics evolve in discrete time.  Bayesian inference using Markov chain Monte Carlo techniques in this model is feasible when the time step is big enough.  It was efficiently applied to a system with 13 time steps \cite{ahmed10}.  Increasing the time granularity of the model dramatically increases the number of latent variables making inference prohibitively expensive.  On the other hand, \citet{wang08} presented a model where the word distribution per topic evolve in continuous time.  They used variational Bayes methods for fast inference.  A big limitation of their model is that it uses a predefined and fixed number of topics that does not evolve over time.  In this thesis I develop and use a topic model that is a mixture of these two models where word distribution per topic and the number of topics evolve in continuous-time.

The need for the dynamic continuous-time topic model I am developing is evident in the mass media business where news stories are published round-the-clock\footnote{Agence France-Presse (AFP) releases on average one news story every 20 seconds (5000 per day) while Reuters releases 800,000 English-language news stories annually.  See: \texttt{http://www.afp.com/en/agency/afp-in-numbers} and \texttt{http://www.infotoday.com/it/apr01/news6.htm}}.  To provide the news reader with a broad context and rich reading experience, many news outlets provide a list of related stories from news archives to the ones they present.  As these archives are typically huge, manual search for these related stories is infeasible.  Having the archived stories categorized into geographical areas or news topics such as politics and sports may help a little;  a news editor can comb through a category looking for relevant stories.  However, relevant stories may cross category boundaries.  Keyword searching the entire archive may not be effective either; it returns stories based on keyword mentions not the topics the stories cover.  A dynamic continuous-time topic model can be efficiently used to find relevant stories.  It can track a topic over time even as a story develops and the word distribution associated with its topic evolves.  It can detect the number of topics discussed in the news over time and fine-tune the model accordingly.

\begin{shaded}
  \paragraph{Goal:} To develop a continuous-time topic model in which the word distribution per topic and the number of topics evolve in continuous time.
\end{shaded}

\begin{figure}
  \centering
  \includegraphics[width=0.8\linewidth]{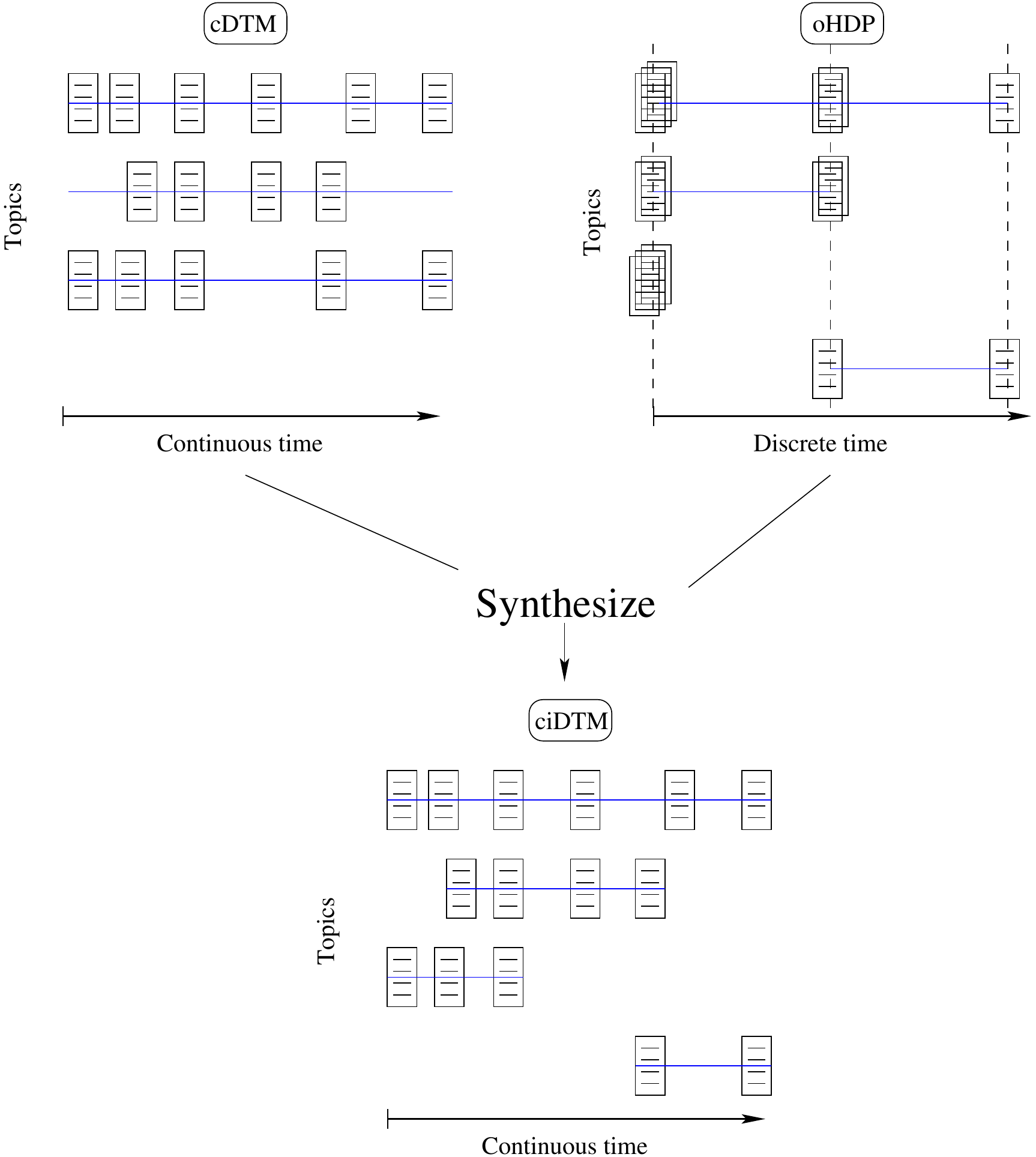}
  \caption{Top left: the continuous-time dynamic topic model (cDTM) has a continuous-time domain; word and topic distributions evolve in continuous time.  The number of topics in this model is fixed though.  This may lead to having two separate topics being merged into one topic which was the case in the first topic from below.  Top right: the online hierarchical Dirichlet process (oHDP) based topic model evolves the number of topics over time.  The five documents belonging to the first and second topics from below were associated with their respective topics and were not merged into one topic which was the case with cDTM on the left.  However, the model has a discrete-time domain which practically limits the degree of time granularity that we can use;  very small time steps will make the model prohibitively expensive to inference.  Bottom: The continuous-time infinite dynamic topic model (ciDTM) is a mixture of the oHDP and cDTM.  It has a continuous-time domain like cDTM, and its number of topics evolve over time as in oHDP.  It overcomes limitations of both models regarding evolution of the number of topics and time-step granularity.}
  \label{fig:cidtm_block}
\end{figure}

\subsection{Central thesis}
A continuous-time topic model with an evolving number of topics and a dynamic word distribution per topic (ciDTM) can be built using a Wiener process to model the dynamic topics in a hierarchical Dirichlet process \cite{TehJorBea2006}.  This model cannot be efficiently emulated by the discrete-time infinite dynamic topic model when the topics it models evolve at different speeds.  Using a fine time granularity to fit the fastest evolving topic would make inference in this system impractical.  On the other hand, ciDTM cannot be emulated by a continuous-time dynamic topic model as it would require the use of a fixed number of topics which is a model parameter.  Apart from the problem of finding an appropriate value for the parameter to start with, this value remains fixed over time leading to the potential problem of having multiple topics merge into one topic or having one topic split into multiple topics to keep the overall number of topics fixed.

For the problem of creating news timelines, ciDTM is more efficient and less computationally expensive than using a discrete-time unbounded-topic model \cite{ahmed10} with a very small time unit (epoch), and more accurate than using a continuous-time bounded-topic model \cite{wang08}.  This is illustrated in Figure~\ref{fig:cidtm_block}.

\subsection{Technical objectives}
A news timeline is characterized by a mixture of topics.  When a news story arrives to the system, it should be placed on the appropriate timeline, if such one exists, or a new timeline will be created.  A set is created of related stories that may not belong to the same timeline as the arriving story.  In this process, care should be taken to avoid some pitfalls:  If the dynamic number of topics generated by the model becomes too small, some news timelines may get conflated and distances between news stories may get distorted.  If the dynamic number of topics becomes too large then number of variational parameters of the model will explode and the inference algorithm will become prohibitively expensive to run.

\begin{figure}
  \centering
  \includegraphics[width=\linewidth]{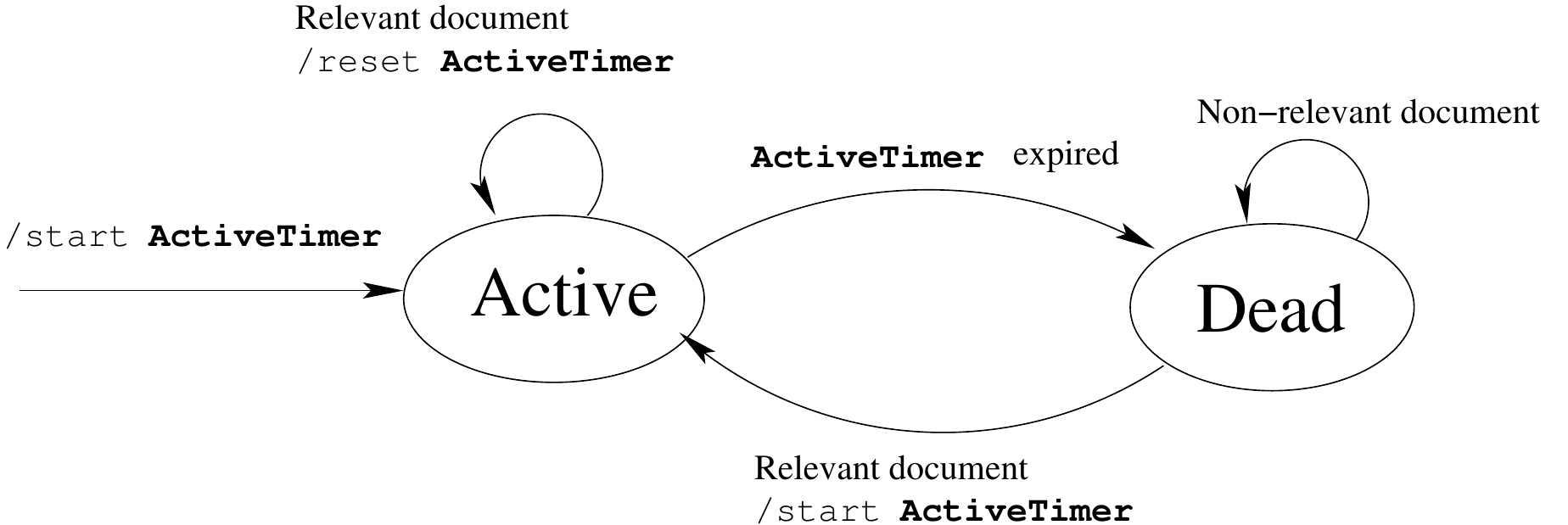}
  \caption{Topic state-transition diagram:  Oval shaped nodes represent states with their names typed inside of them.  Arrows represent transitions from one state to another, or to itself.  Each arrow has an \texttt{Event / Action} describing the event that triggered the transition, and the action taken by the system in response to that transition.  The start state is represented by an arrow with no origin pointing to the state.}
  \label{fig:topic_state_transition}
\end{figure}

In a topic model system that receives text stream, a topic has two dormancy states: \emph{\texttt{Active}} and \emph{\texttt{Dead}}.  Transition between these two states is illustrated in a state-transition diagram in Figure~\ref{fig:topic_state_transition}.  When a new topic is born, it enters the \texttt{Active} state and a timer (\texttt{ActiveTimer}) starts.  The timer is reset whenever a document relevant to the topic is received and processed by the system.  When the timer expires, the topic transitions to the \texttt{Dead} state.  The topic remains in this state as long as the system receives and processes non-relevant documents.  If a relevant document is received and processed by the system, the topic transitions to the \texttt{Active} state, and the timer (\texttt{ActiveTimer}) is started.

\section{Existing approaches: atemporal models}
The volume of digitized knowledge has been increasing with an unprecedented rate recently.  The past decade has witnessed the rise of projects like JSTOR \cite{taylor01}, Google Books \cite{darnton09}, the Internet Archive \cite{green02} and Europeana \cite{gradmann09} that scan and index printed books, documents, newspapers, photos, paintings and maps.  The volume of data that is born in digital format is even more rapidly increasing.  Whether it is mainly unstructured user-generated content such as in social media, or generated by news portals.

New innovative ways were developed for categorizing, searching and presenting the digital material \cite{cafarella04, giles98, ley02}.  The usefulness of these services to the end user, which are free in most cases, is evident by their ever increasing utilization in learning, entertainment and decision-making.

A document collection can be indexed and keyword search can be used to search for a document in it.  However, most keyword search algorithms are context-free and lack the ability to categorize documents based on their topic or find documents related to one of interest to us.  The large volume of document collections precludes manual techniques of annotation or categorization of the text.

The process of finding a set of documents related to a document at hand can be automated using topic models \cite{blei11}.  Most of these existing models are atemporal and perform badly in finding old relevant stories because the word collections that identifies the topics that these stories cover change over time and the model does not account for that change.  Moreover, these models assume the number of topics is fixed over time, whereas in reality this number is constantly changing.  Some famous examples of these models that were widely used in the past are latent semantic analysis (LSA) \cite{lsa98}, probabilistic LSA (pLSA) \cite{plsa99} and latent Dirichlet allocation (LDA) \cite{blei03}.

\subsection{Latent semantic analysis}
Latent semantic analysis (LSA) is a deterministic technique for analyzing the relationship between documents or between words \cite{lsa98}.  Like most topic models it treats a document as a bag of words; a document is a vector of term frequency ignoring term sequence.  It is built on the idea that words that tend to appear in the same document together tend to carry similar meanings.  In this document analysis technique, which is sometimes known as latent semantic indexing (LSI), the very large, noisy and sparse term-document frequency matrix gets its rank lowered using matrix factorization techniques such as singular value decomposition.  In the lower-rank approximation matrix, each document is represented by a number of concepts in lieu of terms.

LSA is known for its inability to distinguish the different meanings of a polysemy word and place synonym words closer together in the semantic space;  Different occurrences of a polysemous word could have different meanings based on its context.  However, since LSA treats each word as a point in space, that point will be the semantic average of the different meanings of the word in the document.  Different words with the same meaning (synonyms) represent another challenge to LSA in information retrieval applications.  If a document is made out of the synonyms of terms found in another document, the correlation between these two documents over their terms using LSA will not be as high as we would expect it to be.

\subsection{Probabilistic latent semantic analysis}
To overcome some of the LSA limitations, we can condition the occurrence of a word with a document on a latent topic variable.  These will lead us to the probabilistic latent semantic analysis (pLSA).  This model overcomes the discrepancy of the assumed joint Gaussian distribution between terms and documents in LSA and the observed Poisson distribution between them.  However, pLSA is not a proper generative model for new documents, and the number of parameters in its model grows linearly with the number of documents.  To overcome the generative process issue, a Dirichlet prior can be used for the document topic distribution, leading us to the latent Dirichlet allocation, which is the basis for topic models.

\subsection{Latent Dirichlet allocation and topic models}
Topic models are probabilistic models that capture the underlying semantic structure of a document collection based on a hierarchical Bayesian analysis of the original text \cite{blei03, blei09}.  By discovering word patterns in the collection and establishing a pattern similarity measure, similar documents can be linked together and semantically categorized based on their topic.

\begin{figure}
  \centering
  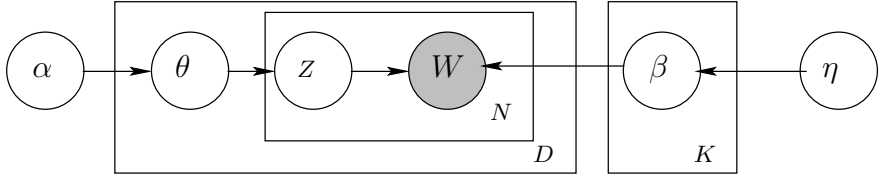  
  \caption{Latent Dirichlet allocation (LDA) graphical model.  The rectangles, known as plates in this notation, represent replication.  Plate with multiplicity $D$ denotes a collection of documents.  Each of these documents is made of a collection of words represented by plate with multiplicity $N$.  Plate with multiplicity $K$ represents a collection of topics.  The shaded node $W$ represents a word which is the only observed random variable in the model.  The non-shaded nodes represent latent random variables in the model;  $\alpha$ is the Dirichlet prior parameter for topic distribution per document.  $\theta$ is a topic distribution for a document, while $Z$ is the topic sampled from $\theta$ for word $W$.  $\beta$ is a Markov matrix giving the word distribution per topic, and $\eta$ is the Dirichlet prior parameter used in generating that matrix.  Shaded nodes are observed random variables, while non-shaded nodes are latent random variables.  Nodes are labeled with the variable they represent.  The rectangles are known as plates in this notation and they represent replication.  For nested plates, an inner plate has multiplicity equal to the replication value shown on its lower-right corner times the multiplicity of its parent.}
  \label{fig:lda_graphical_model}
\end{figure}

A graphical model for the latent Dirichlet allocation is shown in Figure~\ref{fig:lda_graphical_model} \citep{blei11}.  The rectangles, known as plates in this notation, represent replication.  Plate with multiplicity $D$ denotes a collection of documents.  Each of these documents is made of a collection of words represented by plate with multiplicity $N$.  Plate with multiplicity $K$ represents a collection of topics.  The shaded node $W$ represents a word which is the only observed random variable in the model.  The non-shaded nodes represent latent random variables in the model;  $\alpha$ is the Dirichlet prior parameter for topic distribution per document.  $\theta$ is a topic distribution for a document, while $Z$ is the topic sampled from $\theta$ for word $W$.  $\beta$ is a Markov matrix giving the word distribution per topic, and $\eta$ is the Dirichlet prior parameter used in generating that matrix.

The reason for choosing the Dirichlet distribution to model the distribution of topics in a document and to model the distribution of words in a topic \cite{buntine09estimating} is because the Dirichlet distribution is convenient distribution on the simplex \cite{blei03}, it is in the exponential family \cite{wainwright08}, has finite dimensional sufficient statistics \cite{buntineJ05}, and is conjugate to the multinomial distribution \cite{rice2001}.  These properties help developing an inference procedure and parameter estimation as pointed out by \citet{blei03}.

\begin{figure}
  \centering
  \includegraphics[angle=90, width=0.5\textheight]{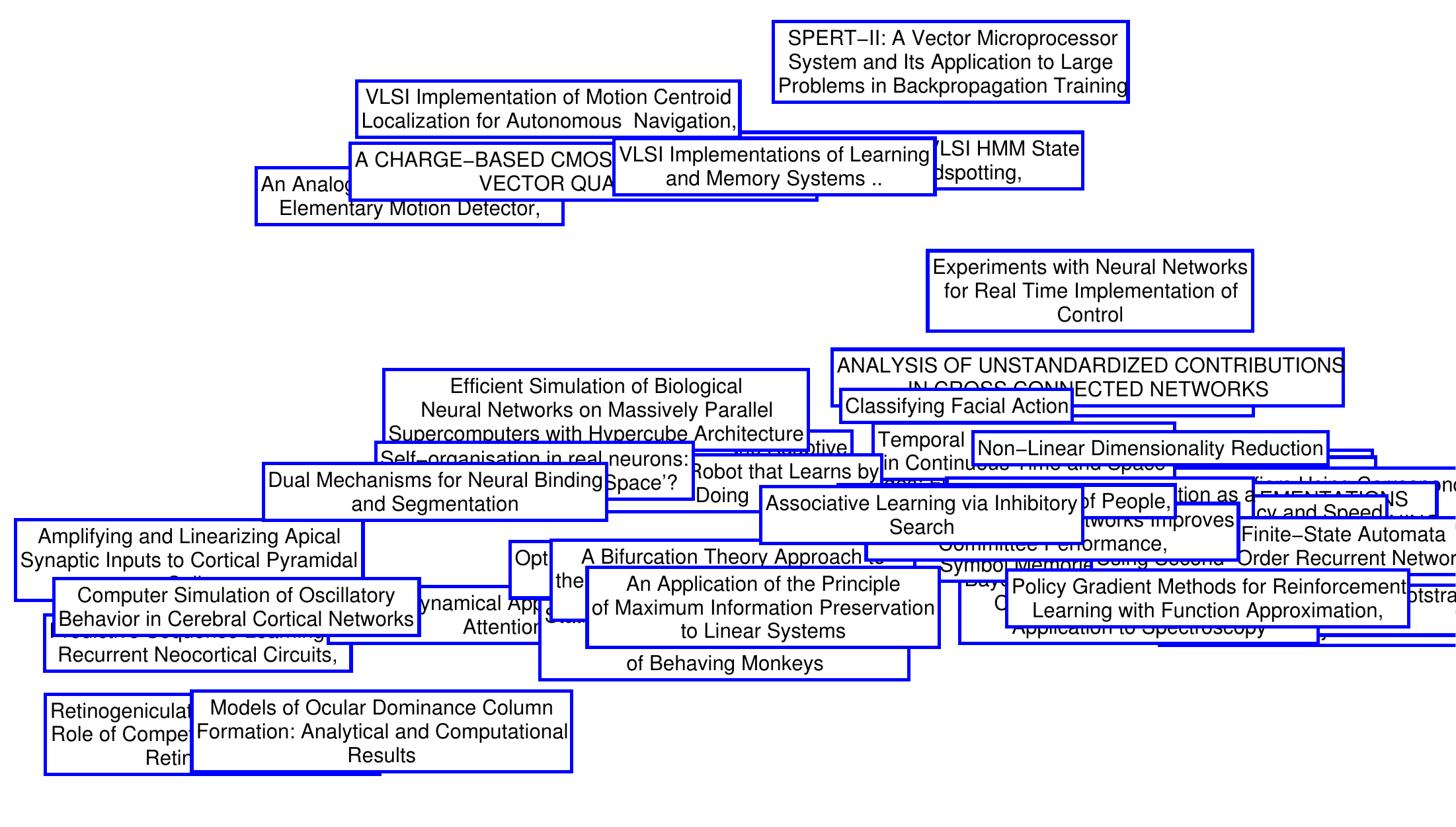}  
  \caption{Visualization of a subset of 50 documents in two dimensions from the NIPS dataset.  Each rectangle represents a document with its title inside of the rectangle.  Distance between two documents is directly proportional to their similarity.  The symmetrized Kullback-Leibler divergence is used to evaluate the distance between pairs of documents.}
  \label{fig:lda_nips_docs}
\end{figure}

\section{Significance}
Even though news tracker and news aggregator systems have been used for a few years at a commercial scale for web news portals and news websites, most of them only provide relevant stories from the near past.  It is understood that this is done to limit the number of relevant stories but this at the same time casts doubt over the performance of these systems when they try to dig for related stories that are a few years old and therefore have different word distributions for the same topic.

Topic models not only help us automate the process of text categorization and search, they enable us to analyze text in a way that cannot be done manually.  Using topic models we can see how topics evolve over time \cite{ahmed10}, and how different topics are correlated with each other \cite{blei05}, and how this correlation changes over time.  We can project the documents on the topic space and take a bird's eye view to understand and analyze their distribution.  We can zoom-in to analyze the main themes of a topic, or zoom-out to get a broader view of where this topic sits among other related topics.  It should be noted that topic modeling algorithms do all that with no prior knowledge of the existence or the composition of any topic, and without text annotation.
 
The successful utilization of topic modeling in text encouraged researchers to explore other domains of applications for it.  It has been used in software analysis \cite{linstead09} as it can be used to automatically measure, monitor and better understand software content, complexity and temporal evolution.  Topic models were used to ``improve the classification of protein-protein interactions by condensing lexical knowledge available in unannotated biomedical text into a semantically-informed kernel smoothing matrix'' \cite{polajnar09}.  In the field of signal processing, topic models were used in audio scene understanding \cite{kim09}, where audio signals were assumed to contain latent topics that generate acoustic words describing an audio scene.  It has been used in text summarization \cite{das09}, building semantic question answering systems \cite{celi09}, stock market modeling \cite{doyle09} and music analysis \cite{hu09}.

Using variational methods for approximate Bayesian inference in the developed hierarchical Dirichlet allocation model for the dynamic continuous-time topic model will facilitate inference in models with a higher number of latent topic variables.

Other than the obvious application for the timeline creation system in retrieving a set of documents relevant to a document at hand, topic models can be used as a visualization technique.  The user can view the timeline at different scale levels and understand how different events temporally unfolded.

\cleardoublepage
\chapter{Bayesian models and inference algorithms}
Evaluation of the posterior distribution $p(\mathbf{Z|X})$ of the set of latent variables $\mathbf{Z}$ given the observed data variable set $\mathbf{X}$ is essential in topic modeling applications \cite{bishopPattern}.  This evaluation is infeasible in real-life applications of practical interest due to the high number of latent variables we need to deal with \cite{breslow93}.  For example, the time complexity of the junction tree algorithm is exponential in the size of the maximal clique in the junction tree.  Expectation evaluation with respect to such a highly complex posterior distribution would be analytically intractable \cite{wainwright08}.

Just like the case with many engineering problems when finding an exact solution is not possible or too expensive to obtain, we resort to approximate methods.  Even in some cases when the exact solution can be obtained, we might favor an approximate solution because the benefit of reaching the exact solution does not justify the extra cost spent to reach it.  When the nodes or node clusters of the graphical model are almost conditionally independent, or when the node probabilities can be determined by a subset of its neighbors, an approximate solution will suffice for all practical purposes \cite{murphy99}.  Approximate inference methods fall broadly under two categories, stochastic and deterministic \cite{koller1999general}.

Stochastic methods, such as Markov Chain Monte Carlo (MCMC) methods, can theoretically reach exact solution in limited time given unlimited computational power \cite{gilks2005markov}.  In practice, the quality of the approximation obtained is tied to the available computational power.  Even though these methods are easy to implement and therefore widely used, they are computationally expensive.

Deterministic approximation methods, like variational methods, make simplifying assumptions on the form of the posterior distribution or the way the nodes of the graphical model can be factorized \cite{jordan1999introduction}.  These methods therefore could never reach an exact solution even with unlimited computational resources.

\section{Introduction}
As I stated earlier, the main problem in graphical model applications is finding an approximation for the posterior distribution $p(\mathbf{Z}|\mathbf{X})$ and the model evidence $p(\mathbf{X})$, where $\mathbf{X}=\{\mathbf{x_1,\ldots,x_N}\}$ is the set of all observed variables and $\mathbf{Z}=\{\mathbf{z_1,\ldots,z_N}\}$ is the set of all latent variables and model parameters.  $p(\mathbf{X})$ can be decomposed using \cite{bishopPattern}:

\begin{equation}
  \ln p(\mathbf{X}) = \mathcal{L}(q) + \text{KL}(q||p)
\end{equation}

where

\begin{gather}
  \mathcal{L}(q) = \int q(\mathbf{Z})\ln\left\{\frac{p(\mathbf{X,Z})}{q(\mathbf{Z})}\right\}d\mathbf{Z} \label{varLowerBound}\\
  \text{KL}(q||p) = - \int q(\mathbf{Z})\ln\left\{\frac{p(\mathbf{Z|X})}{q(\mathbf{Z})}\right\}d\mathbf{Z}
\end{gather}

Our goal is to find a distribution $q(\mathcal{Z})$ that is as close as possible to $p(\mathbf{Z|X})$.  To do so, we need to minimize the Kullback-Leibler (KL) distance between them \cite{kullback1951information}.  Minimizing this measure while keeping the left-hand side value fixed means maximizing the lower bound $\mathcal{L}(q)$ on the log marginal probability.  The approximation in finding such a distribution arises from the set of restrictions we put on the family of distributions we pick $q$ from.  This family of distributions has to be rich enough to allow us to include a distribution that is close enough to our target posterior distribution, yet the distributions have to be tractable \cite{bishopPattern}.

This problem can be transformed into a non-linear optimization problem if we use a parametric distribution $p(\mathbf{Z}|\omega)$, where $\omega$ is its set of parameters.  $\mathcal{L}(q)$ becomes a function of $\omega$ and the problem can be used using non-linear optimization methods such as Newton or quasi-Newton methods \cite{shanno1970conditioning}.

\subsection{Factorized distributions}
Instead of restricting the form of the family of distributions we want to pick $q(\mathbf{Z})$ from, we can make assumptions on the way it can be factored.  We can make some independence assumptions \cite{wainwright08}.  Let us say that our set of latent variables $\mathbf{Z}$ can be factorized according to \cite{jordan1999introduction}:

\begin{equation}
  q(\mathbf{Z}) = \prod_{i=1}^Mq_i(\mathbf{Z}_i) \label{factDistro}
\end{equation}

This approximation method is known in the physics domain by \emph{mean field theory} \cite{barabasi1999mean}.

To maximize the lower bound $\mathcal{L}(q)$ we need to minimize each one of the factors $q_i\mathbf{Z}_i$.  We can do so by substituting \eqref{factDistro} into \eqref{varLowerBound} to get the following \cite{bishopPattern}:

\
\begin{align}
  \mathcal{L}(q) &= \int \prod_i q_i \left\{ \ln p(\mathbf{X,Z}) - \sum_i\ln q_i \right\} d\mathbf{Z}\\
                 &= \int q_j \left\{\int \ln p(\mathbf{X,Q}) \prod_{i\ne j}q_i d\mathbf{Z}_i \right\} - \int q_j \ln q_j d\mathbf{Z}_j + c\\
                 &= \int q_j \ln \tilde{p}(\mathbf{X,Z}_j)d\mathbf{Z}j -  \int q_j \ln q_j d\mathbf{Z}_j + c \label{negativeKL}
\end{align}

where

\begin{equation}
  \ln\tilde{p}(\mathbf{X,Z}_j) = \mathbb{E}_{i\ne j} [\ln p(\mathbf{X,Q})] + c
\end{equation}

and

\begin{equation}
  \mathbb{E}_{i\ne j}[\ln p(\mathbf{X,Z})] = \int \ln p(\mathbf{X,Z})\prod_{i\ne j} q_i d\mathbf{Z}_i
\end{equation}

Where $\mathbf{E}_{i\ne j}$ is the expectation with respect to $q$ over $z_i$ such that $i\ne j$.

Since \eqref{negativeKL} is a negative KL divergence between $q_j(Z_j)$ and $\tilde{p}(\mathbf{X,Z}_j)$, we can maximize $\mathcal{L}(q)$ with respect to $q_j(\mathbf{Z}_j)$ and obtain \cite{bishopPattern}:

\begin{equation}
  \label{eq:var_fact}
  \ln q_j^{\star}(\mathbf{Z}_j) = \mathbb{E}_{i\ne j} [\ln p(\mathbf{X,Z})] + c
\end{equation}

To get rid of the constant we can take the exponential of both sides of this equation to get:

\begin{equation}
  q_j^{\star}(\mathbf{Z}_j) = \frac{\exp(\mathbb{E}_{i\ne j}[\ln p(\mathbf{X,Z})])}{\int \exp(\mathbb{E}_{i\ne j}[\ln p(\mathbf{X,Q})])d\mathbf{Z}_j}
\end{equation}

\subsubsection{Example}
For illustration, we are going to take a simple example for the use of factorized distribution in the variational approximation of parameters of a simple distribution \cite{mackay2003}.  Let us take the univariate Gaussian distribution over $x$.  In this example, we will infer the posterior distribution of its mean ($\mu$) and precision ($\tau=\sigma^{-2}$), given a data set $\mathcal{D}=\{x_1,\ldots,x_N\}$ of i.i.d. observed values with a likelihood:

\begin{equation}
  \label{eq:data_likelihood}
  p(\mathcal{D}|\mu,\tau) = \left(\frac{\tau}{2\pi} \right)^{N/2} 
                  \exp \left\{ -\frac{\tau}{2}\sum_{n=1}^N (x_n - \mu)^2 \right\}
\end{equation}

with a Gaussian and Gamma conjugate prior distributions for $\mu$ and $\tau$ as follows:

\begin{align}
  \label{eq:gaussian_priors}
  p(\mu|\tau) &= \mathcal{N}(\mu|\mu_0, (\lambda_0\tau)^{-1})\\
  p(\tau) &= \text{Gamma}(\tau|a_0,b_0)
\end{align}

By using factorized variational approximation, we can assume the posterior distribution factorizes according to:

\begin{equation}
  \label{eq:gaussian_fact}
  q(\mu,\tau) = q_{\mu}(\mu)q_{\tau}(\tau)
\end{equation}

We can evaluate the optimum for each of the factors using \eqref{eq:data_likelihood}.  The optimal factor for the mean is:

\begin{equation}
  \ln q^{\star}_{\mu}(\mu) = -\frac{\mathbb{E}_{\tau}}{2} \left\{ \lambda_0(\mu - \mu_0)^2 + \sum_{n=1}^N(x_n-\mu)^2 \right\} + c
\end{equation}

Which is the Gaussian $\mathcal{N}(\mu|\mu_N,\lambda_N^{-1})$ with mean and precision given by:

\begin{align}
  \mu_N     &= \frac{\lambda_0 \mu_0 + N\bar{x}}{\lambda_0 + N}\\
  \lambda_N &= (\lambda_0 + N) \mathbb{E}[\tau] \label{eq:gaussian_mu}
\end{align}

And the optimal factor for the precision is:

\begin{equation}
  \ln q^{\star}_{\tau}(\tau) = (a_0 - 1 + \frac{N}{2}) \ln \tau - b_0\tau
      -\frac{\tau}{2}\mathbb{E}_{\mu}\left[ \sum_{n=1}^N(x_n-\mu)^2 + \lambda_0(\mu - \mu_0)^2\right] + c
\end{equation}

Which is a gamma distribution $\text{Gamma}(\tau|a_N,b_N)$ with parameters:

\begin{align}
  a_N &= a_0 + \frac{N}{2}\\
  b_N &= b_0 + \frac{1}{2}\mathbb{E}_{\mu}\left[ \sum_{n=1}^N (x_n-\mu)^2 + \lambda_0(\mu-\mu_0)2 \right] \label{eq:gaussian_tau}
\end{align}

As we can see from \eqref{eq:gaussian_mu} and \eqref{eq:gaussian_tau}, the optimal distributions for the mean and precision factors depend on expectations function of the other variable.  Their values can be evaluated iteratively by first assuming an initial value, lets say for $\mathbb{E}[\tau]$ and use this value to evaluate $q_{\mu}(\mu)$.  Given this value, we can use it to extract $\mathbb{E}[\mu]$ and $\mathbb{E}[\mu^2]$ and use them to evaluate $q_{\tau}(\tau)$.  This value can in turn be used to extract a new revised value for $\mathbb{E}_{\tau}$ which can be used to update $q_{\mu}(\mu)$ and continue the cycle until their values converge to the optimum values.

\subsection{Variational parameters}
Variational methods transform a complex problem into a simpler form by decoupling the degrees of freedom of this problem by adding variational parameters.  For example, we can transform the logarithm function as follows \cite{jordan1999introduction}:
\begin{equation}
  \log(x)=\min_{\lambda}\{\lambda x - \log \lambda - 1\}
\end{equation}

Here I introduced the variational parameter $\lambda$  Which we are trying its value that minimizes the function.

\begin{figure}
  \centering
  \resizebox{0.9\linewidth}{!}{\input{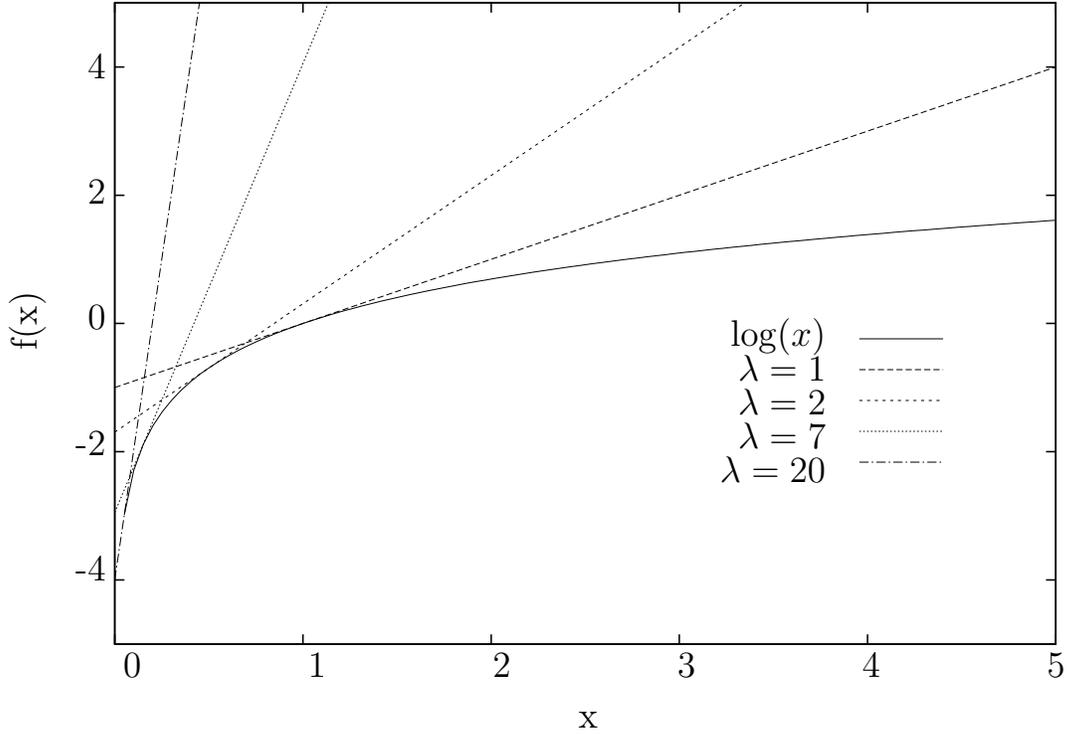}}
  \caption{Variational transformation of the logarithmic function.  The transformed function ($\lambda x - \log \lambda - 1$) forms an upper found for the original function.\label{var_log}}
\end{figure}

As we can see in Figure~\ref{var_log}, for different values of $\lambda$, there is a tangent line to the concave logarithmic function, and the set of lines formed by varying over the values of $\lambda$ forms a family of upper bounds for the logarithmic function.  Therefore,
\begin{equation}
  \log(x) \le \lambda x - \log \lambda - 1 \qquad \forall \lambda
\end{equation}

\section{Variational inference}
In this section, I use variational inference to find an approximation to the true posterior of the latent topic structure \cite{xing2002generalized}; The topic distribution per word, the topic distribution per document, and the word distribution over topics.

I use variational Kalman filtering \citep{kalman1960} in continuous time for this problem.  The variational distribution over the latent variables can be factorized as follows:

\begin{align}\nonumber
  \label{eq:var_factorization}
  &q(\beta_{1:T, z_{1:T}, 1:N}, \theta_{1:T} | \hat{\beta},\phi,\gamma) =\\\nonumber
      &\qquad\qquad\prod_{k=1}^{K}q(\beta_{1,k},\ldots,\beta_{T,k}|\hat{\beta}_{1,k},\ldots,\hat{\beta}_{T,k}) \times \\
      &\qquad\qquad \prod_{t=1}^T{\left( q(\theta_t|\gamma_t) \prod_{n=1}^{N_t} q(z_{t,n}|\phi_{t,n}) \right)}
\end{align}

Where $\beta$ is the word distribution over topics and $\beta_{1:T, z_{1:T}, 1:N}$ is the word distribution over topics for time $1:T$, topic $z_{1:T}$ and word index $1:N$, where $N$ is the size of the dictionary.

In equation \eqref{eq:var_factorization}, $\gamma_t$ is a Dirichlet parameter at time $t$ for the multinomial per document topic distribution $\theta_t$, and $\phi_{t,n}$ is a multinomial parameter at time $t$ for word $n$ for the topic $z_{t,n}$.  $\{\hat{\beta}_{1,k},\ldots,\hat{\beta}_{T,K}\}$ are Gaussian variational observations for the Kalman filter \cite{wang08}.

In discrete-time topic models, a topic at time $t$ is represented by a distribution over all terms in the dictionary including terms not observed at that time instance.  This leads to high memory requirements specially when the time granularity gets finer.  In my model, I use sparse variational inference \cite{girolami2001variational} in which a topic at time $t$ is represented by a multinomial distribution over terms observed at that time instance;  variational observations are only made for observed words.  The probability of the variational observation $\hat{\beta}_{t,w}$ given $\beta_{t,w}$ is Gaussian \cite{wang08}:

\begin{equation}
  \label{eq:var_forward_main}
  p(\hat{\beta}_{t,w} | \beta_{t,w}) = \mathcal{N}(\beta_{t,w},\hat{v}_t)
\end{equation}

I use the forward-backward algorithm \citep{rabiner1993} for inference for the sparse variational Kalman filter.  The variational forward distribution $p(\beta_{t,w}|\hat{\beta}_{1:t,w})$ is Gaussian \cite{blei06}:

\begin{equation}
  p(\beta_{t,w} | \hat{\beta}_{1:t,w}) = \mathcal{N}(m_{t,w},V_{t,w})
\end{equation}

where

\begin{align}
  \label{eq:var_forward_mean}
  m_{t,w} &= \mathbb{E}(\beta_{t,w}|\hat{\beta}_{1:t,w})\nonumber\\
         &= \left(\frac{\hat{v}_t}{V_{t-1,w} + v\Delta_{s_t} + \hat{v}_t}\right)m_{t-1,w}\nonumber\\
         &\quad+ \left( 1 - \frac{\hat{v}_t}{V_{t-1,w} + v\Delta_{s_t} + \hat{v}_t}\right)\hat{\beta}_{t,w}\\
  V_{t,w} &= \mathbb{E}((\beta_{t,w}-m_{t,w})^2|\hat{\beta}_{1:t,w})\nonumber\\
         &= \left( \frac{\hat{v}_t}{V_{t-1,w} + v\Delta_{s_t} + \hat{v}_t} \right)(V_{t-1,w} + v\Delta_{s_t})
\end{align}

Similarly, the backward distribution $p(\beta_{t,w}|\hat{\beta}_{1:T,w})$ is Gaussian \cite{blei06}:

\begin{equation}
  \label{eq:var_back_main}
  p(\beta_{t,w}|\hat{\beta}_{1:T,w}) = \mathcal{N}(m_{t,w},V_{t,w})
\end{equation}

where

\begin{align}
  \label{eq:var_back_mean}
  \widetilde{m}_{t-1} &= \mathbb{E}(\beta_{t-1}|\hat{\beta}_{1:T})\nonumber\\
                &= \left(\frac{v\Delta_{s_t}}{V_{t-1,w} + v\Delta_{s_t}} \right)m_{t-1}\nonumber\\
                &\quad+ \left( 1 - \frac{v\Delta_{s_t}}{V_{t-1,w} + v\Delta_{s_t}} \right) \widetilde{m}_{t,w}\\
  \widetilde{V}_{t-1,w} &= \mathbb{E}((\beta_{t-1} - \widetilde{m}_{t-1})^2|\hat{\beta}_{1:T})\nonumber\\
                      &= V_{t-1,w} + \left( \frac{V_{t-1}}{V_{t-1} + v\Delta_{s_t}}\right)^2 \left( \widetilde{V}_t - V_{t-1} - v\Delta_{s_t} \right)
\end{align}

The likelihood of the observations has a lower bound defined by:

\begin{align}
  \label{eq:var_lower_obs}
  \mathcal{L}(\hat{\beta}) \ge &\sum_{t=1}^T \mathbb{E}_q\left[ \log p(w_t|\beta_t) - \log p(\hat{\beta}_t|\beta_t) \right]\nonumber\\
                               &+\sum_{t=1}^T\log q(\hat{\beta}_t|\hat{\beta}_{1:t-1})
\end{align}

where

\begin{align}
  \label{eq:6}
  \mathbb{E}_q\log p(w_t|\beta_t)    =&\sum_{w}n_{t,w}\mathbb{E}_q\left( \beta_{t,w} - \log\sum_w\exp(\beta_{t,w}) \right)\nonumber\\
                                  \ge&\sum_{w}n_{t,w}\widetilde{m}_{t,w}\nonumber\\
                                     &-n_t \log \sum_w\exp(\widetilde{m}_{t,w} + \widetilde{V}_{t,w}/2)\\
  \mathbb{E}_q\log p(\hat{\beta}_t|\beta_t) &= \sum_w \delta_{t,w}\mathbb{E}_q\log q(\hat{\beta}_{t,w}|\beta_{t,w})\\
  \log q(\hat{\beta}_t|\hat{\beta}_{1:t-1})  &= \sum_{w} \delta_{t,w} \log q(\hat{\beta}_{t,w}|\hat{\beta}_{1:t-1,w})
\end{align}

where $\delta_{t,w}$ is the Dirac delta function \cite{dirac1992principles} and it is equal to 1 iff $\hat{\beta}_{t,w}$ is in the variational observations.  $n_{t,w}$ is the number of words in document $d_t$, and $n_t=\sum_wn_{t,w}$.


\cleardoublepage
\chapter{Existing solutions and their limitations}
Several studies have been done to account for the changing latent variables of the topic model.  \citet{xing05} presented a dynamic logistic-normal-multinomial and logistic-normal-Poisson models that he used later as building blocks for his models.  \citet{wang06} presented a non-Markovian continuous-time topic model in which each topic is associated with a continuous distribution over timestamps.  \citet{blei06} proposed a dynamic topic model in which the topic's word distribution and popularity are linked over time, though the number of topics was fixed.  This work was picked up by other researchers who extended this model.  In the following I describe some of these extended models.

\section{Temporal topic models}
Traditional topic models which are manifestations of graphical models model the occurrence and co-occurrence of words in documents disregarding the fact that many document collections cover a long period of time.  Over this long period of time, topics which are distributions over words could change drastically.  The word distribution for a topic covering a branch of medicine is a good example of a topic that is dynamic and evolves quickly over time.  The terms used in medical journals and publications change over time as the field develops.  Learning the word distribution for such a topic using a collection of old medical documents would not be good in classifying new medical documents as the new documents are written using more modern terms that reflect recent medical research directions and discoveries that keep advancing every day.  Using a fixed word distribution for such a topic, and for many other topics that usually evolve in time, would result in wrong document classification and wrong topic inference.  The error could become greater over the course of time as the topic evolves more and more and the distance between the original word distribution that was learned using the old collection of documents and the word distribution of the same topic in a more recent document collection for the same field becomes greater.  Therefore, there is a strong need to add a temporal model to such topic models to reflect the changing word distribution for topics in a collection of documents to improve topic inference in a dynamic collection of documents.

\section{Topics over time}
\begin{figure}
  \centering
  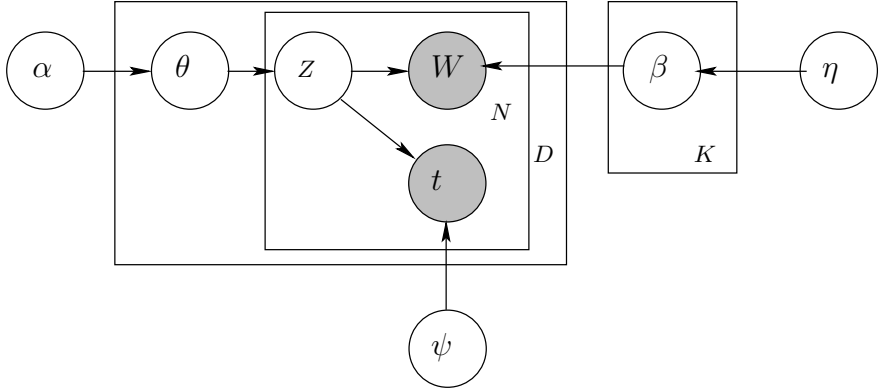  
  \caption{Topics over time graphical model.  In this view, one timestamp $t$ is sampled from a Beta distribution per word in a document.  All words in a single document have the same timestamp.  All the symbols in this figure follow the notation given in Figure~\ref{fig:lda_graphical_model} and $\psi$ is the Beta distribution prior.}
  \label{fig:tot_gm}
\end{figure}

Several topic models were suggested that add a temporal component to the model.  I will refer to them in this dissertation as temporal topic models.  These temporal topic models include the Topics over Time (TOT) topic model \cite{wang06}.  This model directly observes document timestamp.  In its generative process, the model generates word collection and a timestamp for each document.  In their original paper, \citeauthor{wang06} gave two alternatives views for the model.  The first one has a generative process in which for each document a multinomial distribution $\theta$ is sampled from a Dirichlet prior $\alpha$.  And then from that multinomial $\theta$ one timestamp $t$ is sampled for that document from a Beta distribution, and one topic $z$ is sampled for each word $w$ in the document.  Another multinomial $\beta$ is sampled for each topic in the document from a Dirichlet prior $\eta$, and a word $w$ is sampled from that multinomial $\beta$ given the topic $z$ sampled for this document.  This process which seems natural to document collections in which each document has one timestamp contrasts the other view which the authors presented and is given in Figure~\ref{fig:tot_gm}.  In this view which the authors adopted in their model, the generative process is similar to the generative process presented earlier for the first view, but instead of sampling one timestamp $t$ from a Beta distribution for each document, one timestamp is sampled from a Beta distribution for each word $w$ in the document.  All words in the same document however have the same timestamp.  The authors claim that this view makes it easier to implement and understand the model.

It is to be noted that the word distribution per topic in this TOT model is fixed over time though.  ``TOT captures changes in the occurrence (and co-occurrence conditioned on time) of the topics themselves, not changes of the word distribution of each topic.'' \cite{wang06}  The authors argue that evolution in topics happens by the changing occurrence and co-occurrence of topics as two co-occurring topics would be equivalent to a new topic that is formed by merging both topics, and losing the co-occurrence would be equivalent to splitting that topic into two topics.

In this TOT model, topic co-occurrences happen in continuous-time and the timestamps are sampled from a Beta distribution.  A temporal topic model evolving in continuous-time has a big advantage over a discrete-time topic model.  Discretization of time usually comes with the problem of selecting a good timestep.  Picking a large timestep leads to the problem of having documents covering a large time period over which the word distributions for the topics covered in these documents evolved significantly used in learning these distributions, or are inferenced using one fixed distribution over that period of time.  Picking a small timestep would complicate inference and learning as the number of parameters would explode as the timestep granularity increases.  Another problem that arises with discretization is that it does not account for varying time granularity over time.  Since topics evolve at different paces, and even one topic may evolve with different speeds over time, having a small timestep at one point in time to capture the fast dynamics of an evolving topic may be unnecessary later in time when the topic becomes stagnant and does not evolve as quickly.  Keeping a fine grained timestep at that point will make inference and learning slower as it increases the number of model parameters.  On the other hand, having a coarse timestep at one point in time when a topic does not evolve quickly may be suitable for that time but would become too big of a timestep when the topic starts evolving faster in time, and documents that fall within one timestep would be inferenced using the fixed word distributions for the topics not reflecting the change that happened to these topics.  To take this case to the extreme, a very large timestep covering the time period over which the document collection exists would be equivalent to using a classical latent Dirichlet allocation model that has no notion of time at all.

It is to be noted that the TOT topic model uses a fixed number of topics.  This limitation has major implications because not only do topics evolve over time, but topics get born, die, and reborn over time.  The number of active topics over time should not be assumed to be fixed.  Assuming a fixed number could lead to having topics conflated or merged in a wrong way.  Assuming a number of topics that is greater than the actual number of topics in document collection at a certain point in time causes the actual topics to be split over more than one topic.  In an application that classifies news articles based on the topics they discuss, this will cause extra classes to be created and makes the reader distracted between two classes that cover the same topic.  On the other hand, having a number of topics that is smaller than the actual number of topics covered by a collection of documents makes the model merge different topics into one.  In the same application of new articles classification, this leads to having articles covering different topics appearing under the same class.  Article classes could become very broad and this is usually undesirable as the reader relies on classification to read articles based on his/her focused interest.

In the TOT model, exact inference cannot be done.  \citet{wang06} resorted to Gibbs sampling in this model for approximate inference.  Since a Dirichlet which is a conjugate prior for the multinomial distribution is used, the multinomials $\theta$ and $\phi$ can be integrated out in this model.  Therefore, we do not have to sample $\theta$ and $\phi$.  This makes the model more simple, faster to simulate, faster to implement, and less prone to errors.  The authors claim that because they use a continuous Beta distribution rather than discretizing time, sparsity would not be a big concern in fitting the temporal part of the model.  In the training phase or learning the parameters of the model, every document has a single timestamp, and this timestamp is associated with every word in the document.  Therefore, all the words in a single document have the same timestamp.  Which is what we would naturally expect.  However, in the generative graphical model presented in Figure~\ref{fig:tot_gm} for the topics over time topic model, one timestamp is generated for each word in the same document.  This would probably result in different words appearing in the same document having different timestamp.  This is typically something we would not expect, because naturally, all words in a single document have the same timestamp because they were all authored and released or published under one title as a single publication unit or publication entity.  In this sense, the generative model presented in Figure~\ref{fig:tot_gm} is deficient as it assigns different timestamps to words within the same document.  The authors of the paper that this model was presented in argue that this deficiency does not distract a lot from this model and it still remains a powerful in modeling large dynamic text collections.  

An alternative generative process for the topics over time was also presented by the authors of this model.  In this alternative process, one timestamp is generated for each document using rejection sampling or importance sampling from a mixture of per topic Beta distributions over time with mixture weight as the per document $\theta_d$ over topics.

Instead of jointly modeling co-occurrence of words and timestamps in document collections, other topic models relied on analyzing how topics change over time by dividing the time covered by the documents into regions for analysis or by discretizing time.  \citet{grif04finding} used atemporal topic model to infer the topic mixtures of the proceedings of the National Academy of Sciences (PNAS).  They then ordered the documents in time based on their timestamps and assigned them to different time regions and analyzed their topic mixtures over time.  This study does not infer or learn the timestamps of documents and merely analyzes the topics learned using a simple latent Dirichlet allocation model \cite{blei03}.  Instead of learning one topic model for the entire document collection and then analyzing the topic structure of the documents in the collection, \citet{wang05} first divided the documents into consecutive time regions based on their timestamps.  They then trained a different topic model for each region and analyzed how topics changed over time.  This model has several limitations: First, the alignment of topics from one time region to the next is hard to do and would probably be done by hand, which is hard to do even with relatively small number of topics.  Second, the number of topics was held constant throughout time even at times when the documents become rich in context and they naturally contain more topics, or at times when the documents are not as rich and contain relatively less number of topics.  The model does not account for topics dying out and others being born.  Third, finding the correct time segment and number of segments is hard to do as it typically involves manual inspection of the documents.  The model however benefited from the fact that different models for documents in adjacent time regions are similar and the Gibbs sampling parameters learned for one region could be used as a starting point for learning parameters for the next time region \cite{song05}.

In their TimeMines system, \citet{swan00} generated a topic model that assigns one topic per document for a collection of news stories used to construct timelines in a topic detection and tracking task.

The topics over time topic model is a temporal model but not a Markovian one.  It does not make the assumption that a topic state at time $t+1$ is independent on all previous states of this topic except for its state at time $t$.  \citet{sarkar05}  analyze the dynamic social network of friends as it evolves over time using a Markovian assumption.  \citet{nodelman02} developed a Continuous-time Bayesian network (CTBN) that does not rely on time discretization.  In their  model, a Bayesian network evolves based on a continuous-time transition model using a Markovian assumption.  \citet{kleinberg03} created a model that relies on the relative order of documents in time instead of using timestamps.  This relative ordering may simplify the model and may be suitable for when the documents are released on a fixed or near fixed time interval but would not take into account the possibility that in some other applications like in news streams, the pace at which new stories are released and published varies over time.

\section{Discrete-time infinite dynamic topic model}
\citet{ahmed10} proposed a solution that overcomes the problem of having a fixed number of topics.  They proposed an infinite Dynamic Topic Model (iDTM) that allows for an unbounded number of topics and an evolving representation of topics according to a Markovian dynamics.  They analyzed the birth and evolution of topics in the NIPS community based on conference proceedings.  Their model evolved topics over discrete time units called epochs.  All proceedings of a conference meeting fall into the same epoch.  This model does not suit many applications as news articles production and tweet generation is more spread over time and does not usually come out in bursts.

In many topic modeling applications, such as for discussion forums, news feeds and tweets, the time duration of an epoch may not be clear.  Choosing too coarse a resolution may render invalid the assumption that documents within the same epoch are exchangeable.  Different topics and storylines will get conflated, and unrelated documents will have similar topic distribution.  If the resolution is chosen to be too fine, then the number of variational parameters of the model will explode with the number of data points.  The inference algorithm will become prohibitively expensive to run.  Using a discrete time dynamic topic model could be valid based on assumptions about the data.  In many cases, the continuity of the data which has an arbitrary time granularity prevents us from using a discrete time model.

To summarize: in streaming text topic modeling applications, the discrete-time model given above is brittle.  An extension to continuous-time will give it the needed flexibility to account for change in temporal granularity.

\subsection*{Statistical model}
\begin{figure}
  \centering
  \resizebox{\linewidth}{!}{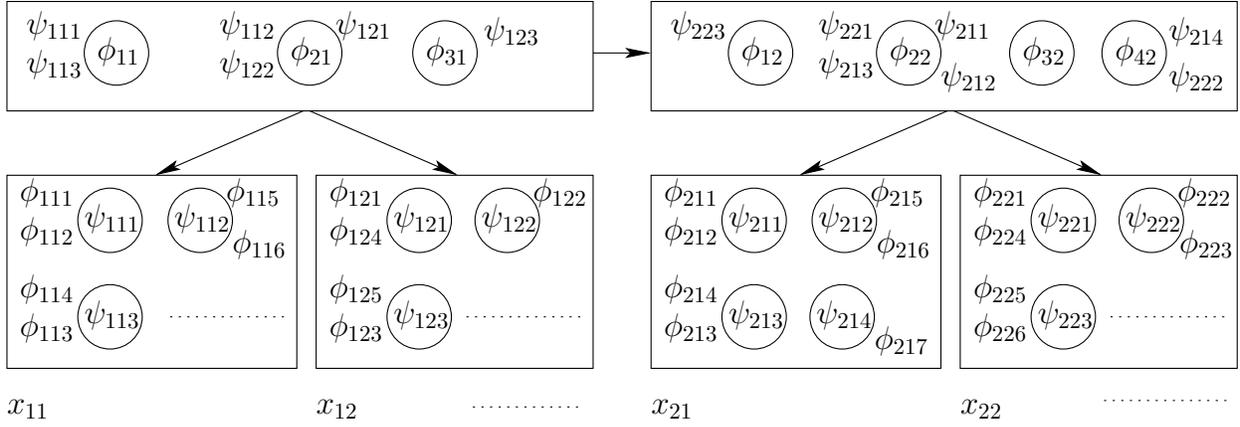}
  \caption{A graphical model representation of the infinite dynamic topic model using plate notation.}
  \label{fig:idtm}
\end{figure}

Figure~\ref{fig:idtm} shows a graphical model representation for an order one recurrent Chinese restaurant franchise process (RCRF).  The symbols in this figure follow the same notation used for LDA in Figure~\ref{fig:lda_graphical_model} In RCRF, documents in each epoch are modeled using a Chinese restaurant franchise process (CRFP).  The menus that the documents were sampled from at different timesteps are tied over time.

For a given Dirichlet Process (DP), $DP(G_0,\alpha)$, with a base measure $G_0$, and a concentration parameter $\alpha$, $G \sim DP(G_0,\alpha)$ is a Dirichlet distribution over the parameter space $\theta$.  By integrating out $G$, $\theta$ follows a Polya urn distribution \cite{blackwell:polya} or a recurrent Chinese restaurant process \cite{ahmed10}
\begin{equation}
  \theta_i|\theta_{1:i-1},G_0,\alpha \sim \sum_k\frac{m_k}{i-1+\alpha}\delta(\phi_k)+\frac{\alpha}{i-1+\alpha}G_o
\end{equation}
where, $\phi_{1:k}$ are the distinct values of $\theta$, and $m_k$ is the number of parameter $\theta$ with $\phi_k$ value.  A Dirichlet process mixture model (DPM) can be built using the given DP on top of a hierarchical Bayesian model.

One disadvantage of using RCRP is that in it each document is generated using a single topic.  This assumption is unrealistic specially in some domains like in modeling news stories where each document is typically a mixture of different topics.  To overcome this limitation, I can use a hierarchical Dirichlet process (HDP) in which each document can be generated from multiple topics.

To add temporal dependence in our model, we can use the temporal Dirichlet process mixture model (TDPM) proposed in \cite{AhmedX08} which allows unlimited number of mixture components.  In this model, $G$ evolves as follows \cite{ahmed10}:
\begin{gather}
  G_t|\phi_{1:k},G_o,\alpha \sim DP(\zeta,D)\\
  \zeta = \alpha + \sum_km'_{kt}\\
  D = \sum_k\frac{m'_{kt}}{\sum_lm'_{lt}+\alpha}\delta(\phi_k) + \frac{\alpha}{\sum_lm'_{lt}+\alpha}G_0\\
  m'_{kt} = \sum_{\delta=1}^{\Delta} \exp^{-\delta/\lambda}m_{k,t-\delta}
\end{gather}
where $m'_{kt}$ is the prior weight of component $k$ at time $t$, $\Delta$, $\lambda$ are the width and decay factor of the time decaying kernel.  For $\Delta=0$, this TDPM will represent a set of independent DPMs at each time step, and a global DPM when $\Delta=T$ and $\lambda=\infty$.  As we did with the time-independent DPM, we can integrate out $G_{1:T}$ from our model to get a set of parameters $\theta_{1:t}$ that follows a Poly-urn distribution:
\begin{equation}
  \theta_{ti}|\mathbf{\theta_{t-1:t-\Delta}},\theta_{t,1:i-1},G_0,\alpha \propto \sum_k(m'_{kt}+m_{kt})\delta(\phi_{kt})+\alpha G_0
\end{equation}

In this process, topic popularity at epoch $t$ depends on its use at this epoch $m_{kt}$, and its use in the previous $\Delta$ epochs, $m'_{kt}$.  Therefore, after $\Delta$ epochs of not being used, a topic can be considered dead.  This makes sense as higher-order models will require more epochs to pass by without using a topic before it is considered dead.  This is because the length of the chain is longer and the effect of the global menu of topics carries along for more epochs.  On the other hand, in lower-order models the effect of the global menu of topics can only be passed along for less epochs than in the previous case.

By placing this model on top of a hierarchical Dirichlet process as indicated earlier, we can tie all random measures $G_d$, from which the multinomially distributed parameters $\mathbf{\theta_d}$ are drawn for each $\mathbf{w_d}$ document in a topic model scheme by modeling $G_0$ as a random measure sampled from $DP(\gamma, H)$.  By integrating $G_d$ out of this model, we get the Chinese restaurant franchise process (CRFP) \cite{blei06};
\begin{gather}
  \theta_{di}|\theta_{d,1:i-1},\alpha,\boldsymbol{\psi} \sim \sum_{b=1}^{b=B_d}\frac{n_{db}}{i-1+\alpha}\delta_{\psi_{db}} + \frac{\alpha}{i-1+\alpha}\delta_{\psi_{db^{new}}}\\
\psi_{db^{new}}|\boldsymbol{\psi},\gamma \sim \sum_{k=1}^K \frac{m_k}{\sum_{l=1}^Km_l+\gamma}\delta_{\phi_k} + \frac{\gamma}{\sum_{l=1}^Km_l+\gamma}H
\end{gather}
where $\psi_{db}$ is topic $b$ for document $d$, $n_{db}$ is the number of words sampled from it, $\psi_{db^{new}}$ is a new topic, $B_d$ is the number of topics in document $d$, and $m_k$ is the number of documents sharing topic $\phi_k$.

We can make word distributions and topic trends evolve over time if we tie together all hyper-parameter base measures $G_0$ through time.  The model will now take the following form:
\begin{equation}
  \theta_{tdi}|\theta_{td,1:i-1},\alpha,\boldsymbol{\psi_{t-\Delta:t}} \sim \sum_{b=1}^{b=B_d}\frac{n_{tdb}}{i-1+\alpha}\delta_{\psi_{tdb}} + \frac{\alpha}{i-1+\alpha}\delta_{\psi_{tdb^{new}}} \label{posterior}
\end{equation}
\begin{align}
  \psi_{tdb^{new}}|\boldsymbol{\psi},\gamma &\sim \sum_{k:m_{kt}>0} \frac{m_{kt}+m_{kt}'}{\sum_{l=1}^{K_t}m_{lt}+m_{lt}'+\gamma}\delta_{\phi_{kt}} \nonumber \\
&+  \sum_{k:m_{kt}=0} \frac{m_{kt}+m_{kt}'}{\sum_{l=1}^{K_t}m_{lt}+m_{lt}'+\gamma}\mathrm{P}(.|\phi_{k,t-1}) \nonumber \\
&+ \frac{\gamma}{\sum_{l=1}^{K_t}m_{lt}+m_{lt}'+\gamma}H \label{likelihood}
\end{align}
where $\phi_{kt}$ evolves using a random walk kernel like in \cite{blei06}:
\begin{gather}
   H = N(0,\sigma I) \label{baseMeasure}\\
   \phi_{k,t}|\phi_{k,t-1} \sim N(\phi_{k,t-1},\rho I)\\
   w_{tdi}|\phi_{kt} \sim \mathcal{M}(L(\phi_{kt}))\\
   L(\phi_{kt}) = \frac{\exp(\phi_{kt})}{\sum_{w=1}^W\exp(\phi_{ktw})}
\end{gather}

We can see from \eqref{posterior}, \eqref{likelihood}, \eqref{baseMeasure} the non-conjugacy between the base measure and the likelihood.

\begin{table}[!h]
  \centering
  \begin{tabular}{c l}
    \hline
    Symbol & Definition\\
    \hline
    $DP(.)$  & Dirichlet process\\
    $G_0$    & base measure\\
    $\alpha$ & concentration parameter\\
    $\theta$ & Dirichlet distribution parameter space\\
    $\phi_{1:k}$ & Distinct values of $\theta$\\
    $m_k$    & number of parameter $\theta$ with $\phi_k$ value\\
    $m_{kt}'$ & prior weight of component $k$ at time $t$\\
    $\Delta$ & width of time decaying kernel\\
    $\lambda$& decay factor of a time decaying kernel\\
    $w$      & word in a document\\
    $H$      & base measure of the DP generating $G_0$\\
    $\gamma$ & concentration parameter for the DP generating $G_0$\\
    $\psi_{tdb}$   & topic $b$ for document $d$ at time $t$\\
    $n_{tdb}$      & number of words sampled from $\psi_{tdb}$\\
    $\psi_{tdb^{new}}$   & a new topic\\
    \hline
  \end{tabular}
  \caption{Table of notations for iDTM}
\end{table}

The model given above is suitable for discrete time topic models with evolving number of topics.  The topic trends and the word distribution of topics will evolve in discrete time though.

Evolving topics in continuous time can be achieved by using a Brownian motion model \cite{peres10} that models the natural parameters of a multinomial distribution for the words over topics.  A Dirichlet distribution can be used to model the natural parameters of multinomial distribution of the topics given the words of the parameter.

\section{Online hierarchical Dirichlet processes}
\label{subsec:ohdp}
Traditional variational inference algorithms are suitable for some applications in which the document collection to be modeled is known before model learning or posterior inference takes place.  If the document collection changes, however, the entire posterior inference procedure has to be repeated to update the learned model.  This clearly incurs the additional cost of relearning and re-analyzing a potentially huge volume of information specially as the collection grows over time.  This cost could become very high and a compromise should be made if this traditional variational inference algorithm is to be used about having an up-to-date model against saving computational power.

Online inference algorithms do not require several passes over the entire dataset to update the model which is a requirement for traditional inference algorithms.  \citet{sato2001} introduced an online variational Bayesian inference algorithm that gives variational inference algorithms an extra edge over their MCMC counterparts.

Traditional variational inference algorithms approximate the true posterior over the latent variables by suggesting a simpler distribution that gets refined to minimize its Kullback-Leibler (KL) distance to the true posterior.  In online variational inference, this optimization is done using stochastic approximation \cite{sato2001, wang11}.

\subsection*{Statistical model}
At the top level of a two level hierarchical Dirichlet process (HDP) a Dirichlet distribution $G_0$ is sampled from a Dirichlet process (DP).  This distribution $G_0$ is used as the base measure for another DP at the lower level from which another Dirichlet distribution $G_j$ is drawn.  This means that all the distributions $G_j$ share the same set of atoms they inherited from their parent with different atom weights.  Formally put:

\begin{align}
  \label{eq:15}
  G_0 &\sim DP(\gamma, H)\\
  G_j &\sim DP(\alpha_0, G_0)
\end{align}

Where $\gamma$ and $H$ are the concentration parameter and base measure for the first level DP, $\alpha_0$ and $G_0$ are the concentration parameter and base measure for the second level DP, and $G_j$ is the Dirichlet distribution sampled from the second level DP.

In a topic model utilizing this HDP structure, a document is made of a collection of words, and each topic is a distribution over the words in the document collection.  The atoms of the top level DP are the global set of topics.  Since the base measure of the second level DP is sampled from the first level DP, then the sets of topics in the second level DP are subsets of the global set of topics in the first level DP.  This ensures that the documents sampled from the second level process share the same set of topics in the upper level.  For each document $j$ in the collection, a Dirichlet $G_j$ is sampled from the second level process.  Then, for each word in the document a topic is sampled then a word is generated from that topic.

In Bayesian non-parametric models, variational methods are usually represented using a stick-breaking construction.  This representation has its own set of latent variables on which an approximate posterior is given \cite{blei04, teh07, kurihara07}.  The stick-breaking representation used for this HDP is given at two levels: corpus-level draw for the Dirichlet $G_0$ from the top-level DP, and a document-level draw for the Dirichlet $G_j$ from the lower-level DP.  The corpus-level sample can be obtained as follows:

\begin{align}
  \label{eq:16}
  \beta_k' &\sim Beta(1, \gamma)\\
  \beta_k  &=\beta_k'\prod_{l=1}^{k-1}(1-\beta_l')\\
  \phi_k   &\sim H\\
  G_0      &=\sum_{k=1}^\infty \beta_k\delta_{\phi_k}
\end{align}

where $\gamma$ is a parameter for the Beta distribution, $\beta_k$ is the weight for topic $k$, $\phi_k$ is atom (topic) $k$, $H$ is the base distribution for the top level DP, and $\delta$ is the Dirac delta function.

The second level (document-level) draws for the Dirichlet $G_j$ are done by applying Sethuraman stick-breaking construction of the DP again as follows:

\begin{align}
  \label{eq:17}
  \psi_{jt} &\sim G_0\\
  \pi_{jt}' &\sim Beta(1, \alpha_0)\\
  \pi_{jt}  &= \pi_{jt}' \prod_{l=1}^{t-1}(1 - \pi_{jl}')\\
  G_j      &=\sum_{t=1}^\infty \pi_{jt}\delta_{\psi_{jt}}
\end{align}

where $\psi_{jt}$ is a document-level atom (topic) and $\pi_{jt}$ is the weight associated with it.

This model can be simplified by introducing indicator variables $\pmb{c_j}$ that are drawn from the wights $\pmb{\beta}$ \cite{wang11}

\begin{equation}
  \label{eq:18}
  c_{jt} \sim Mult(\pmb{\beta})
\end{equation}

The variational distribution is thus given by:

\begin{align}
  \label{eq:14}
  q(\pmb{\beta}', \pmb{\pi}', \pmb{c}, \pmb{z}, \pmb{\phi}) &= q(\pmb{\beta}')q(\pmb{\pi}')q(\pmb{c})q(\pmb{z})q(\pmb{\phi})\\
  q(\pmb{\beta}') &= \prod_{k=1}^{K-1}q(\beta_k'|u_k,v_k)\\
  q(\pmb{\pi}') &= \prod_j \prod_{t=1}^{T-1}q(\pi_{jt}'|a_{jt},b_{jt})\\
  q(\pmb{c}) &= \prod_j \prod_t q(c_{jt}|\varphi_{jt})\\
  q(\pmb{z}) &= \prod_j \prod_n q(z_{jn}|\zeta_{jn})\\
  q(\pmb{\phi}) &= \prod_k q(\phi_k|\lambda_k)
\end{align}

where $\pmb{\beta}'$ is the corpus-level stick proportions and $(u_k,v_k)$ are parameters for its beta distribution, $\pmb{\pi}_j'$ is the document-level stick proportions and $(a_{jt},b_{jt})$ are parameters for its beta distribution, $\pmb{c}_j$ is the vector of indicators, $\pmb{\phi}$ is the topic distributions, and $\pmb{z}$ is the topic indices vector.  In this setting, the variational parameters are $\varphi_{jt}$, $\zeta_{jn}$, and $\lambda_k$.

The variational objective function to be optimized is the marginal log-likelihood of the document collection $\mathcal{D}$ given by \cite{wang11}:

\begin{align}
  \label{eq:19}
  \log p(\mathcal{D}|\gamma, \alpha_0, \zeta) &\ge \mathbb{E}_q[\log p(\pmb{D},\pmb{\beta}', \pmb{\pi}', \pmb{c}, \pmb{z}, \pmb{\phi})] + H(q)\\
  & = \sum_j \{\mathbb{E}_q [\log (p(\pmb{w}_j|\pmb{c}_j, \pmb{z}_j, \pmb{\phi}) p(\pmb{c}_j|\pmb{\beta}') p(\pmb{z}_j|\pmb{\pi}_j') p(\pmb{\pi}_j'|\alpha_0))] \\
  &\qquad+ H(q(\pmb{c}_j)) + H(q(\pmb{z}_j)) + H(q(\pmb{\pi}_j'))\} \\
  &\qquad+ \mathbb{E}_q[\log p(\pmb{\beta}') p(\pmb{\phi})] + H(q(\pmb{\beta}')) + H(q(\pmb{\phi}))\\
  &=\mathcal{L}(q)
\end{align}

Where $H(.)$ is the entropy term for the variational distribution.

\section{Continuous-time dynamic topic model}
\label{subsec:cdtm}
On the other hand, \citet{wang08} proposed a continuous-time dynamic topic model.  That model uses Brownian motion to model \cite{peres10} the evolution of topics over time.  Even though that models uses a novel sparse variational Kalman filtering algorithm for fast inference, the number of topics it samples from is bounded, and that severely limits its application in news feed storyline creation and article aggregation.  When the number of topics covered by the news feed is less than the pre-tuned number of topics set of the model, similar stories will show under different storylines.  On the other hand, if the number of topics covered becomes greater than the pre-set number of topics, topics and storylines will get conflated.

\subsection*{Statistical model}
\begin{figure}[tbh]
  \centering
  \resizebox{0.5\linewidth}{!}{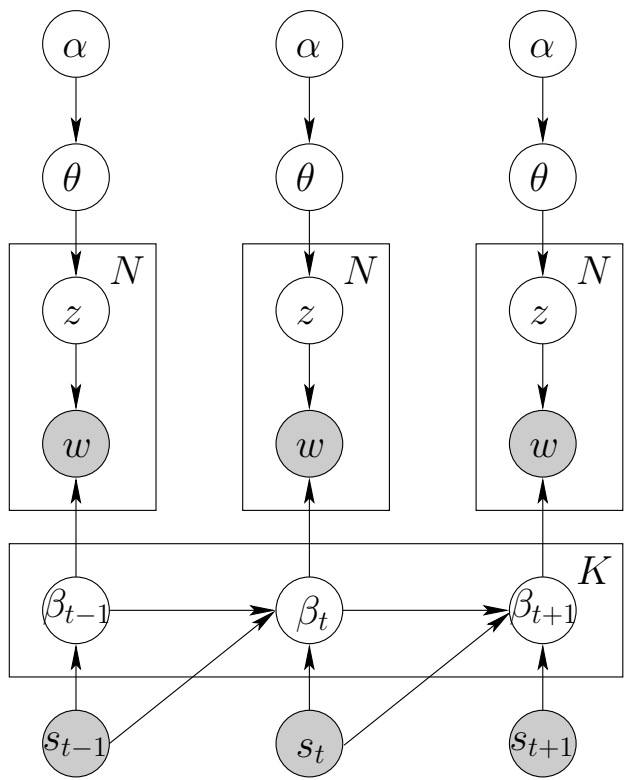}
  \caption{Graphical model representation of the continuous-time dynamic topic model using plate notation.}
  \label{fig:cdtm}
\end{figure}

A graphical model representation for this model using plate notation is given in Figure~\ref{fig:cdtm}.

Let the distribution of the $k^{th}$ topic parameter for word $w$ be:
\begin{gather}
  \beta_{0,k,w} \sim \mathcal{N}(m,v_0)\\
  \beta_{j,k,w}|\beta_{i,k,w},s \sim \mathcal{N}(\beta_{i,k,w},v\Delta_{s_j,s_i})
\end{gather}
where $i$, $j$ are time indexes, $s_i$, $s_j$ are timestamps, and $\Delta_{s_j,s_i}$ is the time elapsed between them.  In this model, the multinomially distributed topic distribution at time $t$ is sampled from a Dirichlet distribution $\theta_t \sim Dir(\alpha)$, and then a topic $z_{t,n}$ is sampled from a multinomial distribution parametrized by $\theta_t$, $z_{t,n} \sim \text{Mult}(\theta_t)$.

To make the topics evolve over time, we define a Wiener process \cite{peres10}  $\{X(t), t \ge 0\}$ and sample $\beta_t$ from it.  The obtained unconstrained $\beta_t$ can then be mapped on the simplex.  More formally:
\begin{gather}
  \beta_{t,k}|\beta_{t-1,k},s \sim \mathcal{N}(\beta_{t-1,k},v\Delta_{st}I)\\
  \pi(\beta_{t,k})_w = \frac{\exp(\beta_{t,k,w})}{\sum_w\exp(\beta_{t,k,w})}\\
  w_{t,n} \sim \text{Mult}(\pi(\beta_{t,z_t,n}))
\end{gather}
where $\pi(.)$ maps the unconstrained multinomial natural parameters to its mean parameters, which are on the simplex.

The posterior, which is the distribution of the latent topic structure given the observed documents, is intractable.  We resort to approximate inference.  For this model, sparse variational inference presented in \cite{wang08} could be used.

\begin{table}[tbh]
  \centering
  \begin{tabular}{c l}
    \hline
    Symbol & Definition\\
    \hline
    $B_d$    & number of topics in document $d$\\
    $\mathcal{N}(m,v_0)$ & a Gaussian distribution with mean $m$ and variance $v_0^2$\\
    $\beta_{i,k,w}$  & distribution of words over topic $k$ at time $i$ for word $w$\\
    $s_i$    & timestamp for time index $i$\\
    $\Delta_{s_j,s_i}$  & time duration between $s_i$ and $s_j$\\
    $Dir(.)$      & Dirichlet distribution\\
    $z_{t,n}$      & topic $n$ sampled at time $t$\\
    $I$      & identity matrix\\
    Mult(.)  & multinomial distribution\\
    $\pi(.)$ & mapping function\\
    \hline
  \end{tabular}
  \caption{Table of notations for cDTM}
\end{table}


\cleardoublepage
\chapter{Continuous-time infinite dynamic topic model}
\begin{shaded}
\noindent In this chapter I describe my own contribution, the continuous-time infinite dynamic topic model.
\end{shaded}
\section{Dim sum process}
I propose building a topic model called the \emph{continuous-time infinite dynamic topic model} (ciDTM) that combines the properties of 1) the continuous-time Dynamic Topic Model (cDTM), and 2) the online Hierarchical Dirichlet Process model (oHDP).

I will refer to the stochastic process I develop that combines the properties of these two systems as the \emph{Dim Sum Process}.


Dim sum is a style of Chinese food.  One important feature of dim sum restaurants is the freedom given to the chef to create new dishes depending on seasonal availability and what he thinks is auspicious for that day.  This leeway given to the cook leads to change over time of the ingredient mixture of the different dishes the restaurant serve to better satisfy the customers taste and suit seasonal changes and availability of these ingredients.

The generative process of a dim sum process is similar to that of a Chinese restaurant franchise process \cite{TehJorBea2006} in the way customers are seated in the restaurant and the way dishes are assigned to tables, but differs in that the ingredient (mixture) of the dishes evolves in continuous time.

When the dim sum process is used in a topic modeling application, each customer corresponds to a word and the restaurant represents a document.  A dish is mapped to a topic and a table is used to group a set of customers to a dish.

The generative process of the dim sum process proceeds as follows:

\begin{itemize}
  \item Initially the dim sum restaurant (document) is empty.
  \item The first customer (word) to arrive sits down at a table and orders a dish (topic) for his table.
  \item The second customer to arrive has two options.  1) She can sit at a new table with probability $\alpha/(1+\alpha)$ and orders food for her new table, or 2) she cang sit at the occupied table with probability $1/(1+\alpha)$ and eat from the food that has been already ordered for that table.
  \item When the $n-1$ customer enters the restaurant, he can sit down at a new table with probability $\alpha/(n + \alpha)$, or he can sit down at table $k$ with probability $n_k/(n + \alpha)$ where $n_k$ is the number of customers currently sitting at table $k$.
\end{itemize}
    
Higher values of $\alpha$ leads to higher number of occupied tables and dishes (topics) sampled in the restaurant (document).  This model can be extended to a franchise restaurant setting where all the restaurants share one global menu from which the customers order.  To do so, each restaurant samples its parameter $\alpha$ and its local menu from a global menu.  This global menu is a higher level Dirichlet process.  This two level Dirichlet process is known in the literature as the Chinese restaurant franchise process (CRFP) \cite{TehJorBea2006}.

The main difference between a dim sum process and a CRFP is in the global menu.  This global menu is kept fixed over time in the CRFP setting, whereas it evolves in continuous time in the dim sum process using a Brownian motion model \cite{peres10}.

\begin{figure}
  \centering
  \resizebox{0.8\linewidth}{!}{\input{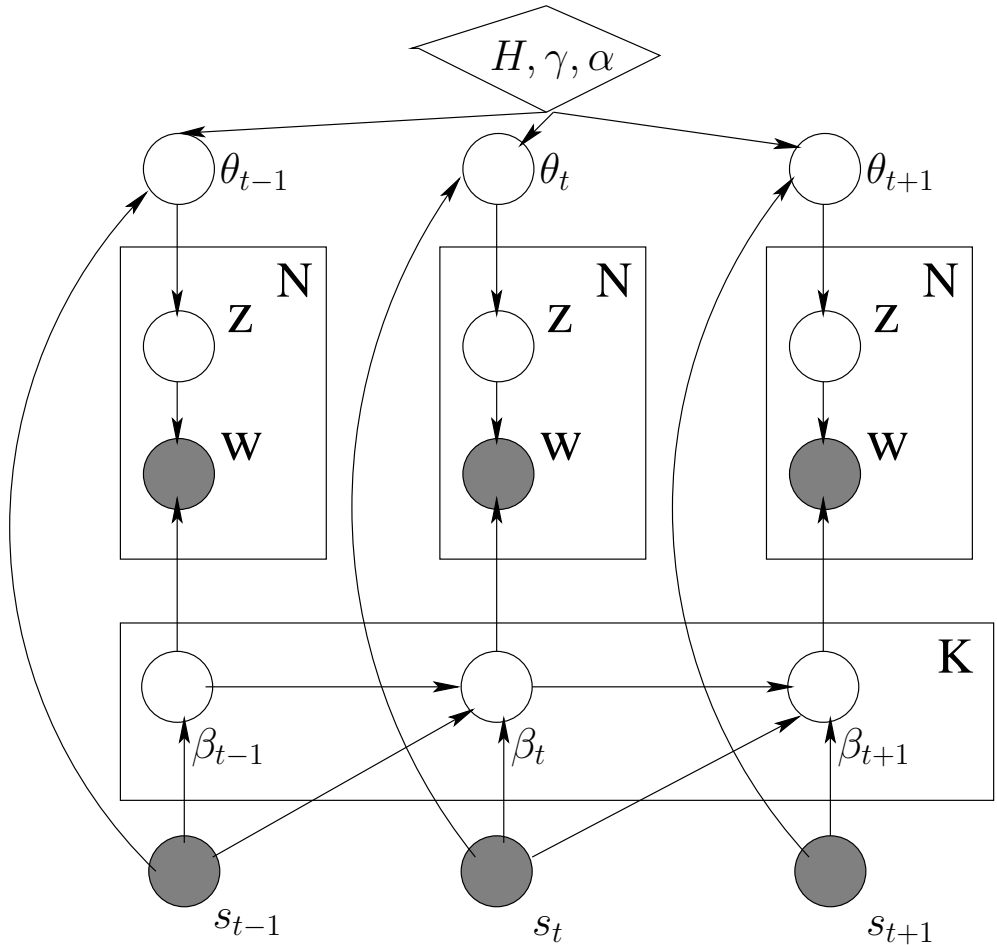}}
  \caption{Continuous-time infinite dynamic topic model.  None-shaded nodes represent unobserved latent variables, shaded nodes are observed variables, diamonds represent hyperparameters, and plates represent repetition.  Symbols in this figure follow the notation in Figure~\ref{fig:idtm} and $H$ is the base distribution for the upper level DP, $\gamma$ and $\alpha$ are the concentration parameters for the upper level, and lower level DPs respectively, and $s_t$ is the timestamp for document at time $t$.\label{cidtm}}
\end{figure}

The dim sum process can be described using plate notation as shown in Figure \ref{cidtm}.  This model combines the two models in that it gets rid of the highest level time specific hierarchical Dirichlet process.  It implicitly ties the base measures $G_0$ across all documents.  This measure which ensures sharing of the topics through time and over documents, is sampled from a $DP(\gamma,H)$, and it has been integrated out from our model.  Note that I reduced the hierarchical structure of the model one level, and it now represents a Chinese restaurant process instead of a recurrent Chinese restaurant process.

The other modification I make to the model is to evolve its topic distribution using a Brownian motion model \cite{peres10}, similar to the one used in the continuous-time dynamic topic model.

The true posterior is not tractable; we have to resort to approximate inference for this model. Moreover, due to non-conjugacy between the distribution of words for each topic and the word probabilities, I cannot use collapsed Gibbs sampling.  Instead I plan to use variational methods for inference.

\begin{figure}
  \centering
  \includegraphics[width=\linewidth]{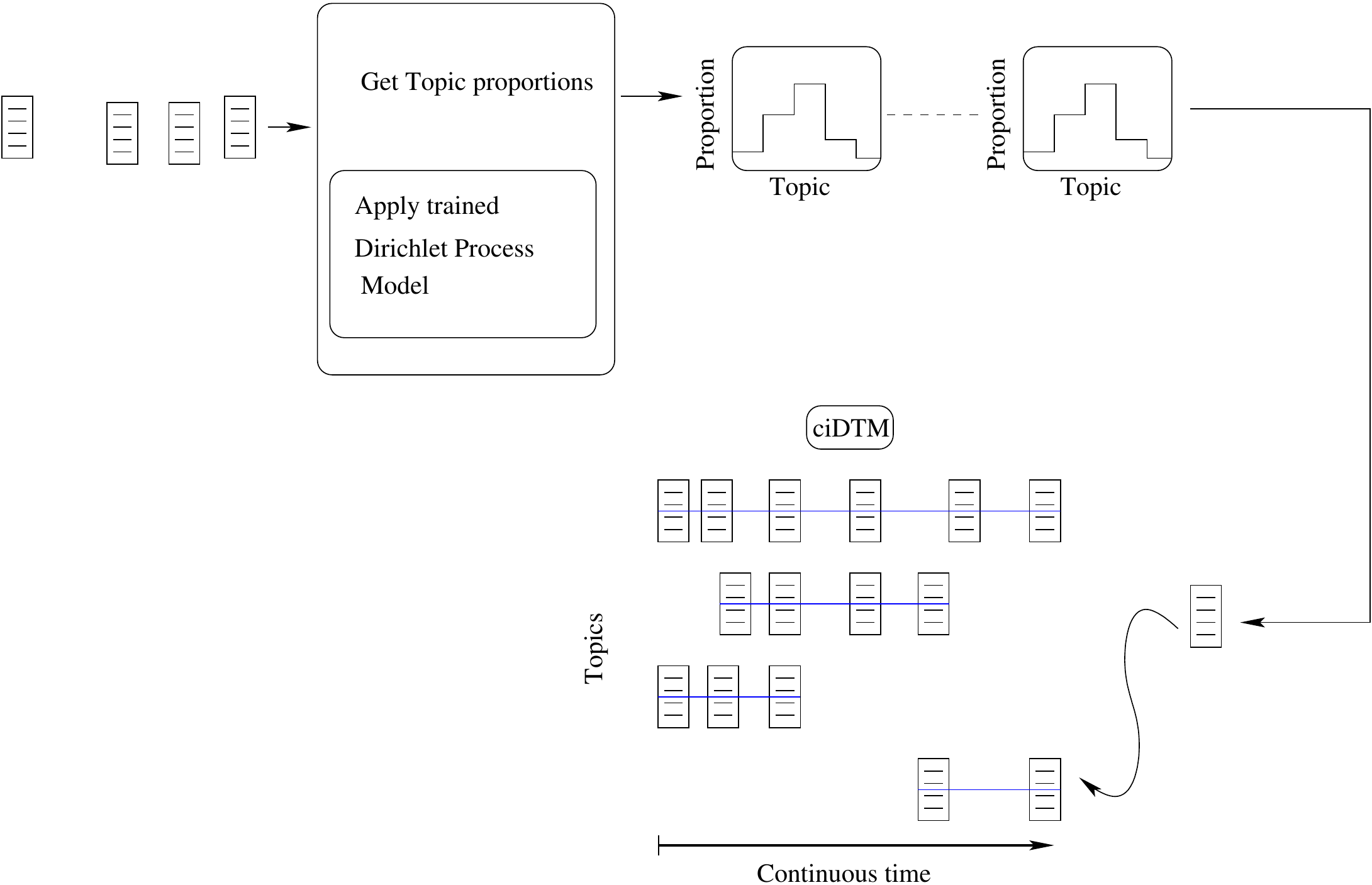}
  \caption{A block diagram describing the workflow for the continuous-time infinite dynamic topic model working on a single document at a time.\label{cidtm_block}}
\end{figure}

Figure~\ref{cidtm_block} shows a block diagram for the continuous-time infinite dynamic topic model operating on a single document at a time.

\section{Mathematical model}
The continuous-time infinite dynamic topic model (ciDTM) is a mixture of the continuous-time dynamic topic model presented earlier in Section \ref{subsec:cdtm}, and the online hierarchical Dirichlet process (oHDP) model presented in Section \ref{subsec:ohdp}.

A generative process for this ciDTM using dim sum process proceeds as follows:

We build a two level hierarchical Dirichlet process (HDP) like the one presented in Section \ref{subsec:ohdp}.

At the top level of a two level hierarchical Dirichlet process (HDP), a Dirichlet distribution $G_0$ is sampled from a Dirichlet process (DP).  This distribution $G_0$ is used as the base measure for another DP at the lower level from which another Dirichlet distribution $G_j$ is drawn.  This means that all the distributions $G_j$ share the same set of atoms they inherited from their parent with different atom weights.  Formally put:

\begin{align}
  \label{eq:ch5:15}
  G_0 &\sim DP(\gamma, H)\\
  G_j &\sim DP(\alpha_0, G_0)
\end{align}

Where $\gamma$ and $H$ are the concentration parameter and base measure for the first level DP, $\alpha_0$ and $G_0$ are the concentration parameter and base measure for the second level DP, and $G_j$ is the Dirichlet distribution sampled from the second level DP.

In a topic model utilizing this HDP structure, a document is made of a collection of words, and each topic is a distribution over the words in the document collection.  The atoms of the top level DP are the global set of topics.  Since the base measure of the second level DP is sampled from the first level DP, then the sets of topics in the second level DP are subsets of the global set of topics in the first level DP.  This ensures that the documents sampled from the second level process share the same set of topics in the upper level.  For each document $j$ in the collection, a Dirichlet $G_j$ is sampled from the second level process.  Then, for each word in the document a topic is sampled then a word is generated from that topic.

In Bayesian non-parametric models, variational methods are usually represented using a stick-breaking construction.  This representation has its own set of latent variables on which an approximate posterior is given \cite{blei04, teh07, kurihara07}.  The stick-breaking representation used for this HDP is given at two levels: corpus-level draw for the Dirichlet $G_0$ from the top-level DP, and a document-level draw for the Dirichlet $G_j$ from the lower-level DP.  The corpus-level sample can be obtained as follows:

\begin{align}
  \label{eq:ch5:16}
  \beta_k' &\sim Beta(1, \gamma)\\
  \beta_k  &=\beta_k'\prod_{l=1}^{k-1}(1-\beta_l')\\
  \phi_k   &\sim H\\
  G_0      &=\sum_{k=1}^\infty \beta_k\delta_{\phi_k}
\end{align}

where $\gamma$ is a parameter for the Beta distribution, $\beta_k$ is the weight for topic $k$, $\phi_k$ is atom (topic) $k$, $H$ is the base distribution for the top level DP, and $\delta$ is the Dirac delta function.

The second level (document-level) draws for the Dirichlet $G_j$ are done by applying Sethuraman stick-breaking construction of the DP again as follows:

\begin{align}
  \label{eq:ch:5:17}
  \psi_{jt} &\sim G_0\\
  \pi_{jt}' &\sim Beta(1, \alpha_0)\\
  \pi_{jt}  &= \pi_{jt}' \prod_{l=1}^{t-1}(1 - \pi_{jl}')\\
  G_j      &=\sum_{t=1}^\infty \pi_{jt}\delta_{\psi_{jt}}
\end{align}

where $\psi_{jt}$ is a document-level atom (topic) and $\pi_{jt}$ is the weight associated with it.

This model can be simplified by introducing indicator variables $\pmb{c_j}$ that are drawn from the weights $\pmb{\beta}$ \cite{wang11}

\begin{equation}
  \label{eq:ch5:18}
  c_{jt} \sim Mult(\pmb{\beta})
\end{equation}

The variational distribution is thus given by:

\begin{align}
  \label{eq:ch5:14}
  q(\pmb{\beta}', \pmb{\pi}', \pmb{c}, \pmb{z}, \pmb{\phi_0}) &= q(\pmb{\beta}')q(\pmb{\pi}')q(\pmb{c})q(\pmb{z})q(\pmb{\phi_0})\\
  q(\pmb{\beta}') &= \prod_{k=1}^{K-1}q(\beta_k'|u_k,v_k)\\
  q(\pmb{\pi}') &= \prod_j \prod_{t=1}^{T-1}q(\pi_{jt}'|a_{jt},b_{jt})\\
  q(\pmb{c}) &= \prod_j \prod_t q(c_{jt}|\varphi_{jt})\\
  q(\pmb{z}) &= \prod_j \prod_n q(z_{jn}|\zeta_{jn})\\
  q(\pmb{\phi_0}) &= \prod_k q(\phi_k|\lambda_k)
\end{align}

where $\pmb{\beta}'$ is the corpus-level stick proportions and $(u_k,v_k)$ are parameters for its beta distribution, $\pmb{\pi}_j'$ is the document-level stick proportions and $(a_{jt},b_{jt})$ are parameters for its beta distribution, $\pmb{c}_j$ is the vector of indicators, $\pmb{\phi_0}$ is the topic distributions, and $\pmb{z}$ is the topic indices vector.  In this setting, the variational parameters are $\varphi_{jt}$, $\zeta_{jn}$, and $\lambda_k$.

The variational objective function to be optimized is the marginal log-likelihood of the document collection $\mathcal{D}$ given by \cite{wang11}:

\begin{align}
  \label{eq:ch5:19}
  \log p(\mathcal{D}|\gamma, \alpha_0, \zeta) &\ge \mathbb{E}_q[\log p(\pmb{D},\pmb{\beta}', \pmb{\pi}', \pmb{c}, \pmb{z}, \pmb{\phi_0})] + H(q)\\
  & = \sum_j \{\mathbb{E}_q [\log (p(\pmb{w}_j|\pmb{c}_j, \pmb{z}_j, \pmb{\phi_0}) p(\pmb{c}_j|\pmb{\beta}') p(\pmb{z}_j|\pmb{\pi}_j') p(\pmb{\pi}_j'|\alpha_0))] \\
  &\qquad+ H(q(\pmb{c}_j)) + H(q(\pmb{z}_j)) + H(q(\pmb{\pi}_j'))\} \\
  &\qquad+ \mathbb{E}_q[\log p(\pmb{\beta}') p(\pmb{\phi_0})] + H(q(\pmb{\beta}')) + H(q(\pmb{\phi_0}))\\
  &=\mathcal{L}(q)
\end{align}

Where $H(.)$ is the entropy term for the variational distribution.

I use coordinate ascent to maximize the log-like likelihood given in \eqref{eq:ch5:19}.  Next, given the per-topic word distribution $\pmb{\phi_0}$, I use a Wiener motion process \cite{peres10} to make the topics evolve over time.  I define the process $\{X(t), t \ge 0\}$ and sample $\pmb{\phi}_t$ from it.  The obtained unconstrained $\pmb{\phi}_t$ can then be mapped on the simplex.  More formally:
\begin{gather}
  \pmb{\phi}_{t,k}|\pmb{\phi}_{t-1,k},s \sim \mathcal{N}(\pmb{\phi}_{t-1,k},v\Delta_{st}I)\\
  \pi(\pmb{\phi}_{t,k})_w = \frac{\exp(\pmb{\phi}_{t,k,w})}{\sum_w\exp(\pmb{\phi}_{t,k,w})}\\
  w_{t,n} \sim \text{Mult}(\pi(\pmb{\phi}_{t,z_t,n}))
\end{gather}
where $\pi(.)$ maps the unconstrained multinomial natural parameters to its mean parameters, which are on the simplex.

The posterior, which is the distribution of the latent topic structure given the observed documents, is intractable.  We resort to approximate inference.  For this model, sparse variational inference presented in \cite{wang08} could be used.


\cleardoublepage
\chapter{Testbed development}
In this chapter, I will motivate the need for a continuous-time infinite dynamic topic model and present the dataset/corpus I will be using for my testbed and evaluation of the model I will develop as well as other competing models.  I start by describing the corpus:
\begin{enumerate}
  \item Why do I need the corpus?  
  \item What should this corpus be made of?  
  \item What makes a good corpus?  
  \item How can a good corpus be created?  
  \item Are there corpora that fit my needs?  
  \item Can I modify existing corpora to fit my needs?  
  \item What are the challenges in creating such a corpus?  
  \item Who else could benefit from this corpus if I can publicly publish it?  and
  \item Can I publicly publish this corpus?
\end{enumerate}

After that, I will define the problem of creating timelines for different news stories which has many potential applications especially in the news media industry.  I will follow that by several attempts to solve this problem, at first using a simple topic model, then successively using more advanced models that alleviate some of the shortcomings of the earlier models.

\section{News stories corpus}
In order to test the performance of my ciDTM and create news timelines I need a news corpus.  This corpus should be made of a collection of news stories.  These stories could be collected from news outlets, like newspapers, newswire, transcripts of radio news broadcasts, or crawled from websites of news agencies or newspapers websites.  In my case, I can crawl newspapers websites, news agencies websites, or news websites.  For one of these sources to be considered a valid source for my corpus, each of the news stories they publish should contain: 1) an identification number, 2) story publication date and time, 3) story title, 4) story text body, and 5) a list of related stories.

Many news agencies like Reuters and Associated Press, news websites like BBC News and Huffington Post, and newspaper websites like The Guardian and New York Times all meet these conditions in their published stories.  There are few differences though that make some of them better than the others, and each one of them has its advantages and disadvantages.

\paragraph{News agencies} typically publish more stories per day than the other sources I considered.  This makes the news timeline richer with stories and makes more news timelines.  They tend to publish more follow-up stories than other sources which contributes to the richness of the timeline as well.  They also cover a bigger variety of topics and larger geographical region, usually all world countries, than other sources.  They do not have limitations on the word count of their stories as they are not restricted by page space or other space requirements which make their stories richer in syntax and vocabulary than other sources.

On the other hand, news agencies only publish stories produced by their own journalists, and their own journalists have to follow the agency's guidelines and rules of writing, editing and updating news stories.  This restriction makes the stories more uniform regarding editorial structure, and sometimes they have vocabulary uniformity also.  Even though this uniformity does not affect the journalistic quality of the news stories they publish, it makes the task of news timeline creation easier for the topic model.  The topic model could just learn the set of keywords used for the news stories that fall under a timeline that gets repeated over and over by the same journalist who covers the topic for the news agency over a certain period of time.  The model could be better tested if the stories are written by different journalists belonging to different organizations and following different rules and guidelines.

The process of crawling news agencies news websites is usually easy.  Their web pages are usually well formatted, have few advertisements and little unrelated content.  They are usually well tagged also.  News agencies often tend not to have strong political bias in their news coverage when compared to newspapers and news websites.  This will be reflected in the lack of opinionated adjectives that comes with political bias which affects the vocabulary structure of the news stories.

\paragraph{News websites} typically only exist in electronic form on the web and collect their news stories from different sources.  They purchase stories from different news agencies.  Some of them have their own dedicated journalists and freelance journalists, and some purchase stories from other news websites and newspapers.  Many of the stories they gather from other resources pass through an editorial step in which some sections of the story may be removed for publication space limitations.  In other cases, sections written by their own journalists or collected from other sources could be added to the story, or two stories could be merged to fill up publication space.  This editorial process could lead to syntactic and vocabulary richness in the news stories as the stories belonging to the same timeline would have lots of synonym words, and this synonymy should be learned by the topic model in order to classify the related stories as belonging to the same timeline.

As news websites only exist in electronic form on the web, they tend to make the most of it and be rich in media format; a story could have the main story text, text commentary, audio, video and links to other websites covering the same story.  This richness in media format could help in modeling the document and in placing it on the correct timeline.  This richness makes it harder for automated crawling applications to extract the news story components, such as title, story text, and related stories, from a page crammed with different kinds of content like advertisements, unrelated stories, headers, footers, and such.

News websites tend to have soft publication space limitation;  a news story could run few pages long or be a couple of paragraph.  The list of related stories could be rich in the case of news websites as the media richness discussed earlier usually includes links to news stories published by other sources, let us call them external sources, and by the same news website, let us call them internal sources, and that again helps in better testing the news timeline creation.  Different news stories along the same news timeline will include many synonym words, and the synonymy should be learned by the topic model to identify these stories as belonging to the same timeline.

News websites tend to have more political bias than news agencies, but less bias than newspapers.  This means related news stories they publish would not carry so much at the same opinionated adjectives in their coverage of the story.  It is to be noted that this will usually only apply to the stories written by their own journalists, and not collected or purchased from news agencies.

\paragraph{Newspaper websites} tend to fall in the middle between news agencies websites and news websites regarding news stories diversity, coverage and richness.  Newspapers usually collect news stories from news agencies for regions of the world they do not cover.  They have their own journalists who write according to the newspaper's editorial and political rules and guidelines, and they purchase stories from other newspapers also.  This makes a similar syntactic and vocabulary richness in their content, even though it does not match that of the news websites.  The stories collected or purchased from other sources typically go through an editorial process in which a story may be cut short, extended by adding content from other sources to it, or merged with another story purchased or collected from another sources covering the same topic.  Newspapers usually put on their website all the stories they publish on paper.  They sometimes publish extended versions online, and may also have online exclusive content.  The online exclusive content usually has soft space limitations.  The paper-published content usually has hard space limitations though.  

The process of crawling newspaper websites is usually easier than crawling news websites as the former's web pages usually contain less advertisements, less unrelated content, and less links to other sources.  Newspapers, news web pages are usually well tagged.  Newspaper websites tend to heavily link to their own sources, unlike news websites which link heavily to external sources.  If I am interested in creating a rich news timeline, this will be counted in favor of newspaper websites as I usually crawl web pages with the same format, which means crawling only internal sources not external sources because they usually have different web page format.

Newspapers usually tend to have stronger political bias in their news coverage when compared to news agencies and news websites.  They tend to use more strongly opinionated adjectives.  This has the side effect of having news stories belonging to the same timeline sharing a collection of adjective keywords that makes the process of clustering the news stories into different timelines easier.  This is usually not desirable as this could make the model more reliant on clustering stories based on keywords, instead of using word distribution, and even better, a word distribution that changes over time.

A good text news stories corpus would contain a syntactically and semantically rich collection of stories.  The stories should be collected from different sources and written by different authors who follow different writing and editorial rules and guidelines, or even better if they do not share any set of these standards.  This diversity in vocabulary and syntactic structure will translate to a larger set of synonyms and antonyms being used in the collection.  The topic model is tested on the different relationships among these words and the degree to which it correctly learns these relationship translates to good performance in document classification and correct placement of a news story on its natural timeline.

A good corpus should have a big set of related stories for each news story it contains, if such related stories exist in the corpus.  The bigger the set, the richer the news timeline becomes, and the more chances of success and failure the topic model will have in creating the timeline.  This will generate another challenge in discovering the birth/death/revival of topics.  The longer the timeline gets, the more of this topic life cycle can be detected or missed.

The set of related news stories provided by many news outlets for each of the news stories they release could be either manually created or automatically generated.  In the manual creation process, a news editor sifts through past news stories manually or assisted by search tools and looks for relevant stories.  The relevancy judgment is done by a human.  The number of stories in the set of related stories is usually kept within a certain limit to emphasize the importance and relatedness of the stories in the set.  The set usually includes the more recent five or seven related stories.  A related stories set created manually this way is needed for my corpus to be used in testing the performance of the timeline creation algorithm and the topic detection and tracking algorithm.  This manually created set is called the \emph{gold standard}.  It represents the highest possible standard that any algorithm trying to solve this problem should seek to match.

Not all news outlets use human annotators to judge the relatedness of the news stories to create a set of related stories.  Some of them use algorithms to do this job.  This is usually driven by the need to avoid the cost of having a human annotator.  Related stories generated by such systems, like Newstracker\footnote{BBC links to other news sites: \url{http://www.bbc.co.uk/news/10621663}} system being used by BBC News, cannot be used as a gold standard for my system.  They do not represent the ultimate standard that cannot be surpassed.  Their performance can be improved upon by other systems addressing the same problem.  They can be used as a baseline for the performance of other systems that tries to match or even exceed their performance.

A good corpus should have news stories time stamped with high time granularity.  Some news sources like news agencies publish news stories round-the-clock;  Agence France-Presse (AFP) releases, on average, one story every 20 seconds\footnote{Agence France-Presse (AFP) releases on average one news story every 20 seconds (5000 per day) while Reuters releases 800,000 English-language news stories annually.  See: \url{http://www.afp.com/en/agency/afp-in-numbers} and \url{http://www.infotoday.com/it/apr01/news6.htm}}.  Such a fast-paced publication needs a few minutes' resolution and accurate time stamped stories to correctly place the story on its timeline.

A news timeline typically contains many news stories, usually over ten stories, and extends over a relatively long period of time, several months long.  One set of related news stories cannot be used to create a timeline; it typically contains less than ten related stories, and in many cases, the stories extend over a few days or a week.  To be able to create a timeline, different sets of related news stories have to be chained.  For each story in the set of related stories, I get its set of related stories.  For each one of these stories in turn, I get a set of related stories.  This process can repeat and I can extend the chain as desired.  However, the longer the chain gets, the less related the stories at the end of the chain will be to the original story at the other end of the chain.  This is because at each step along the chain we compare the stories to the current node (story) in the chain, and not the first node (original) that we want to get a set of stories related to it.

I create a news corpus by crawling news websites like the British newspaper \emph{The Guardian} website.  I start by a set of diverse news stories seeds, or links.  This set typically covers a wide variety of topics and geographical areas.  For each one of these links, I use the link as the story identifier.  I crawl the main news story text and title, its release date and time, and the set of related news stories.  I follow the links to the related news stories section to create the set of related news stories.  These related stories were hand-picked by a news editor and therefore I can use them for my gold standard.  I repeat this process until I crawl a predefined number of news stories.

\section{Corpora}
\label{sec:corpus}
For my experiments I use two corpora: a Reuters corpus, and a BBC news corpus.
\subsection{Reuters}
\label{subsec:corpus:reuters}
The Reuters-21578 corpus \cite{reuters21578}\footnote{http://www.daviddlewis.com/resources/testcollections/reuters21578/} is currently the most widely used test collection for text categorization research, and it is available for free online.  The corpus is made of 21578 documents in English that appeared on the Reuters newswire in 1987.  The average number of unique words in a news story in this corpus is 58.4.  It has a vocabulary of 28569 unique words.  This collection of news articles comes in XML format.  Each document comes with:

\paragraph{Date}  The date the story was published.  This date is accurate to milliseconds.  E.g. 26-FEB-1987 15:01:01.79.

\paragraph{Topic}  A manually assigned topic that the news story discusses.  There are 135 topics.  E.g. Cocoa.

\paragraph{Places} A Geographical location where the story took place.  E.g. El-Salvador.

\paragraph{People} Names of famous people mentioned in the corpus.  E.g. Murdoch.

\paragraph{Organizations} Names of organizations mentioned in the corpus.  E.g. ACM.

\paragraph{Exchanges} Abbreviations of the various stock exchanges mentioned in the corpus.  E.g. NASDAQ.

\paragraph{Companies} Names of companies mentioned in the corpus.  E.g. Microsoft.

\paragraph{Title} The title of the news story.  E.g. Bahia cocoa review.

\paragraph{Body} The body of the news story.  E.g. Showers continued throughout the week in the Bahia cocoa zone, ...

In my experiments, I only use \emph{Date} and \emph{Body}.

\subsection{BBC News}
\label{subsec:corpus:bbcn}
The BBC news corpus is made of 10,000 news stories in English I collected myself from the BBC news website\footnote{http://www.bbcnews.com}.  The stories in the corpus cover about two and a half years time period, from April 2010 to November 2012.  The average number of unique words in a news story in this corpus is 189.6 and the corpus has a vocabulary of 64374 unique words.  For each news story I collect:

\paragraph{ID} An identification for the news story.  I use the story web page URL for that.  E.g. http://www.bbc.co.uk/news/world-europe-10912658

\paragraph{Date}  The date the story was published.  This date is accurate to seconds.  E.g. 2010/08/09 15:51:53

\paragraph{Title}  The title of the news story.  E.g.  Death rate doubles in Moscow as heatwave continues

\paragraph{Body}  The body of the news story.  E.g.  Death rate doubles in Moscow as heatwave continues Extreme heat and wildfires have led to ...

\paragraph{Related} The ID of a related news story.  E.g. http://www.bbc.co.uk/news/world-europe-10916011

\section{Evaluation of proposed solution}
To evaluate the performance of the proposed solution I am going to use the per-word log-likelihood of a news story conditioned on the rest of the data.  More formally, the performance measure will be:

\begin{equation}
  \label{eq:20}
  likelihood_{pw} = \frac{\sum_{\pmb{w}\in \mathcal{D}}\log p(\pmb{w}|\mathcal{D}_{-\pmb{w}})}{\sum_{\pmb{w} \in \mathcal{D}} |\pmb{w}|}
\end{equation}

Where $\pmb{w}$ is a document in the collection of documents $\mathcal{D}$,  $|\pmb{w}|$ is the number of words in document $\pmb{w}$, and $\mathcal{D}_{-\pmb{w}}$ is the collection of documents $\mathcal{D}$ minus document $\pmb{w}$.

The use of log-likelihood as an evaluation measure for topic models is widely accepted in the machine learning community \cite{chang09, teh07, wallach09}.  It is a measure of the goodness of fit of the model to the data \cite{banerjee08}.  The better the fit, the more likely the model will label the documents with the right topics.

Using the natural log function is convenient for several reasons.  1) It reduces the chances of running into an underflow problem due to very small likelihood values when simulating the model.  2) It allows us to use summation of terms (computationally cheap) instead of product of terms (more computationally expensive), and 3) The natural log function is a monotone transformation which means that extrema of the log-likelihood function is equivalent to the extrema of the likelihood function \cite{pratt76}.

\begin{figure}
  \centering
  \includegraphics[width=0.9\linewidth]{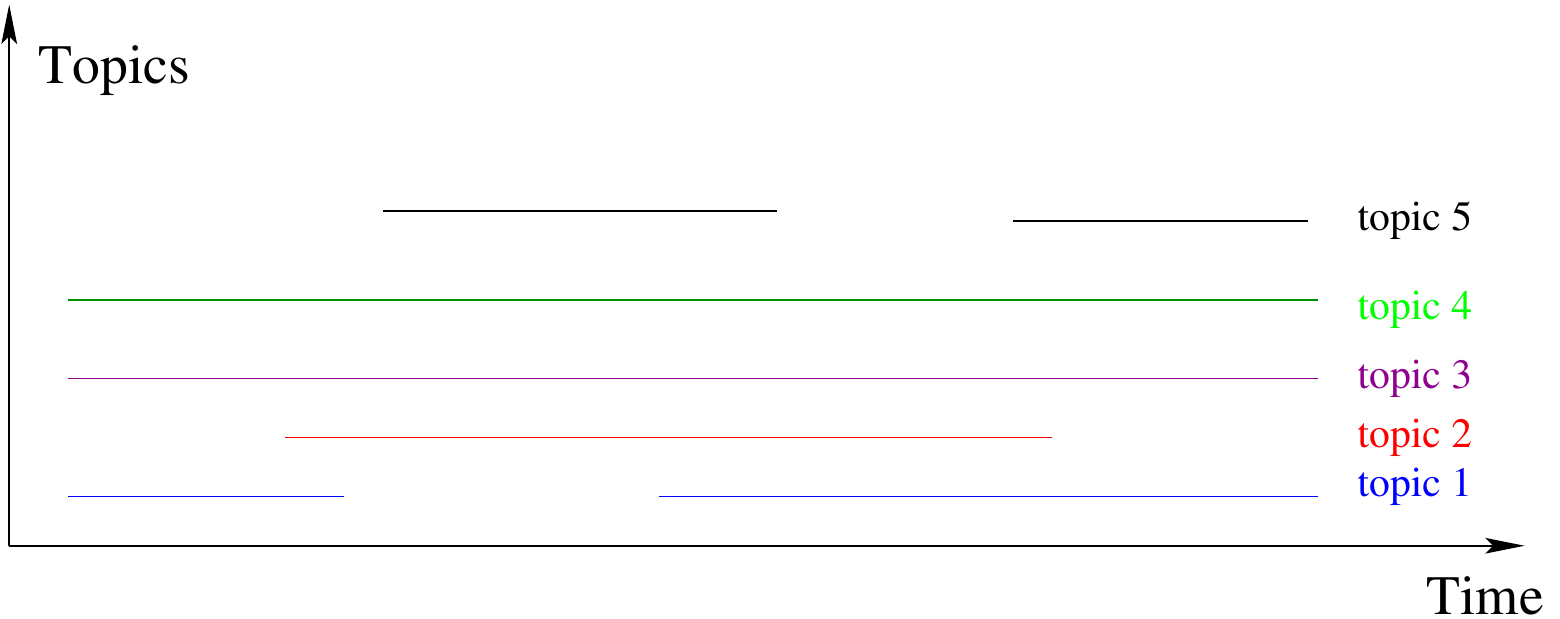}
  \caption{Illustration of the birth/death/resurrection of five topics.  Discontinuity of a topic indicates its death. \label{topicTracking}}
  
\end{figure}

\paragraph{Scalability} A real world news feed contains hundreds of thousands of archived news articles with more articles being generated every minute.  An offline model is expected to be able to build the storyline of a hundred thousand documents in less than a day running on a desktop machine.  An online model, if implemented, should be able to streamline the incoming documents in real time.

\paragraph{Accuracy} The accuracy of the model will be measured with the per-word log-likelihood given in \eqref{eq:20}.  A higher value of the per-word log-likelihood is desirable.  It means that the documents have topics with word distributions that are consistent with the per-topic word distributions the model has learned.


\cleardoublepage
\chapter{Experimental design and results}
I ran experiments to evaluate the performance of my ciDTM model against the continuous-time dynamic topic model (cDTM) presented in Section~\ref{subsec:cdtm}, and the online Hierarchical Dirichlet process presented in Section~\ref{subsec:ohdp}.  To find the best settings for each of the competing models, I evaluated their performance with different practical setting values.  In the next two sections I show the results obtained by running these experiments on the Reuters-21578 and the BBC news corpora presented in Section~\ref{sec:corpus} on page~\pageref{sec:corpus}.

I used the coded provided by \citet{wang08}\footnote{http://www.cs.cmu.edu/~chongw/software/cdtm.tar.gz} to build and run the cDTM system, and used the code provided by \citet{wang11}\footnote{http://www.cs.cmu.edu/~chongw/software/onlinehdp.tar.gz} to build and run the oHDP system.  In my own system, ciDTM, I used some code from the previous two software packages to build and run it.  I used MALLET \cite{McCallumMALLET} also for corpus creation, parsing and filtering.

\section{Reuters corpus}
\begin{figure}[!h]
  \centering
  \includegraphics[width=\textwidth]{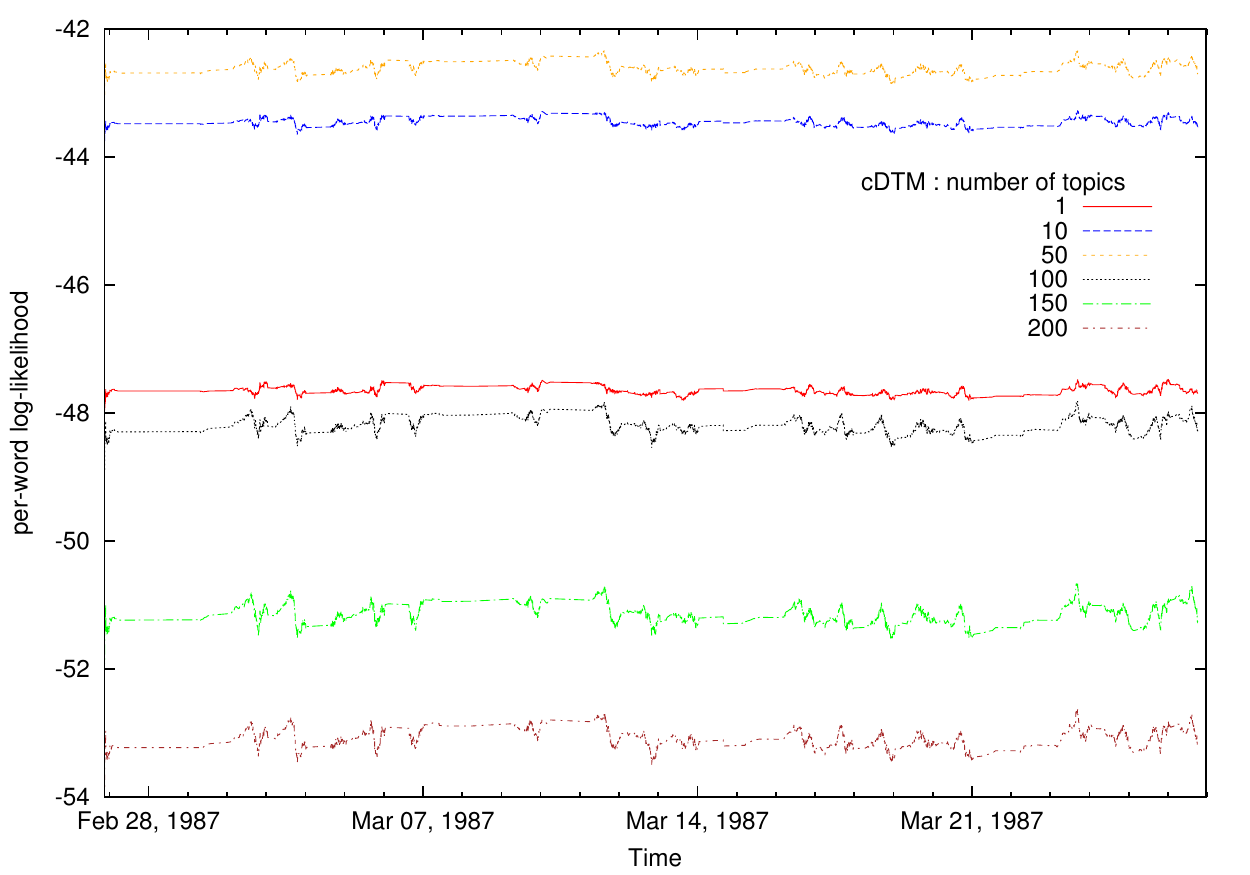}
  \caption{cDTM performance varies with different values for the number of topics using Reuters corpus.  A value of 50 leads to best model fit among the six values tried.  The per-word log-likelihood values shown are the moving average of a set of 100 consecutive documents.  This averaging was needed to smooth out the plot.  Higher values indicate better model fit on this 10,000 news stories corpus.\label{fig:cdtm_likelihood}}.
\end{figure}

\paragraph{cDTM number of topics:}
Since the cDTM has a fixed number of topics, it is expected that its performance will vary with different values of this parameter.  The closer this value is to the average number of topics this model can fit to the data over time the higher the performance will be.  I first evaluated the performance of the cDTM with different values for the number of topics ranging from 1 to 200 topics.  The per-word log-likelihood values for the inferred documents is presented in Figure~\ref{fig:cdtm_likelihood}.  Higher values indicate better model performance.  In all cDTM experiments to follow, the cDTM model was first (unsupervised) trained on a set of 5,000 documents sampled uniformly from the entire corpus of 10,000 documents, then the performance of the trained model was evaluated using the other documents in the corpus.  This guarantees that the documents the model is tested on cover the same time range as the training documents.

Figure~\ref{fig:cdtm_likelihood} shows that among the six values tried, the model performed best with 50 topics. This reflects the moderate richness of the 5,000 Reuters news stories the model was tested on.  By manually inspecting the corpus, I found that most of its documents mainly discuss economy and finance and slightly covers politics.  This corpus is considered somehow limited in the topics it covers as compared to the wide spectrum of topics covered by the stories released by Reuters news agency.

\begin{figure}[!h]
  \centering
  \includegraphics[width=\textwidth]{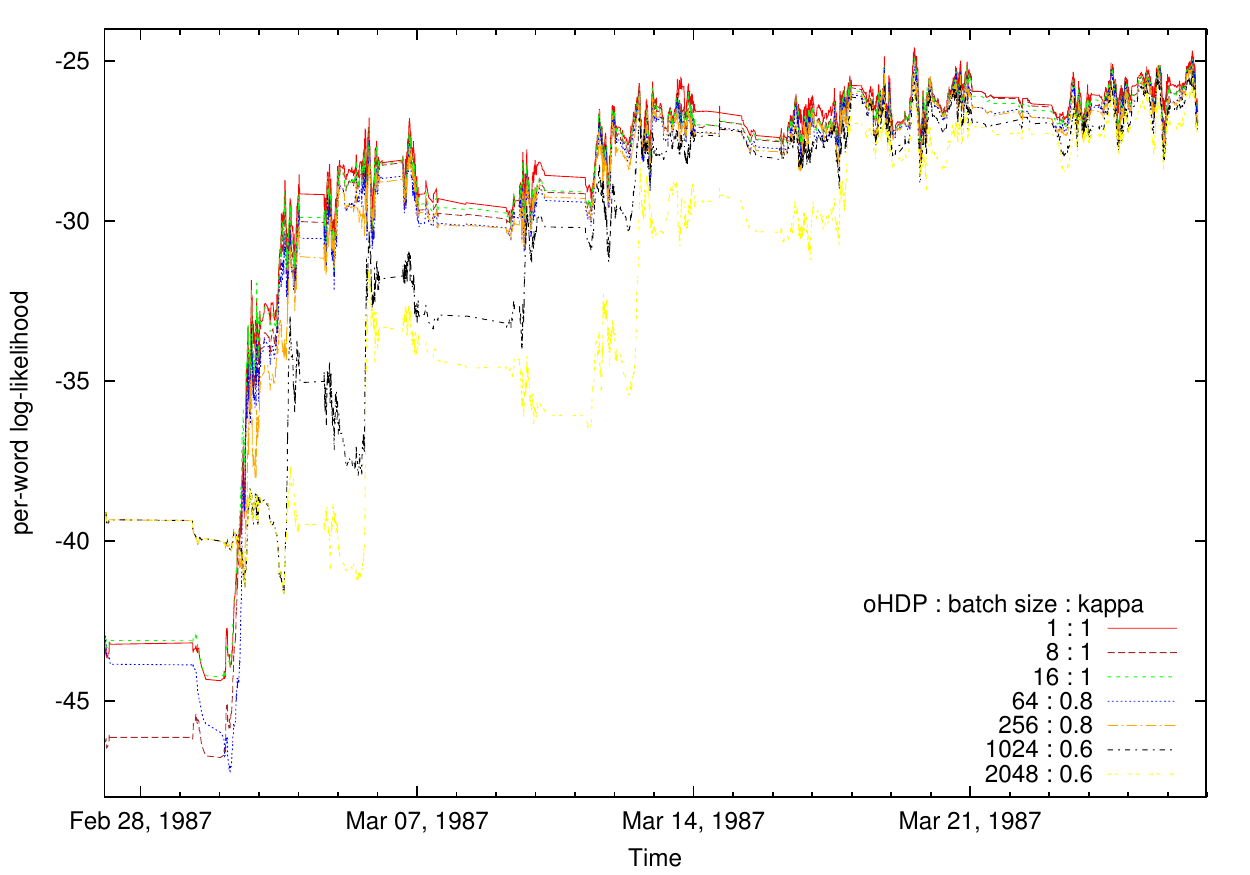}
  \caption{oHDP per-word log-likelihood for different batch size values using Reuters corpus.  The per-word log-likelihood values shown are the moving average of a set of 100 consecutive documents.  This averaging was needed to smooth out the plot.  Higher values indicate better model fit on this 10,000 news stories corpus.\label{fig:ohdp_likelihood}}.
\end{figure}

\paragraph{oHDP batch size} oHDP uses online variational inference to fit its parameters.  For each iteration of the algorithm, it uses a small batch of documents to update these parameters.  I experimented with different batch size and model parameter ($\kappa$) values to determine the best value to use.  These values were suggested by \citet{wang11} as they generated the best results in their case.  Lower $\kappa$ values favor larger batch sizes.  It is to be noted that this batch size value should be relatively small to effectively retain the online inference feature of this algorithm.  Experiment results with seven values are presented in Figure~\ref{fig:ohdp_likelihood}.  The two largest batch sizes, 1024 and 2048, were only included to show the trend of decreasing performance with increasing batch size.  It is not used later on because such a large batch size requires longer period of waiting time to collect the documents and update the system making the system out-of-date in the meantime.

For the oHDP model, I use the same values suggested by \citet{wang11} and given in Table~\ref{tbl:ohdp}.

\begin{table}[!h]
\centering
\begin{tabular}{rl}
  \hline
  $T$        & 300\\
  $K$        & 20\\
  $\alpha$   & 1\\
  $\zeta$    & 0.01\\
  $\gamma$   & 1\\
  \hline
\end{tabular}
\caption{Setting values used for oHDP}
\label{tbl:ohdp}
\end{table}

The per-word log-likelihood values for the model is better with smaller batch sizes.  The difference is minor for batches smaller than 64.

\begin{figure}[!h]
  \centering
  \includegraphics[width=\textwidth]{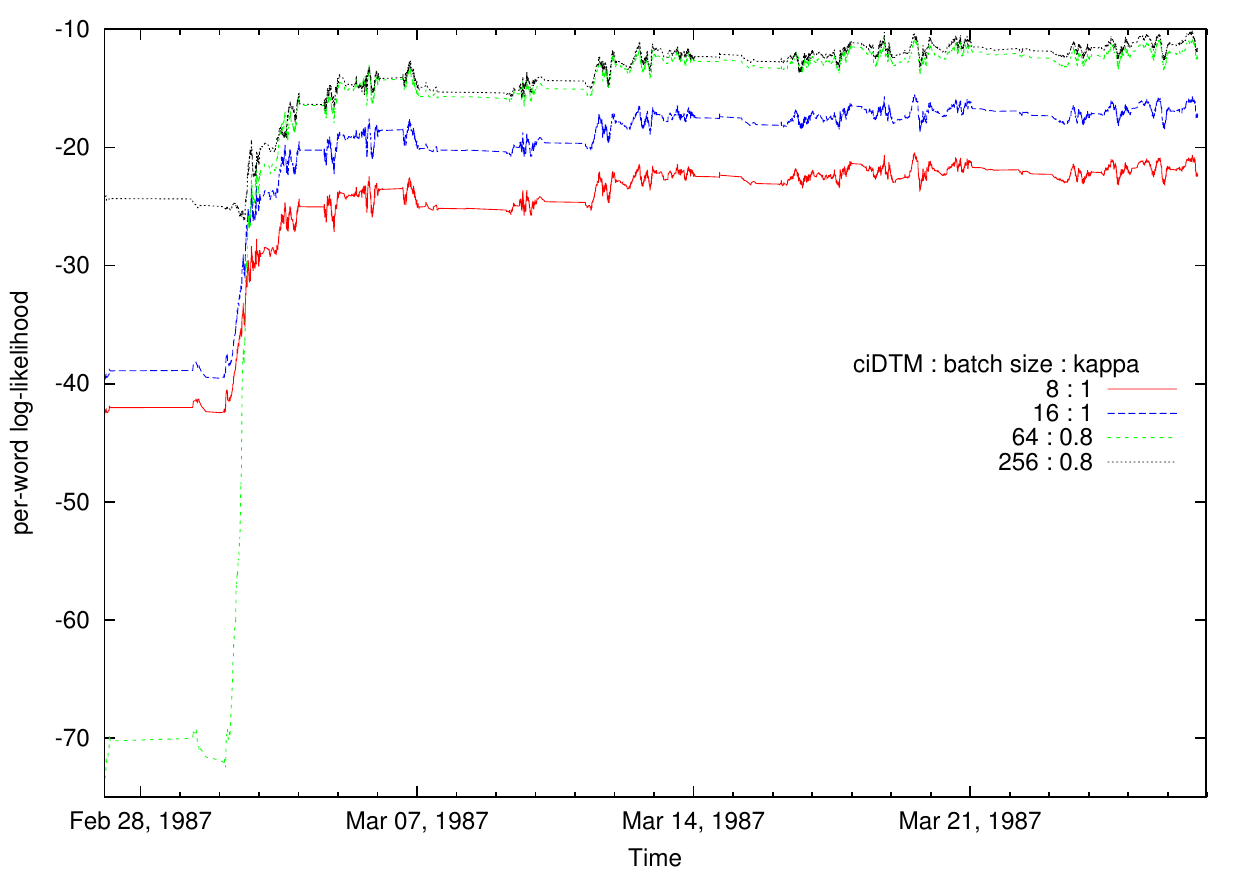}
  \caption{ciDTM per-word log-likelihood for different batch size values using Reuters corpus.  The per-word log-likelihood values shown are the moving average of a set of 100 consecutive documents.  This averaging was needed to smooth out the plot.  Higher values indicate better model fit on this 10,000 news stories corpus.\label{fig:cidtm_likelihood}}.
\end{figure}

\paragraph{ciDTM batch size} The size of the batch of documents ciDTM processes with every iteration of its algorithm has a double effect: 1) it affects the convergence of the online variational inference algorithm as was the case with oHDP above, and 2) it affects the variational Kalman filter used to evolve the per-topic word distribution in continuous-time.  It is expected that larger batches of documents would improve the performance of the model as the document timestamp (arrival time) information and inter-arrival times between documents will be used by the filter to dynamically evolve the per-topic word distribution.  It is to be noted that a batch of size one cannot be used in ciDTM.  The Kalman filter used by the model evolves per-topic word distribution based on documents inter-arrival times in the batch.  The process running the filter starts fresh with every new batch of documents.  Figure~\ref{fig:cidtm_likelihood} shows the results of an experiment in which four different batch size values were tried.

Since this model has the benefit of using an online variational inference algorithm, the batch size should be kept small to maintain this desirable feature.  From the Figure we can see that a point of diminishing return is reached with a batch of size 64.  The performance gained by increasing the batch size from 64 to 256 is negligible.  On the other hand, if documents arrive to the system at a fixed rate.  Then 64 documents will need 4.5 hours to gather, while 256 documents would require 18 hours to collect.  With the pace of 14 documents per hour encountered in this corpus, once a day update for the model would be enough to keep it up-to-date.  I use a batch size of 256 as the optimal batch size for this model.

The ciDTM model parameter values I used are equal to their corresponding parameter values in the cDTM and oHDP models, where applicable.  More specifically, the model parameters I used are:

\begin{table}[!h]
\label{tbl:cidtm}
\centering
\begin{tabular}{rl}
  \hline
  $T$        & 300\\
  $K$        & 20\\
  $\alpha$   & 1\\
  $\zeta$    & 0.01\\
  $\gamma$   & 1\\
  $\alpha_0$ & 0.2\\
  \hline
\end{tabular}
\caption{Setting values used for ciDTM}
\end{table}

\begin{figure}[!h]
  \centering
  \includegraphics[width=\textwidth]{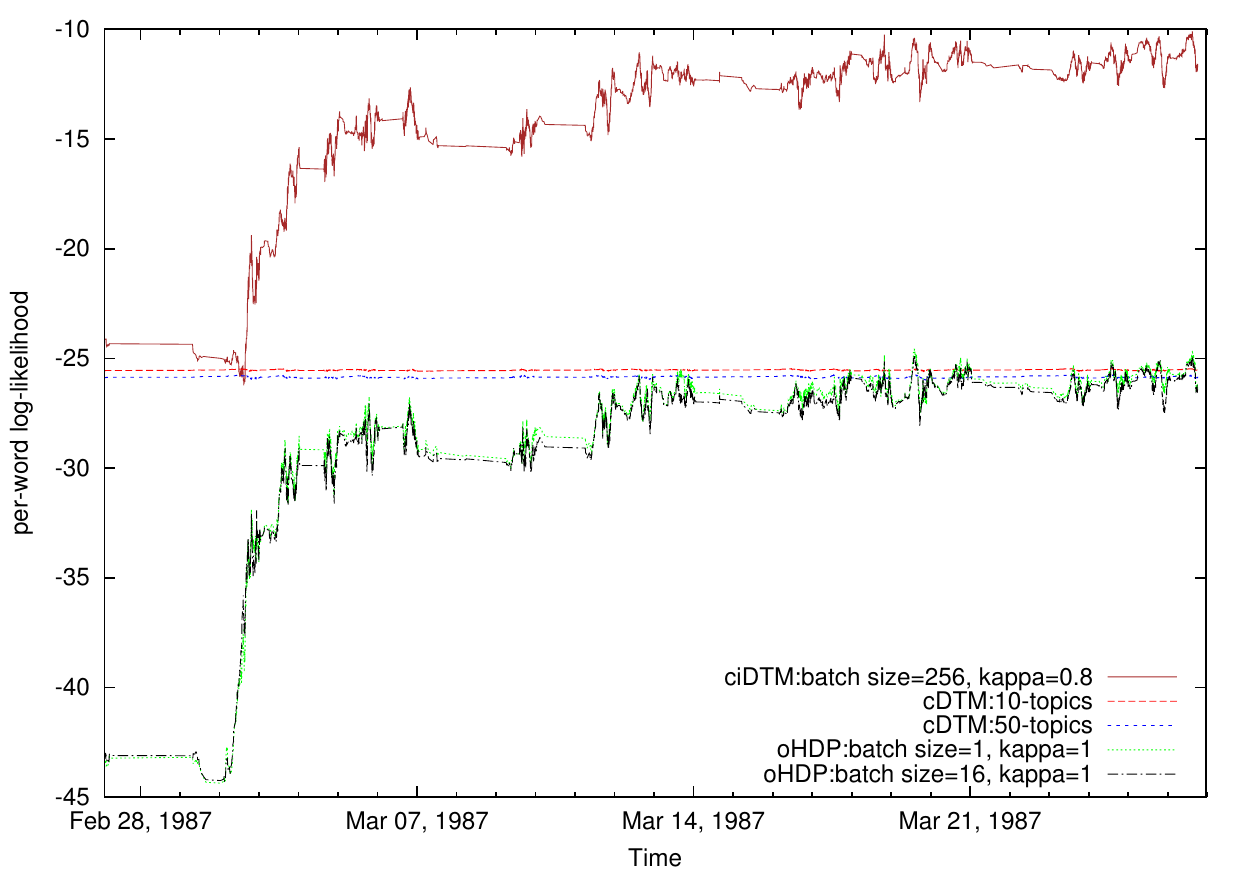}
  \caption{A comparison of ciDTM, oHDP and cDTM using their best setting values discovered earlier using Reuters corpus.  The per-word log-likelihood values shown are the moving average of a set of 100 consecutive documents.  This averaging was needed to smooth out the plot.  Higher values indicate better model fit on this 10,000 news stories corpus.\label{fig:comp_likelihood}}.
\end{figure}

\paragraph{Comparing three models} Figure~\ref{fig:comp_likelihood} shows the per-word log-likelihood performance for the three competing models using their best setting values discovered earlier.  It is clear that ciDTM outperform both cDTM and oHDP by a big margin when tested on this news corpus especially ciDTM with batch sizes greater than 16.

\begin{figure}[!h]
  \centering
  \includegraphics[width=\textwidth]{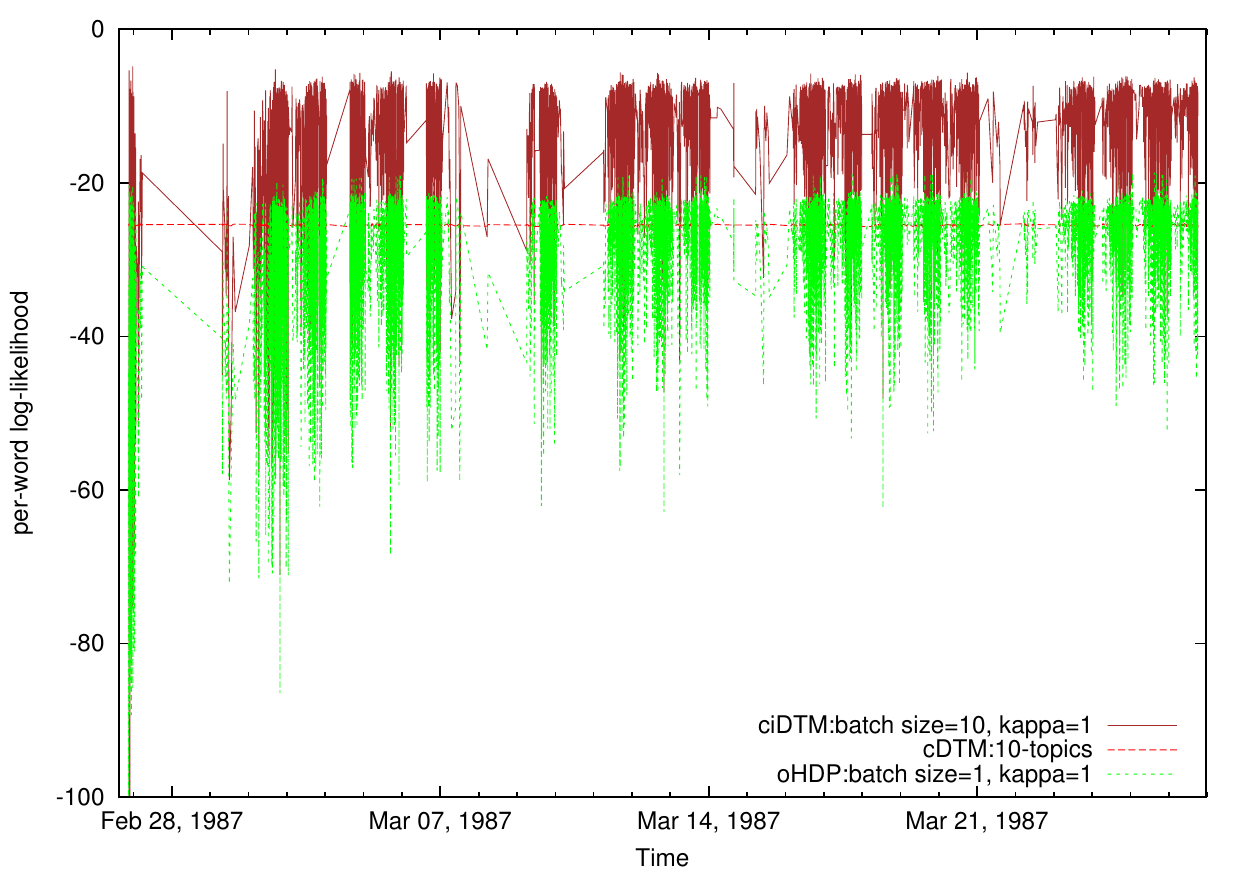}
  \caption{Comparison of the per-word log-likelihood for cDTM, oHDP and ciDTM using 10,000 news stories from the Reuters21578 corpus \cite{reuters21578}.  Higher per-word log-likelihood indicate a better fit of the model to the data.\label{fig:3-models-log-likelihood}}.
\end{figure}

Figure~\ref{fig:3-models-log-likelihood} shows the non-smoothed per-word log-likelihood values for each of the three models with ciDTM with batch size of 10, oHDP with batch size of 1, and cDTM with 10 topics.  ciDTM outperformed the oHDP, while the cDTM performed the words of all three.  The improvement the ciDTM model over oHDP is significant knowing that the value on the independent axis is a log value.  The per-word log-likelihood value is evaluated for each document in the mini batch of documents processed by the ciDTM in each iteration of the algorithm.

Since cDTM relies on an offline learning algorithm.  The per-word log-likelihood obtained is evaluated after the model was trained on the training data set.  This explains the steady state performance of the cDTM algorithm as the word distribution for the topics covered by the corpus were learned by the model first before the log-likelihood values were evaluated.  The fluctuation in the data in the case of ciDTM and oHDP is due to the fact that they are encountering new documents with every time step. As the model learns the word distribution for these topics, the fluctuations starts decreasing over time.  The gap periods in the graph are dormancy periods when no news stories were being published.

\section{BBC news corpus}
\label{sec:results_bbcn}
I repeated the same set of experiments I ran above and used the BBC news corpus presented in Section~\ref{subsec:corpus:bbcn} on page \pageref{subsec:corpus:bbcn}.

\begin{figure}[!h]
  \centering
  \includegraphics[width=\textwidth]{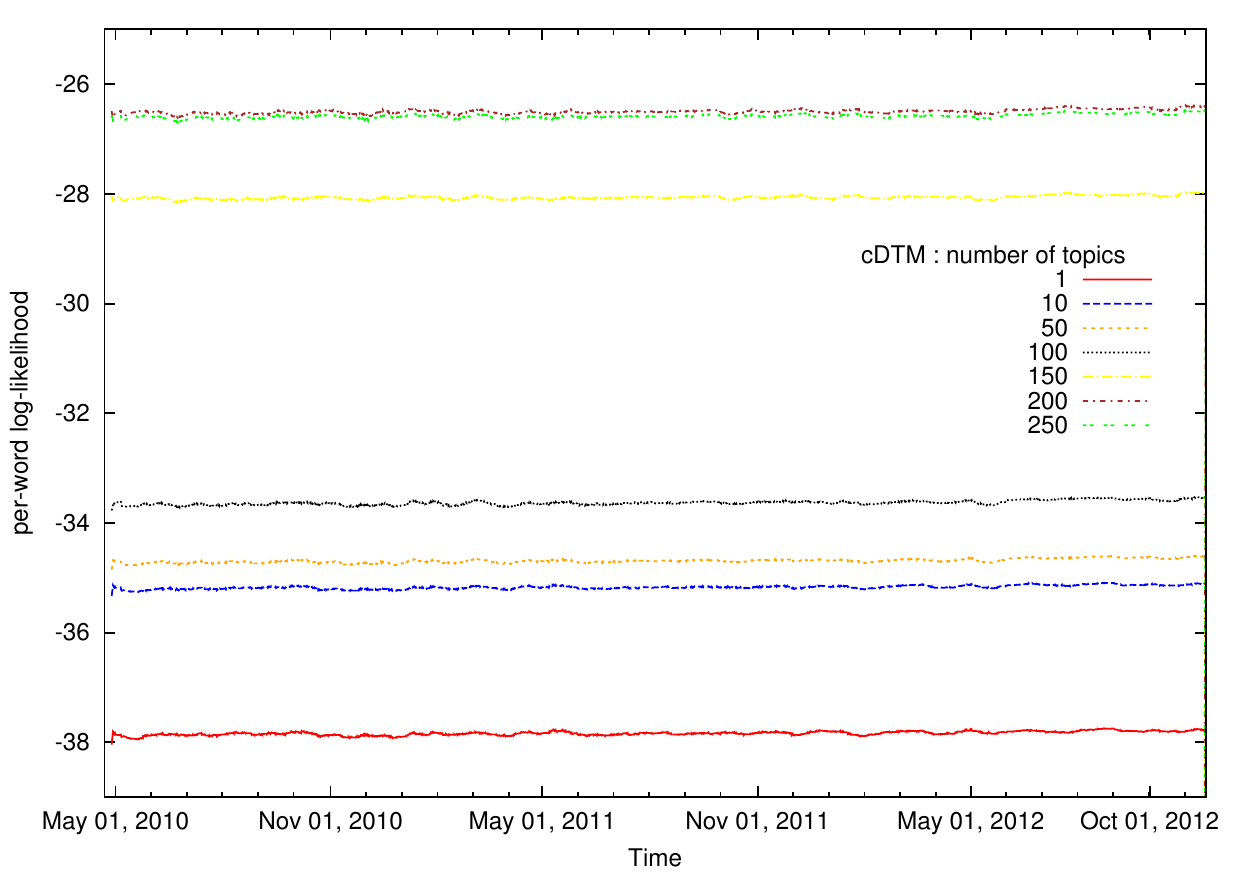}
  \caption{cDTM performance varies with different values for the number of topics using the BBC news corpus.  A value of 200 leads to best model fit among the seven values tried.  The per-word log-likelihood values shown are the moving average of a set of 100 consecutive documents.  This averaging was needed to smooth out the plot.  Higher values indicate better model fit on this 10,000 news stories corpus.\label{fig:cdtm_like_bbcn}}
\end{figure}

\paragraph{cDTM number of topics}  The cDTM model was trained on a set of 5,000 documents sampled uniformly from the 10,000 document corpus.  The trained model was then used to infer the documents' topic mixture and the topics' word mixture for the other half of the corpus.  Figure~\ref{fig:cdtm_like_bbcn} shows the per-word log-likelihood value obtained using different values for the number of topics.

The per-word log-likelihood performance of this model improved as the number of topics used increased.  This improvement reached its peak with a model of 200 topics.  Higher number of topics hurt the model performance as can be seen in the 250 topics model case which performs slightly worse than the 200 topics model.  The reason this model reached its peak performance with 200 topics with this corpus while the same model reached a peak performance with 50 topics with the Reuters corpus can be explained by the properties of these two corpora.  The BBC news corpus has a higher vocabulary richness and the relatively lengthy documents as compared to the Reuters news corpus.  This BBC news corpus has a vocabulary of 64374 unique words and an average document length of 189.6 unique words.  The Reuters news corpus has a vocabulary of 28569 unique words and an average document length of 58.4 unique words.

\begin{figure}[!h]
  \centering
  \includegraphics[width=\textwidth]{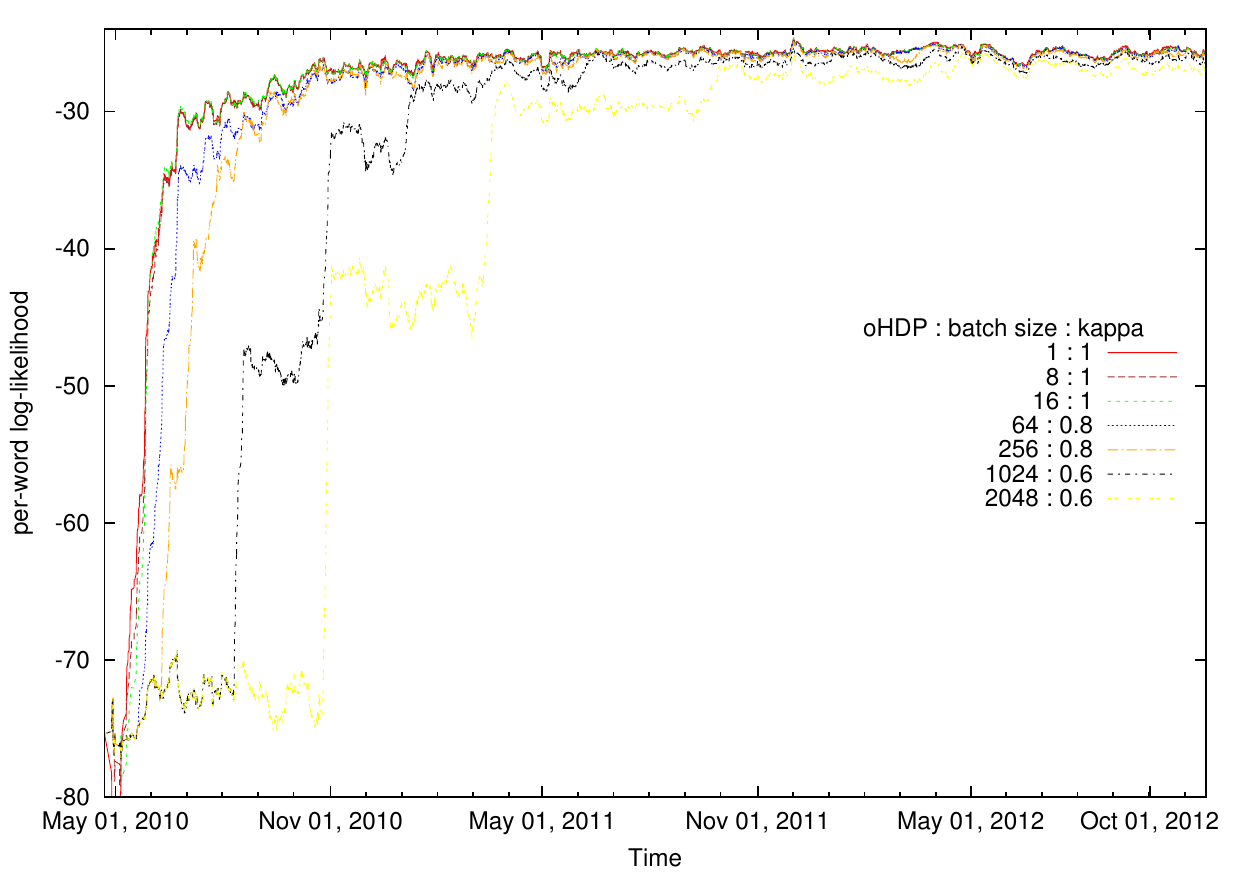}
  \caption{oHDP per-word log-likelihood for different batch size values using the BBC news corpus.  A value of 200 leads to best model fit among the seven values tried.  The per-word log-likelihood values shown are the moving average of a set of 100 consecutive documents.  This averaging was needed to smooth out the plot.  Higher values indicate better model fit on this 10,000 news stories corpus.\label{fig:ohdp_like_bbcn}}
\end{figure}

\paragraph{oHDP batch size}  I experimented with different batch size and $\kappa$ values for the oHDP model.  I used the values suggested by \citet{wang11} and other values that follow the same trend of these values.  \citet{wang11}  found that higher $\kappa$ values favor smaller batch sizes.

Figure~\ref{fig:ohdp_like_bbcn} show the results of running these experiments.  The trend obtained is similar to the one obtained with the Reuters corpus.  The best per-word log-likelihood was obtained using a batch of size 1.  As the batch size increased, the performance tended to consistently drop.

\begin{figure}[!h]
  \centering
  \includegraphics[width=\textwidth]{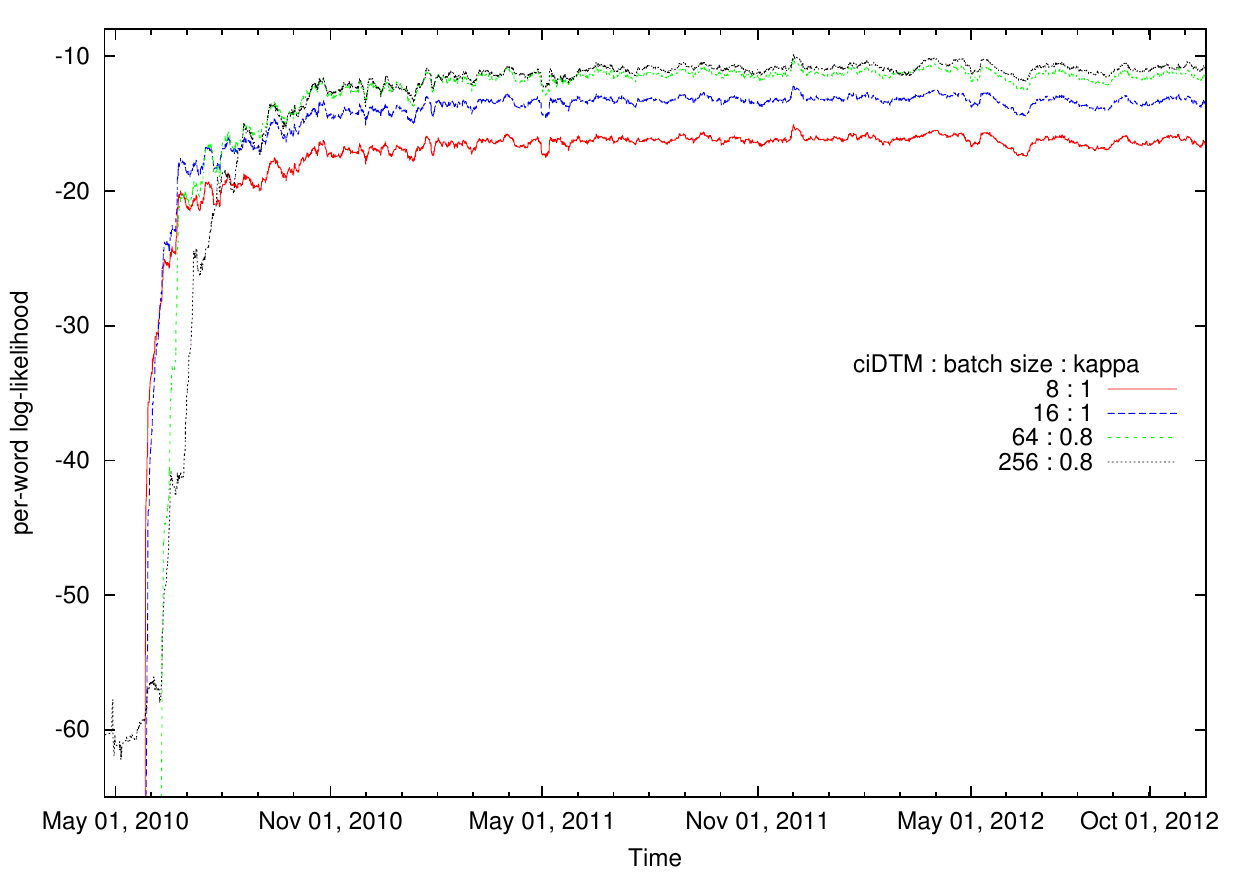}
  \caption{ciDTM per-word log-likelihood for different batch size values using the BBC news corpus.  The per-word log-likelihood values shown are the moving average of a set of 100 consecutive documents.  This averaging was needed to smooth out the plot.  Higher values indicate better model fit on this 10,000 news stories corpus.\label{fig:cidtm_like_bbcn}}
\end{figure}

\paragraph{ciDTM batch size}  The size of the batch of documents the ciDTM model processes in each iteration affects the convergence of the inference algorithm and the variational Kalman filter used to evolve the per-topic word distribution in continuous time.

Figure~\ref{fig:cidtm_like_bbcn} shows how the per-word log-likelihood performance of the model changes with different batch size values.  The trend shown mirrors the one obtained using the Reuters corpus.  The model performance improves as the batch size increases.  A point of diminishing returns is reached around a value of 64 for the batch size.  Minor performance gain obtained using higher batch size values is outweighted by the longer periods separating model updates.

\begin{figure}[!h]
  \centering
  \includegraphics[width=\textwidth]{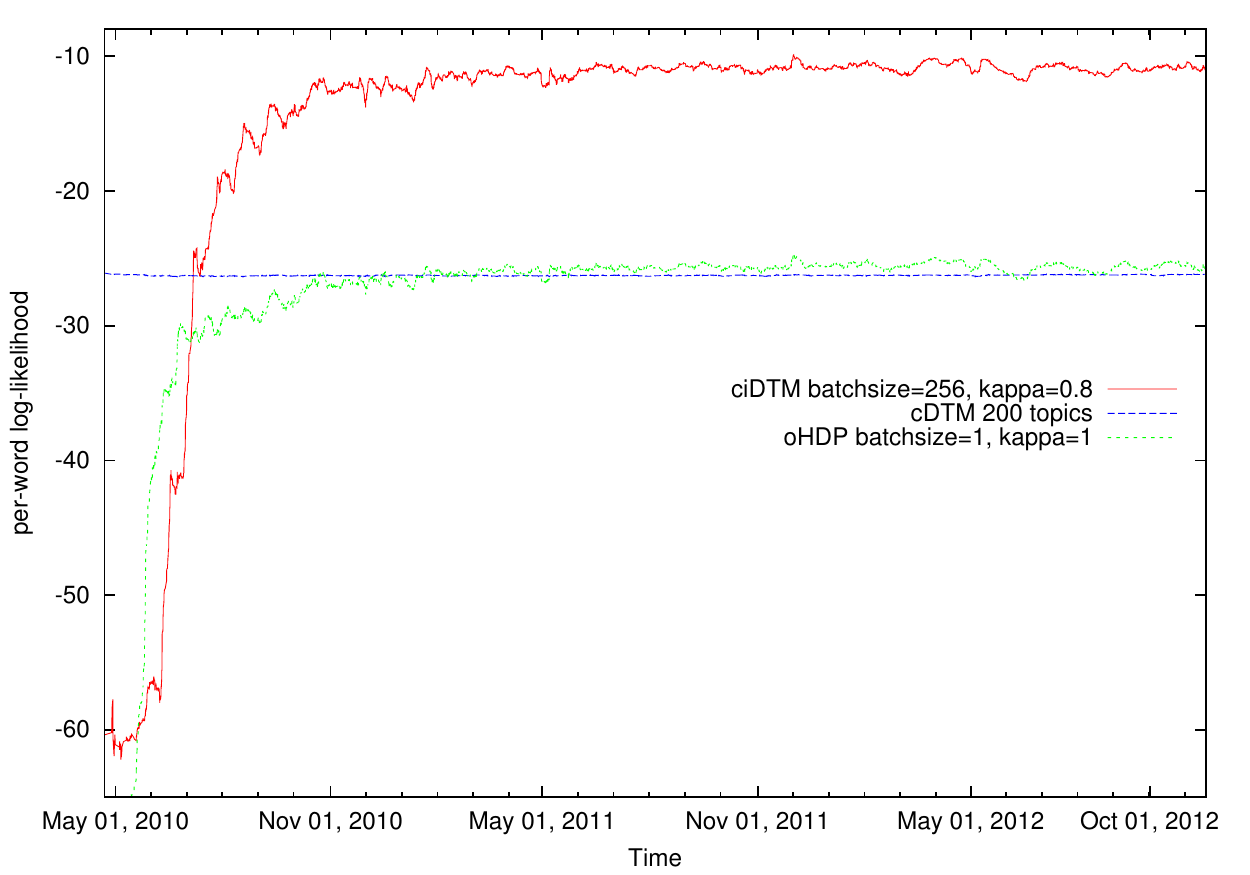}
  \caption{A comparison of ciDTM, oHDP and cDTM using their best setting values discovered earlier using the BBC news corpus.  The per-word log-likelihood values shown are the moving average of a set of 100 consecutive documents.  This averaging was needed to smooth out the plot.  Higher values indicate better model fit on this 10,000 news stories corpus.\label{fig:comp_like_bbcn}}
\end{figure}

\paragraph{Comparing three models}  Figure~\ref{fig:comp_like_bbcn} shows the per-word log-likelihood performance for the cDTM, oHDP and ciDTM models using their best setting values discovered by running the previous experiments.  ciDTM outperforms both oHDP and cDTM by a big margin.  The steady state performance of cDTM and oHDP is about the same.  It took oHDP about 2,000 documents to reach this stage.

\subsection{Why ciDTM outperforms cDTM}
\begin{shaded}
ciDTM outperforms cDTM because ciDTM evolves the number of topics per document while cDTM uses a fixed number of topics per document.

\paragraph{Limitations of cDTM}  Because cDTM uses a fixed and predefined number of topics per document, it does not allow for topic discovery.  When cDTM infers the topic mixture of a document that discusses a new topic, it has to label the document with the predefined set of topics.  The new topic will then get absorbed by that predefined set of topics and a new topic will \emph{not} be learned.  There will be no single topic that all the words in that document can be sampled from.  This will directly reduce the per-word log-likelihood of sampling the words of the document given a topic which is the measure I use to evaluate the performance of the three competing models.

\paragraph{Advantage of ciDTM }  ciDTM evolves the number of topics per document to reflect the number of topics discovered so far in the corpus and to reflect the richness of the document.  As more documents get processed by ciDTM, more topics will be discovered and their word distribution will be learned using the words in these documents.  Discovering new topics instead of absorbing them into a set of predefined topics directly improves the per-word log-likelihood of sampling the words in a document given a single topic which is the measure I use to evaluate the performance of the three competing models.  This is why ciDTM achieves a per-word log-likelihood value higher than that of cDTM.
\end{shaded}

\subsection{Why ciDTM outperforms oHDP}
\begin{shaded}
ciDTM outperforms oHDP because ciDTM evolves the per-topic word distribution in continuous-time using document timestamps while oHDP only relies on document ordering and does not use document timestamps.

\paragraph{Advantage of ciDTM}  Because ciDTM knows how much time elapsed between documents that belong to the same topic, it evolves the topic word distribution accordingly to reflect the real change in that topic word distribution.  The longer the time period separating two documents covering the same topic gets, the more drastic the change to the topic word distribution over that period will become.  This direct proportional relationship between the length of the time period and the degree of change to the topic word distribution is plausible and can be seen in real life.  The closeness of the inferred topic word distribution to the real topic word distribution translates to a higher per-word log-likelihood value for the model because it helps ciDTM better fit the model to the data.

\paragraph{Limitations of oHDP}  Since oHDP is unaware of the time period separating documents, its change to the topic word distribution is unaffected by the length of this time period. If this period gets big enough, the real topic word distribution will drastically change and oHDP will not change its topic word distribution to match it.  This is because oHDP relies on document ordering to make this change and is not aware of the length of that time period.  The big distance between the inferred topic word distribution and the real topic word distribution will lead to a lower per-word log-likelihood value for oHDP.
\end{shaded}

\section{Scalability}
A practical topic model system that infers the topic structure of news stories arriving in real time from a news outlet like a news agency should be able to process these documents in real time.  All the three topic model systems I experimented with earlier, that is, the continuous-time dynamic topic model (cDTM), the online hierarchical Dirichlet process based topic model (oHDP), and the continuous-time infinite dynamic topic model (ciDTM), each one of them was able to train on the entire set of 10,000 documents in less than six hours.  The topic inference process (the test phase of the  model) in which the topic mixtures of news stories arriving in real time from a news outlet takes much less time to finish than the training phase.  The cDTM model was able to finish this task for a test set of 5,000 documents in less then 5 minutes.  Because oHDP and ciDTM are online learners, the learning and inference phases alternated.  The time spent running the test phase was included in the overall run time for these models and was not measured separately.  By manual inspection I found that the time spent on the test phase is very negligible compared to the time spent on the training phase.

Given that a news agency like Agence France-Presse (AFP) releases on average 5,000 news stories per day that is equivalent to one news story every 20 seconds, and assuming that the running time complexity of all three systems is monotonically increasing, all the three systems presented can train on the entire set of documents released by AFP and then infer the topic mixture of a set of documents of the same size in real time.

Taking into account that the cDTM system is an offline learner, with every newly arriving document or set of documents it has to be retrained from scratch.  As time passes by, the number of released documents that the model needs to train on increases linearly with time and at some point in the future the system will not be able to train on all released documents in time before new batch of documents that requires analysis arrives to the system from the news outlet.  This lag will keep increasing linearly with time.  To have the cDTM work in real time a compromise should be made.  The system could maintain a limited history of the most recently released documents to train on.  This history should be limited so that the cDTM system could train on it and then infer the topic mixture in real time.  This solution will have a negative side effect on the performance of the system as the per-topic word distribution evolved by the system will not reflect the status of the per-topic word distribution of documents outside the limited history.  The system will ignore older documents because it does not have enough time to train on it.

The oHDP and ciDTM systems are online learners.  They can train on every news story arriving live from a news agency like AFP due to their fast online learning algorithm.  They can infer the topic mixture of news stories as their arrive to the system as well.  The effect of the per-topic word distribution seen in a document will be retained by the system and carried on to affect the topic composition of documents arriving to the system in some arbitrary point in the future as long as the system is up and running processing documents.  These status of these two systems will reflect the entire history of documents they received.

\begin{figure}[!h]
  \centering
  \includegraphics[width=\linewidth]{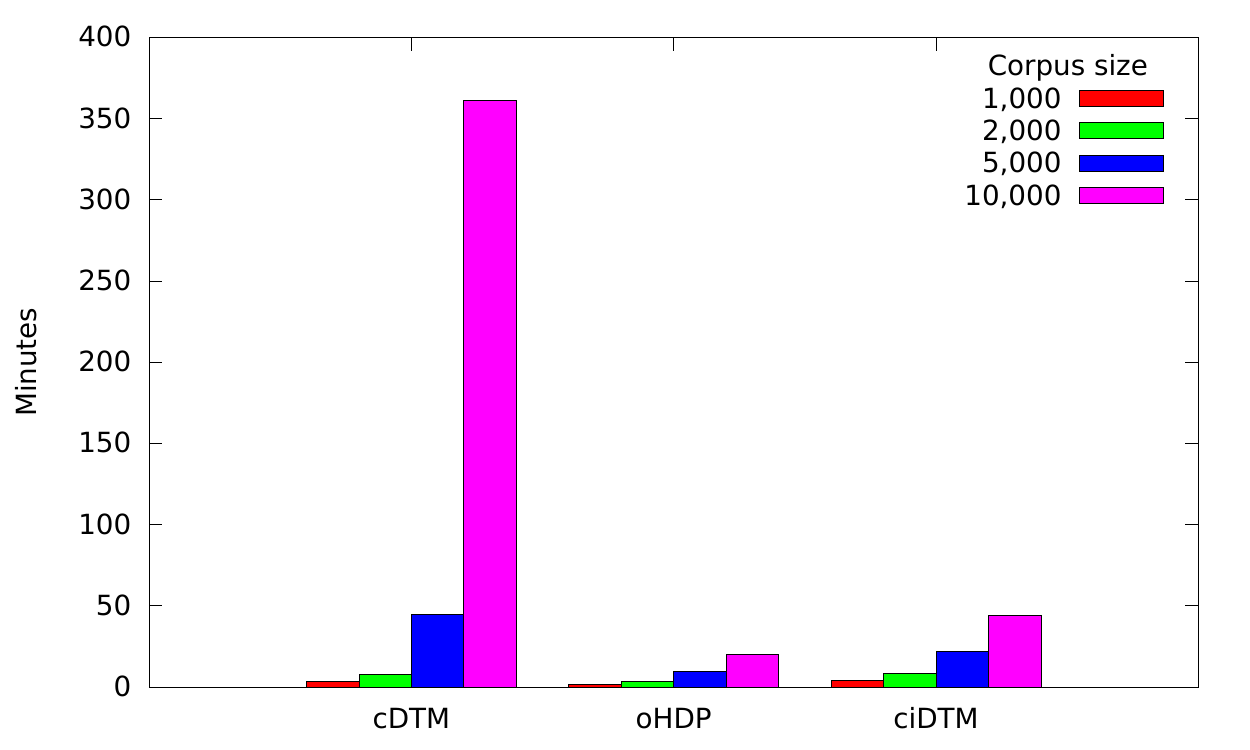}
  \caption{The wall clock running time in minutes for cDTM (10 topics), oHDP (batch size = 1) and ciDTM (batch size = 256) using subsets of the Reuters news corpus with different number of documents.  These systems were run on a PC with 16GB of RAM and 3.4 GHz processor.}
  \label{fig:reuters_runtime}
\end{figure}

To illustrate the time requirements of the three systems and how they scale up I measured the wall clock time it takes each one of them to train on subsets of the Reuters corpus.  I used subsets of size 1,000; 2,000; 5,000 and 10,000.  The inference (testing) time was very negligible compared to training time (about 1 to 100 ratio) for the three systems.  For each one of the three systems I used the settings that resulted in the best performance when experimented with the Reuters news corpus earlier.  I used a cDTM model with 50-topics, an oHDP model with a batch size of 1 and a ciDTM model with a batch size of 256.  The results are presented in Figure~\ref{fig:reuters_runtime}.

\subsection{Scalability of ciDTM and oHDP vs cDTM}
As shown in the figure, the cDTM does not scale up well with the size of the corpus.  It becomes practically unfeasible to use it in real life applications with corpora of average size ($\sim$10,000 documents).  This growth in running time that is almost exponential can be attributed to the fixed number of topics in the model.

\begin{shaded}
The reason cDTM does not scale up well with the size of the corpus is because the number of variables in the variational distribution over the latent variables is exponential in the number of topics and corpus size \cite{wang08}.  Since cDTM needs to tune the variational distribution variables to minimize the KL distance between the true posterior and the variational posterior, the time it needs to do so will grow linearly with the number of these variables which in turn grows exponentially with the size of the corpus.

On the other hand oHDP and ciDTM scale up well with the size of the corpus.  This graceful scalability can be attributed to two factors: 1) the online variational inference algorithm that scales linearly with the number of documents, and 2) the number of topics that varies from one document to another.  It is to be noted however that the oHDP based topic model is faster than the ciDTM.  This is due to the extra time needed by ciDTM to run and sample from the Brownian motion model \cite{peres10} that evolves the per-topic word distribution over time.
\end{shaded}

\section{Timeline construction}
For the timeline construction task, I manually selected 61 news stories from the BBC news corpus covering the Arab Spring events since its inception in January 2011 until the near present (November 2012).  I will call this set of news stories the Arab Spring set.  Discovering such a topic is challenging as the events that fall under it are geographically scattered across the middle east and geographical names in the news stories will not give much clue to the topic model.  More importantly, the events associated with this topic evolve rapidly over a short period of time and the set of vocabulary associated with it changes as well.

The cDTM is not suitable for this task.  It assigns a fixed number of topics to every document in the corpus.  If this number is high, the topic we are trying to discover will be split over more than one topic, and if the number is low, the topic will be merged with other topics.  Moreover, since the number of topics typically vary from one document to another, finding one fixed value for it will always be inferior to other approaches that evolve it dynamically.

I trained the oHDP based topic model and the ciDTM on the entire BBC news corpus and inferred the topics associated with each document in the Arab Spring set.  Then, I manually inspected the inferred topics and found the one that corresponds to the Arab Spring events.  Both systems where trained using their best settings found in earlier experiments on the BBC news corpus in Section~\ref{sec:results_bbcn}.

\begin{figure}[!h]
  \centering
  \includegraphics[width=\linewidth]{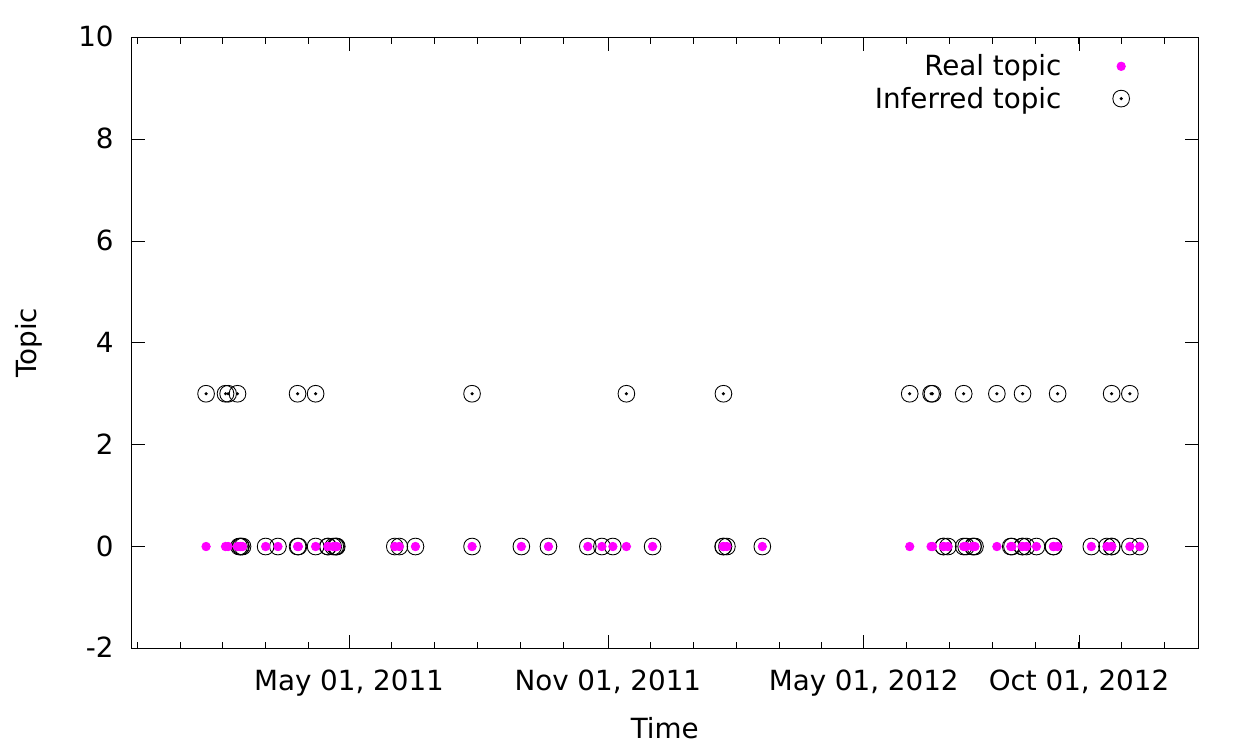}
  \caption{oHDP timeline construction for the Arab Spring topic over 61 documents covering this topic.  The independent axis is the time, while the dependent axis is the topic index.  It was found that topic 0 represents the Arab Spring events topic.  A pink dot (\textcolor{magenta}{$\bullet$}) represents the real topic that should be discovered, an black circle with a small dot ($\odot$) represents the topic/s that was discovered for the document.  A black circle with a pink dot in the middle of it (\rlap{\textcolor{magenta}{$\bullet$}}$\odot$) is a real topic that was successfully discovered by the system.}
  \label{fig:ohdp_timeline}
\end{figure}

Figure~\ref{fig:ohdp_timeline} shows the inferred topics for documents in the Arab Spring set using oHDP.  The notation I used in this Figure is described in its caption.  After the start of the Arab Spring in early January of 2011, the documents discussing its events were assigned topic 3. This topic in the oHDP model corresponds to accidents and disasters.  One month later, the model was able to infer the correct topic and associate it with the documents of the Arab Spring events.  The performance of the system was steady from mid February 2011 until June 2012.  After that, the topic evolved so fast after the Egyptian president assumed office in late June 2012.  The oHDP was unable to evolve the word distribution associated with that topic quickly enough to keep track of the topic.  We can see that after that date the model unable to infer the correct topic for 3 documents in this set.  The documents were associated with the accidents and disasters topic instead.  Overall, the model was unable to infer the correct topic for 10 out of 61 documents in the Arab Spring set.

\begin{table}[!h]
  \centering
  \begin{tabular}{lrccr}
    \cline{3-5}
    &&\multicolumn{2}{c}{Inferred}&\\
    \cline{3-4}
             &&AS &non-AS &Accuracy\\
             \hline\hline
   \multirow{2}{10mm}{True} &AS &51 &10   &0.836\\
                        &non-AS &0  &13   &1.000\\
                     \cline{5-5}
                     &&&&0.865\\
 \hline\hline
  \end{tabular}
  \caption{Confusion matrix for timeline constuction using oHDP.  AS stands for Arab Spring topic, and non-AS is non Arab Spring topic.  The true class is presented in rows (True AS, and non-AS), and Inferred class in columns (Inferred AS and non-AS).}
  \label{tab:timeline_confusion_ohdp}
\end{table}

Table~\ref{tab:timeline_confusion_ohdp} shows the confusion matrix for the oHDP topic assignment to news stories belonging to the Arab Spring timeline.  oHDP failed to correctly tag 10 out of 61 stories as belonging to that topic.  However, it successfully labeled all 13 stories which had non Arab Spring topic components with the correct topic label.  This gives oHDP a recall score of 83.6\% and a precision of 100\%.

\begin{figure}[!h]
  \centering
  \includegraphics[width=\linewidth]{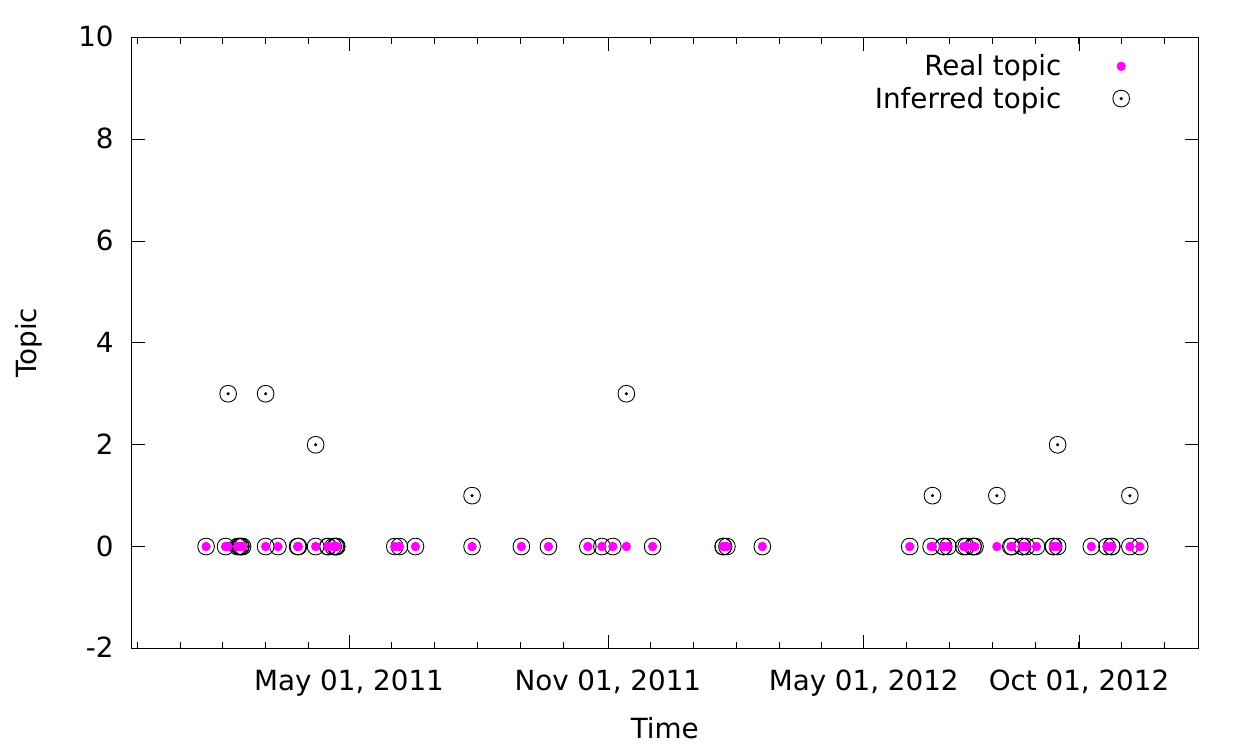}
  \caption{ciDTM timeline construction for the Arab Spring topic over 61 documents covering this topic.  The independent axis is the time, while the dependent axis is the topic index.  It was found that topic 0 represents the Arab Spring events topic.  A pink dot (\textcolor{magenta}{$\bullet$}) represents the real topic that should be discovered, an black circle with a small dot ($\odot$) represents the topic/s that was discovered for the document.  A black circle with a pink dot in the middle of it (\rlap{\textcolor{magenta}{$\bullet$}}$\odot$) is a real topic that was successfully discovered by the system.}
  \label{fig:cidtm_timeline}
\end{figure}

Figure~\ref{fig:cidtm_timeline} shows the inferred topics for documents in the Arab Spring set using ciDTM.  This Figure uses the same notation as the previous Figure and it is described in its caption also.  Topic 0 represents the Arab Spring topic in this model.  At first glance at the Figure, we notice that the ciDTM was able to infer the correct topic for the first document in the Arab Spring set.  This can be explained by the fact that ciDTM trains on a batch of 256 documents.  The model was able to train on a big batch of documents that included some Arab Spring set documents and learn the new topic word distribution before inferring their topics.  If that explanation is valid then this advantage is not intrinsic of the ciDTM as the oHDP based model could behave similarly by increasing its batch size.

We notice that some documents in this set were assigned multiple topics with this model.  Besides topic 0, which corresponds to the Arab Spring events, topics 1, 2 or 3 sometimes appear together with topic 0 in the same document.  By inspecting these topics, I found that topic 1 corresponds to economy and finance, topic 2 corresponds to health and medicine, and topic 3 corresponds to accidents and disasters.  Some of the early documents in the Arab Spring set were assigned topic 3.  This reflects the violence that marred the early days of the Egyptian revolution in late January 2011.  The association of topic 2 with the set documents may reflect the mention of the injured protesters at that time.  Later documents in this set are more associated with topic 1.  This can be explained by the volume of published news stories discussing the Egyptian government efforts to recover from the economical and financial damages which the revolution inflicted on the country.

\begin{table}[!h]
  \centering
  \begin{tabular}{lrccr}
    \cline{3-5}
    &&\multicolumn{2}{c}{Inferred}&\\
    \cline{3-4}
             &&AS &non-AS &Accuracy\\
             \hline\hline
   \multirow{2}{10mm}{True} &AS &57 &4    &0.934\\
                        &non-AS &0  &13   &1.000\\
                     \cline{5-5}
                     &&&&0.946\\
 \hline\hline
  \end{tabular}
  \caption{Confusion matrix for timeline construction using ciDTM.  AS stands for Arab Spring topic, and non-AS is non Arab Spring topic.  The true class is presented in rows (True AS, and non-AS), and Inferred class in columns (Inferred AS and non-AS).}
  \label{tab:timeline_confusion_cidtm}
\end{table}

Table~\ref{tab:timeline_confusion_cidtm} shows the confusion matrix for the ciDTM topic assignment to news stories belonging to the Arab Spring timeline.  ciDTM shows about 10\% improvement in accuracy over the oHDP based topic model due to a 10\% higher true positive rate.  ciDTM achieves a recall score of 93.4\% and a 100\% precision.  ciDTM has a higher recall power without sacrificing precision.

\subsection{Why ciDTM is better than oHDP in timeline construction}
\begin{shaded}
ciDTM is better than oHDP in timeline construction because ciDTM evolves the per-topic word distribution in continuous time while oHDP only relies on document ordering and does not make use of document timestamps in evolving the per-topic word distribution. 
\end{shaded}

This can be seen by comparing the timelines constructed for the Arab Spring topic using oHDP based topic model in Figure~\ref{fig:ohdp_timeline} and using ciDTM in Figure~\ref{fig:cidtm_timeline}.

oHDP and ciDTM were able to detect the birth of a new topic, which is the Arab Spring topic, soon after it started in January of 2011.  The true positive rate was high for both models (oHDP=0.91, ciDTM=0.94) in the first part of this timeline (from January 2011 to March 2012).  In the second part of this timeline (from June 2012 to November 2012) the true positive rate for oHDP sharply dropped to a value of 0.74 while it slightly dropped to a value of 0.92 for ciDTM.

In what follows I explain why ciDTM maintained a high true positive rate value in the second part of the Arab Spring topic timeline while the true positive rate for oHDP dropped sharply.

\begin{shaded}
The corpus used to train both models had no documents with the Arab Spring topic in the period of March 2012 to June 2012.  

\paragraph{Limitations of oHDP}  Because oHDP evolves the per-topic word distribution based on document ordering and does not use document timestamps for that it was unaware of the long time period (3 months) that separated the last document in the first part of the timeline and the first document in the second part of the timeline.  Because of that, oHDP did not evolve the Arab Spring topic word distribution to reflect the changes in vocabulary of the news stories covering this topic.  The actual word distribution for the Arab Spring topic significantly changed and the oHDP did not change its Arab Spring topic word distribution to match it.  This is because oHDP was unaware of the long time period (March 2012 to June 2012 [3 months]) in which no documents covering this topic were included in the corpus.  Because of the big difference between the actual Arab Spring topic word distribution and the oHDP Arab Spring topic word distribution, oHDP failed to correctly label many news stories with the Arab Spring topic in the second part of the timeline after the long period of topic dormancy.  This mislabeling resulted in a sharp drop in the true positive rate value for oHDP in the second part of the timeline when compared to the true positive rate value it achieved in the first part.

\paragraph{Advantage of ciDTM}  On the other hand ciDTM uses document timestamps to evolve the per-topic word distribution.  When ciDTM infers the topic mixture for the first document in the second part of the Arab Spring timeline (June 2012 to November 2012) it evolves the Arab Spring topic word distribution to reflect the long period (March 2012 to June 2012 [3 months]) of dormancy of this topic.  The word distribution of the Arab Spring topic evolved by ciDTM to reflect this long time period of dormancy is closer to the actual Arab Spring topic.  Because of this closeness, ciDTM was able to correctly label news stories with the Arab Spring topic in the second part of the timeline.  This successful labeling resulted in a slight drop in the true positive rate for ciDTM in the second part of the Arab Spring timeline when compared to the true positive rate value it achieved in the first part of the timeline.
\end{shaded}


\bibdata{references}
\bibliography{bibliography}
\bibliographystyle{plainnat}
\addcontentsline{toc}{chapter}{Bibliography}

\appendix
\cleardoublepage
\chapter{Probability distributions}
\label{cha:prob-distr}

In this Appendix, I briefly introduce some of the most important probability distributions in Bayesian data modeling.

\section{Uniform}
This is a simple distribution for a continuous time variable $x$.  The probability that this variable takes any value in its finite domain ($[a,b]$) is constant and can be defined by:

\begin{equation}
  \label{eq:uniform}
  \text{U}(x|a,b) = \frac{1}{b-a}
\end{equation}

\section{Bernoulli}
This distribution is for a single binary variable $x\in\{0,1\}$, can be thought of as representing the result of flipping a coin.  The parameter $\mu \in [0,1]$ gives the probability of success, which is having $x=1$.  More formally:

\begin{equation}
  \label{eq:bernoulli}
  \text{Bern}(x|\mu) = \mu^x (1-\mu)^{1-x}
\end{equation}

\section{Binomial}
If a Bernoulli trial (such as coin flipping) is repeated $N$ times, where the probability of success is $\mu \in [0,1]$, then the probability distribution of the number of successes ($x$) is given by:

\begin{equation}
  \label{eq:binomail}
  \text{Bin}(x|\mu,N) = {N \choose x}\mu^x(1-\mu)^{N-x}
\end{equation}

\section{Multinomial}
The multinomial distribution is a generalization of the binomial distribution where the Bernoulli distribution giving the probability of success is replaced with a categorical distribution.  In this case, there are $K$ possible outcomes for the discrete variable $\mathbf{x}=\{x_1\ldots,x_K\}$ with probabilities $\mathbf{\mu}=\{\mu_1,\ldots,\mu_K\}$ such that $\sum_k\mu_k=1$.

For $N$ observations, if $\mathbf{m}=\{m_1,\ldots,m_K\}$, where $m_k$ is the number of times outcome $x_k$ is observed. Then, $\mathbf{m}$ follows a multinomial distribution with parameters $\mathbf{\mu}$ and $N$.  This distribution can be defined as follows:

\begin{equation}
  \label{eq:multinomial}
  \text{Mult}(\mathbf{m} | \mathbf{\mu}, N) = \frac{N!}{m_1! \ldots m_K!}\prod_{k=1}^{K}\mu_{k}^{m_k}
\end{equation}

The categorical and multinomial distributions are often conflated in computer science literature.  The outcome of a categorical distribution is sometimes represented as a binary 1-of-$K$ vector, indicating which one of the $K$ outcomes was observed.  In this notation, which is adopted by Bishop \citep[p. 690]{bishopPattern}, the categorical distribution represents a multinomial distribution over a single observation.

\section{Beta}
This is a distribution over a continuous variable $x\in[0,1]$ and is often used to represent the probability of a binary event with two parameters $a>0$ and $b>0$.  This Beta distribution is the conjugate prior for the Bernoulli distribution where $a$ and $b$ can be interpreted as the effective prior number of success and failure observations, respectively.  This distribution can be defined by:

\begin{equation}
  \label{eq:beta}
  \text{Beta}(x|a,b) = \frac{\Gamma(a + b)}{\Gamma(a)\Gamma(b)}x^{a-1}(1-x)^{b-1}
\end{equation}

Where $\Gamma(.)$ is the Gamma function given by:

\begin{align}
  \label{eq:gamma}
  \Gamma(x) &= \int_0^{\infty} u^{x-1}e^{-u}du\\
            &= (x-1)!, \qquad \text{for integer x}
\end{align}

\section{Dirichlet}
The Dirichlet distribution is a multivariate distribution over $K$ random variables $0 \le \mu_k \le 1$, where $k=1,\ldots,K$,

\begin{equation}
  \label{eq:1}
  \text{Dir}(\boldsymbol{\mu|\alpha}) = C(\boldsymbol{\alpha})\prod_{k=1}^K\mu_k^{\alpha_k-1}
\end{equation}

where

\begin{equation}
  \label{eq:2}
  C(\boldsymbol{\alpha}) = \frac{\Gamma(\hat{\alpha})}{\Gamma(\alpha_1)\ldots\Gamma(\alpha_K)}
\end{equation}

and

\begin{equation}
  \label{eq:3}
  \hat{\alpha} = \sum_{k=1}^K \alpha_k.
\end{equation}

Subject to the constraints:

\begin{align}
  \label{eq:4}
       0 \le &\mu_k \le 1\\
  \sum_{k=1}^K &\mu_k = 1.
\end{align}

The Dirichlet distribution forms a conjugate prior with the multinomial distribution and is a generalization of the Beta distribution.  The parameters $\alpha_k$ can be interpreted as the effective number of observations of the corresponding values of the K-dimensional observation vector $\mathbf{x}$.


\cleardoublepage
\chapter{Dirichlet Process}
The Dirichlet Process (DP) is a distribution over distributions,  meaning that a draw from the process gives us a Dirichlet distribution.  A DP denoted by $\text{DP}(G_0,\alpha)$ is parametrized by a base measure $G_0$ and a concentration parameter $\alpha$ which is analogous to the mean and variance of a Gaussian distribution, respectively.  By integrating out $G$, $\theta$ follows a Polya urn distribution \cite{blackwell:polya} or a Chinese Restaurant Process \cite{ahmed10}
\begin{equation}
  \theta_i|\theta_{1:i-1},G_0,\alpha \sim \sum_k\frac{m_k}{i-1+\alpha}\delta(\phi_k)+\frac{\alpha}{i-1+\alpha}G_o
\end{equation}
where, $\phi_{1:k}$ are the distinct values of $\theta$, and $m_k$ is the number of parameter $\theta$ with $\phi_k$ value.  A Dirichlet process mixture model (DPM) can be built using the given DP on top of a hierarchical Bayesian model.

Instead of having a fixed Dirichlet distribution to draw a parameter vector $\theta$ from in a generative model, as shown in Figure \ref{fig:lda_graphical_model}, that vector, which represents a topic proportions vector in a topic model, can be drawn from a Dirichlet distribution $G_d$  that was drawn from a DP.  Such a model can accommodate an infinite number of topics.  To share topics among different documents, the document-specific Dirichlet distributions $G_d$ can be tied together by drawing their base measure $G_0$ from another DP.  More formally: $G_0\sim \text{DP}(H,\gamma)$.  This final construction is known as the Hierarchical Dirichlet Process (HDP).  By integrating $G_d$ out of this model, we get the Chinese restaurant franchise process (CRFP) \cite{blei06};
\begin{gather}
  \theta_{di}|\theta_{d,1:i-1},\alpha,\boldsymbol{\psi} \sim \sum_{b=1}^{b=B_d}\frac{n_{db}}{i-1+\alpha}\delta_{\psi_{db}} + \frac{\alpha}{i-1+\alpha}\delta_{\psi_{db^{new}}}\\
\psi_{db^{new}}|\boldsymbol{\psi},\gamma \sim \sum_{k=1}^K \frac{m_k}{\sum_{l=1}^Km_l+\gamma}\delta_{\phi_k} + \frac{\gamma}{\sum_{l=1}^Km_l+\gamma}H
\end{gather}
where $\psi_{db}$ is topic $b$ for document $d$, $n_{db}$ is the number of words sampled from it, $\psi_{db^{new}}$ is a new topic, $B_d$ is the number of topics in document $d$, and $m_k$ is the number of documents sharing topic $\phi_k$.

The generative process of this model follows a Chinese restaurant model.  In this metaphor, a restaurant represents a document, a customer represents a word, and a dish represents a topic.  Customers sitting at the same table in a Chinese restaurant share the same dish served on that table.   When customer $i$ arrives at a restaurant, she is assigned a table to sit at.  This table could be an empty table with probability $\frac{\alpha}{i-1+\alpha}$, or it could be a table occupied by $n_{db}$ customers with probability $\frac{n_{db}}{i-1+\alpha}$.  If an empty table is assigned, the dish served on this table could either be a new dish that has not been served in any restaurant before with probability $\frac{\gamma}{\sum_{l=1}^K m_l + \gamma}$ where $m_l$ is the number of restaurants that have served dish $l$ before, or the dish served on the table could be a dish that has been served on $m_k$ tables across all restaurants with probability $\frac{m_k}{\sum_{l=1}^K m_l + \gamma}$.  Table assignment and dish selection follows a richer scheme.

To add temporal dependence in our model, we can use the temporal Dirichlet process mixture model (TDPM) proposed in \cite{AhmedX08} which allows unlimited number of mixture components.  In this model, $G$ evolves as follows \cite{ahmed10}:
\begin{gather}
  G_t|\phi_{1:k},G_o,\alpha \sim DP(\zeta,D)\\
  \zeta = \alpha + \sum_km'_{kt}\\
  D = \sum_k\frac{m'_{kt}}{\sum_lm'_{lt}+\alpha}\delta(\phi_k) + \frac{\alpha}{\sum_lm'_{lt}+\alpha}G_0\\
  m'_{kt} = \sum_{\delta=1}^{\Delta} \exp^{-\delta/\lambda}m_{k,t-\delta}
\end{gather}
where $m'_{kt}$ is the prior weight of component $k$ at time $t$, $\Delta$, $\lambda$ are the width and decay factor of the time decaying kernel.  For $\Delta=0$, this TDPM will represent a set of independent DPMs at each time step, and a global DPM when $\Delta=T$ and $\lambda=\infty$.  As we did with the time-independent DPM, we can integrate out $G_{1:T}$ from our model to get a set of parameters $\theta_{1:t}$ that follows a Poly-urn distribution:
\begin{equation}
  \theta_{ti}|\mathbf{\theta_{t-1:t-\Delta}},\theta_{t,1:i-1},G_0,\alpha \propto \sum_k(m'_{kt}+m_{kt})\delta(\phi_{kt})+\alpha G_0
\end{equation}

We can make word distributions and topic trends evolve over time if we tie together all hyper-parameter base measures $G_0$ through time.  The model will now take the following form:
\begin{equation}
  \theta_{tdi}|\theta_{td,1:i-1},\alpha,\boldsymbol{\psi_{t-\Delta:t}} \sim \sum_{b=1}^{b=B_d}\frac{n_{tdb}}{i-1+\alpha}\delta_{\psi_{tdb}} + \frac{\alpha}{i-1+\alpha}\delta_{\psi_{tdb^{new}}}
\end{equation}
\begin{align}
  \psi_{tdb^{new}}|\boldsymbol{\psi},\gamma &\sim \sum_{k:m_{kt}>0} \frac{m_{kt}+m_{kt}'}{\sum_{l=1}^{K_t}m_{lt}+m_{lt}'+\gamma}\delta_{\phi_{kt}} \nonumber \\
&+  \sum_{k:m_{kt}=0} \frac{m_{kt}+m_{kt}'}{\sum_{l=1}^{K_t}m_{lt}+m_{lt}'+\gamma}\mathrm{P}(.|\phi_{k,t-1}) \nonumber \\
&+ \frac{\gamma}{\sum_{l=1}^{K_t}m_{lt}+m_{lt}'+\gamma}H
\end{align}
where $\phi_{kt}$ evolves using a random walk kernel like in \cite{blei06}:
\begin{gather}
   H = N(0,\sigma I)\\
   \phi_{k,t}|\phi_{k,t-1} \sim N(\phi_{k,t-1},\rho I)\\
   w_{tdi}|\phi_{kt} \sim \mathcal{M}(L(\phi_{kt}))\\
   L(\phi_{kt}) = \frac{\exp(\phi_{kt})}{\sum_{w=1}^W\exp(\phi_{ktw})}
\end{gather}

We can see from \eqref{posterior}, \eqref{likelihood}, \eqref{baseMeasure} the non-conjugacy between the base measure and the likelihood.


\cleardoublepage
\chapter{Graphical models}
Probabilistic graphical models are tools that allow us to visualize the independence relationship between random variables in a model and use available graph manipulation algorithms to model complex systems, tune the model parameters, and infer the likelihood of events.

Probabilistic graphical models, or simply graphical models, can be categorized into \emph{directed} graphical models (otherwise known as Bayesian networks), and \emph{undirected} graphical models (called Markov networks, or Markov Random Fields [MRF] as well).

\begin{figure}
  \centering
  \includegraphics[width=0.6\textwidth]{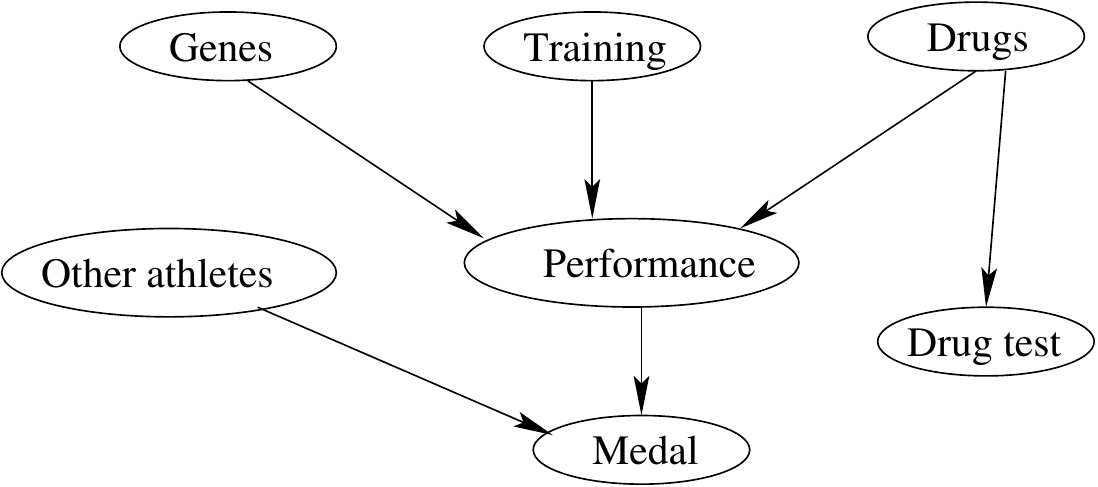}
  \caption{Directed graphical model.\label{dirGM}}
\end{figure}

An example of a directed graphical model is given in Figure \ref{dirGM}.  This graph models the independence/dependence relationship between seven random variables.  The structure of the graph and the relationship between the variables were created using human judgment.  Algorithms for learning the graph structure and parameters do exist \cite{hecherman95learning} but were not used in this simple example.  The nodes in the graph represent random variables and the directed edges show the flow of influence of the random variables.  In this model, the priors \emph{Genes}, \emph{Training}, \emph{Drugs}, define the probability that an athlete would have good genes, train well, and take performance-enhancing drugs, respectively.  These three random variables affect the athlete's \emph{Performance}, which along with the probability that the \emph{Other athletes} would perform well or not affect the probability that our athlete would win a \emph{Medal}.  Drug doping can be tested separately by its effect on the \emph{Drug test}.

\begin{figure}
  \centering
  \includegraphics[width=0.3\textwidth]{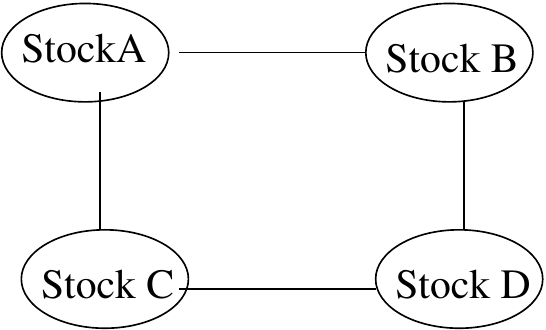}
  \caption{Undirected graphical model.\label{undirGM}}
\end{figure}

An example of an undirected graphical model is given in Figure \ref{undirGM} where the direct interaction between market stocks' values is represented by the graph edges and the nodes themselves represent the value of the market stock.  Since \emph{Stock A} and \emph{Stock D} are not neighbors in the graph they do not affect each other directly.

Simply speaking, a graphical model (directed or undirected) represents a joint distribution $P$ over a set of random variables $\mathcal{X}$.  For inference on a set of variables, we can exhaustively marginalize over the rest of the variables in the model.  However, the inference complexity will increase exponentially with the number of variables and the problem would become intractable.  This problem can be avoided in Bayesian networks (directed graphs) by making use of the independence relationships between the variables as follows:

\begin{defin}
  Given a Bayesian network $\mathcal{G}$ defined over a set of random variables $\mathcal{X}=\{X_1,\ldots,X_n\}$ and a probability distribution $P_{\mathcal{X}}$ over the same set of variables.  Then we say that $P_{\mathcal{X}}$ factorizes according to $\mathcal{G}$ if
  \begin{gather}
    P_{\mathcal{X}} = \prod_{i=1}^nP(X_i|Pa(X_i))
  \end{gather}
\end{defin} 
where $Pa(X_i)$ is the set of parent nodes of $X_i$.  This follows from the effect of the blanket variables that separate variables as follows:

\begin{defin}
  Given a Bayesian network $\mathcal{G}$ defined over a set of random variables $\mathcal{X}=\{X_1,\ldots,X_n\}$.  A variable $X_i$ is independent of its non-descendants given its parents:
  \begin{gather}
    \forall X_i, (X_i\perp nonDescendants | Pa(X_i))
  \end{gather}
\end{defin}

Analogously, for Markov networks:

\begin{defin}
  Given a Markov network structure $\mathcal{H}$ defined over the set of random variables $\mathcal{X}=\{X_i,\ldots,X_n\}$ and a distribution $P_{\mathcal{H}}$ defined over the same space.  We say that the distribution $P_{\mathcal{H}}$ factorizes over $\mathcal{H}$ if there is a set of factors $\pi_1[D_1],\ldots,\pi_m[D_m]$ where $D_i$ is a clique in $\mathcal{H}$ such that:
  \begin{gather}
    P_{\mathcal{H}}=\frac{1}{Z}P'_{\mathcal{H}}(X_i,\ldots,X_n)\\
    P'_{\mathcal{H}} = \prod_{i=1}^m\pi_i[D_i]\\
    Z = \sum_{X_1,\ldots,X_n} P'_{\mathcal{H}}(X_1,\ldots,X_n)
  \end{gather}
\end{defin}
where $P'_{\mathcal{H}}$ is an unnormalized measure, and $Z$ is a normalizing constant.

Independency in Markov networks can be identified using blankets of variables \cite{koller07}:

\begin{defin}
  Given a Markov network $\mathcal{H}$ defined over a set of random variables $\mathcal{X}=\{X_i,\ldots,X_n\}$.  Let $\mathcal{N_H}(X_i)$ be the set of $X_i$ neighbors in the graph.  Then, the \emph{local Markov independencies} associated with $\mathcal{H}$ can be defined as:
  \begin{gather}
    \mathcal{I(H)}=\{(X\perp\mathcal{X}-\mathcal{N_H}(X)-X|\mathcal{N_H(X))}:X\in\mathcal{X}\}
  \end{gather}
\end{defin}

\section{Inference algorithms}
Inference on graphical models could be a \emph{conditional independence query}, or a \emph{Maximum a Priori (MAP)} or its special (and important) case the \emph{Most Probable Explanation (MPE)} among other types of queries.

A conditional independence query finds the probability of assignment of a random variable to a set of values given some evidence:
\begin{gather}
  P(Y| E = e) = \frac{P(Y, e)}{P(e)}
\end{gather}

The MPE query finds the most likely assignment of values to all non-evidence variables.  In other words, it finds the value of $w$ that maximizes $P(w,e)$, where $w$ is the values assigned to $W = \mathcal{X}-E$.

As it turns out, the problem of inference on graphical models is $\mathcal{NP}$-hard at best \cite{freud90}, if no assumptions are made about the structure of the underlying graphical model \cite{cooper90}.  However, exact inference could be done efficiently on low treewidth graphical models.  In many other important models, approximate inference could yield good results.

The problem of inference could be dealt with as an optimization problem, or sampling problem.  The first alternative tries to find a distribution $Q$ that is similar enough to the desired distribution $P$ by searching a set of easy candidate distributions $\mathbf{Q}$ trying to find the one that minimizes the distance measure between $P$ and $Q$.  Among the distance measures that could be used are Euclidean distance, which would require doing hard inference on $P$ taking us back to the main problem, and the entropy based Kullback-Leibler (KL)-divergence measure \cite{kl87}.

\subsection{Example}
Conditional Random Field (CRF) \cite{laf01} models have replaced hidden Markov models (HMMs) and maximum entropy Markov models (MEMM) in many NLP applications such as named-entity recognition \cite{mccallum03ner} and part-of-speech tagging (POS) \cite{sutton04pos} due to their ability to relax strong independence assumptions made in these models.

\begin{figure}
  \centering
  \includegraphics[width=0.9\textwidth]{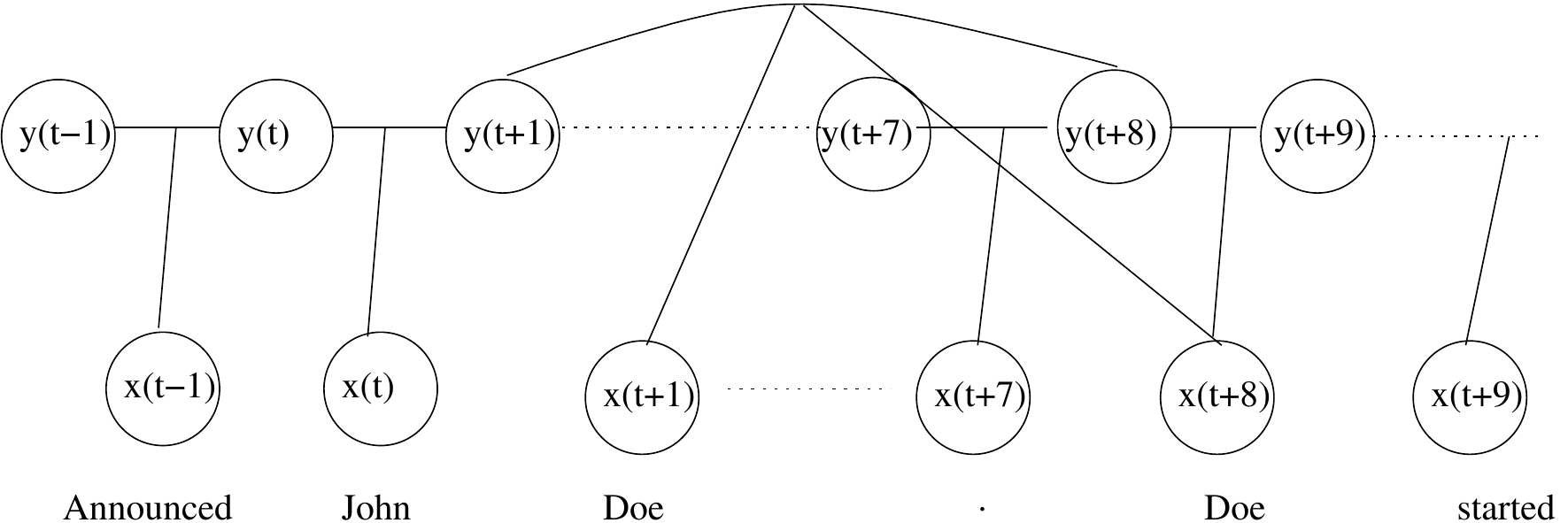}
  \caption{A skip-chain CRF model for an NER task.\label{crfNer}}
\end{figure}

Figure \ref{crfNer} shows a graphical representation for a skip-chain CRF adapted from \cite{mccallum09crfIntro}.  Only the $x$ nodes are observed, and identical words are connected (with \emph{skip edges}) because they are likely to have the same label.  Strong evidence at one end can affect the label at the other endpoint of the edge.  Long range dependencies in a model like this one would be hard to represent using $n$-gram models because they tend to have too many parameters if $n$ is large.

In a skip-chain model, skip-edge creation should depend on the application of the model.  In a POS tagging task, this edge creation could be based on a similarity measure that uses stemming and word roots or edit distance.  In the current example, we are connecting identical capitalized words for the NER task.  Adding so many edges would complicate the model and make approximate inference harder.  In our model, the probability of a label sequence $y$ given an input $x$ could be evaluated by:
\begin{gather}
  p_{\theta}(\mathbf{y|x}) = \frac{1}{Z(\mathbf{x})}\prod_{t=1}^T\Psi_t(y_t, y_{t-1},\mathbf{x}) \prod_{(u,v)\in\mathcal{I}}\Psi_{uv}(y_u,y_v,\mathbf{x})
\end{gather}
where $\mathcal{I}=\{(u,v)\}$ is the set of all skip-edges sequence positions, $\Psi_t$ are the factors  for the linear-chain edges, and $\Psi_{uv}$ are the factors over skip edges (like the one connecting $y_{t+1}$ and $y_{t+1}$ in Figure \ref{crfNer}). And we have:
\begin{gather}
  \Psi_t(y_t,y_{t-1},\mathbf{x})= \exp\left\{\sum_k \lambda_{1k}f_{1k}(y_t,y_{t-1},\mathbf{x},t)\right\}\\
    \Psi_{uv}(y_u,y_v,\mathbf{x})= \exp\left\{\sum_k \lambda_{2k}f_{2k}(y_u,y_v,\mathbf{x},u,v)\right\},
\end{gather}
where $\theta_1=\{\lambda_{1k}\}_{k=1}^{K_1}$ are linear-chain template parameters, and $\theta_2=\{\lambda_{2k}\}_{k=1}^{K_2}$ are the parameters of the skip-template.  The observation functions $q_k(\mathbf{x},t)$ can depend on arbitrary positions of the input string.  They factorize to:
\begin{gather}
  f'_k(y_u,y_v,\mathbf{x},u,v)=1_{\{y_u=\tilde{y}_u\}}1_{\{y_v=\tilde{y}_v\}}q'_k(\mathbf{x},u,v)
\end{gather}

Skip-chain CRFs have been introduced and tested by McCallum and Sutton on a problem of extracting speaker names from seminar announcements \cite{mccallum09crfIntro}.  They reported an improved performance over linear-chain CRFs, which in turn outperforms HMM due to the relaxed independence assumptions between the states and the current observation.  However, skip-chain CRFs fall into the trap of the implicit assumption of independence between current observations $\mathbf{x}$ and the neighbors of the states on both ends of the skip-edge.  For example, in one document, the word \emph{China} representing the country would be linked with a skip-edge to the word China in \emph{The China Daily}, and both could be labeled as COUTRY, or ORGANIZATION in a NER task, even though they are different entities.  A solution to this problem has been proposed by Finkel et al. \cite{finkel05}.  They added richer long-distance factors to the original pairs of words factors.  These factors are useful for adding exceptions to the rule that identical words have the same entity in text.  However, due to the sparsity of the training data this model may fail.

Augmenting the features with neighbors of the states at the ends of the skip-edges may remedy this problem.  This could take the form:
\begin{gather}
      \Psi_{ef}(y_e,y_f,\mathbf{x})= \exp\left\{\sum_k \lambda_{3k}f_{3k}(y_e,y_f,\mathbf{x},e,f)\right\}\\
    \Psi_{gh}(y_g,y_h,\mathbf{x})= \exp\left\{\sum_k \lambda_{4k}f_{4k}(y_g,y_h,\mathbf{x},g,h)\right\},
\end{gather}

Each factor of a skip-edge will be a function of four more states.  This would require the use of different parameter estimation techniques on two stages \cite{finkel05}:  First, we estimate the parameters of the linear-chain CRF, ignoring the skip-edges; then we heuristically select the parameters for the long-distance factors and neighboring states.

Another point worth considering is the precision of the probabilities of the forward-backward belief propagation algorithm.  Their values become too small to be represented within numeric precision.  To solve this problem we could use the logarithmic domain for computation.  Therefore,
\begin{gather}
  \log(ab) = \bigoplus (a + b)
\end{gather}
where
\begin{gather}
  a \oplus b = \log(e^a + e^b),
\end{gather}
or
\begin{gather}
  a \oplus b = a + \log(1 + e^{b-a})
\end{gather}

Even though graphical models give us flexibility in modeling complex systems, these models could easily become intractable if the system input involved a set of rich nonindependent features.  This is typical in many real world Natural Language Processing applications.

Since graphical models are usually used to represent a joint probability distribution $p(\mathbf{y}, \mathbf{x})$, where $\mathbf{y}$ is the attributes of the entity we want to predict, and $\mathbf{x}$ is our observation about this entity.  Modeling the joint distribution for a system with rich setof features is complicated by the fact that it requires modeling the distribution $p(\mathbf{x})$ with its complex dependencies, causing our model to become intractable.  This problem can be solved by modeling the conditional distribution $p(\mathbf{y}|\mathbf{x})$ instead.  This approach in representation is known as Conditional Random Fields (CRFs) \cite{laf01}.


\cleardoublepage
\chapter{Information theory}
\label{cha:information-theory}
Many definitions and equations used in this thesis can be better understood in the context of information theory.  The average amount of information contained in a random variable and the relative entropy measures are useful in understanding how two probability distributions diverge from each other.

\section{Entropy}
The entropy $H(X)$ of a random variable $X$ is defined by \cite{yeung08}:

\begin{equation}
  \label{eq:5}
  H(X) = - \sum_x p(x) \log p(x)
\end{equation}

The base for this logarithm can be chosen to be any number greater than 1.  If the value chosen is 2, the unit for this entropy measure is the \emph{bit}.  If the base is Euler's number $e$, the unit of this entropy measure is \emph{nat}, for natural logarithm.  In this thesis, the base of the logarithm used is the size of the code alphabet, which is 2.  It is to be noted that in computer science a bit is an entity that can take one of two values, 1 or 0.  In information theory, a bit is a unit of measurement for the entropy of a random variable.

It is usually common in information theory to express entropy of a random variable in terms of its expected value.  An alternative definition of the entropy could then given by

\begin{equation}
  \label{eq:7}
  H(X) = -E \log p(X)
\end{equation}

where

\begin{equation}
  \label{eq:8}
  Eg(x) = \sum_x p(x)g(x)
\end{equation}

The entropy $H(X)$ of a random variable $X$ is a functional of the probability distribution $p(x)$.  It measures the average amount of information contained in $X$, or the amount of \emph{uncertainty} removed from $X$ by revealing its outcome.  The value of the entropy depends on $p(x)$ and not the values in its alphabet $\mathcal{X}$.  In information theory, an alphabet $\mathcal{X}$ of a random variable is the set of all values a random variable may take.

To maximize the entropy of random variable we want to maximize the uncertainty of the outcome of that variable.  For a binary random variable $X$ with $\mathcal{X}=\{0,1\}$, this is achieved when both of its outcomes are equally likely to occur, i.e. when $p(0)=p(1)=0.5$.

The joint entropy of two random variables $X$ and $Y$ is defined by

\begin{align}
  \label{eq:9}
  H(X,Y) &= -\sum_{x,y}p(x,y)\log p(x,y) \\
         &= - E \log p(X,Y)
\end{align}

The joint entropy $H(X,Y)$ of two random variables $X$ and $Y$ is given by

\begin{align}
  \label{eq:11}
  H(X,Y) &= -\sum_{x,y} p(x,y)\log p(x,y)\\
         &= -E \log p(X,Y)
\end{align}

In some cases, the outcome of one random variable might give a clue about the outcome of another variable.  In that case, the definition of conditional entropy would be of high interest to us.

For two random variables $X$ and $Y$ the conditional entropy of $Y$ given $X$ is

\begin{align}
  \label{eq:10}
  H(Y|X) &= - \sum_{x,y} p(x,y) \log p(y|x)\\
         &= - E \log p(Y|X)
\end{align}

\section{Mutual information}
For two random variables $X$ and $Y$, the mutual information between $X$ and $Y$ is defined as

\begin{align}
  \label{eq:12}
  I(X;Y) &= \sum_{x,y} p(x,y) \log \frac{p(x,y)}{p(x)p(y)}\\
         &= E \log \frac{p(X,Y)}{p(X)p(Y)}
\end{align}

From this definition, we can see that mutual information is symmetrical in $X$ and $Y$, i.e. $I(X;Y) = I(Y;X)$.

\section{Information divergence}
In many communication applications we want to measure how a probability distribution $p$ differs from another probability distribution $q$ which is defined over the same alphabet.  This measure is usually desired to be a non-negative one.  The \emph{information divergence} measure from $p$ to $q$ can be used in this case and it is defined as

\begin{align}
  \label{eq:13}
  D(p||q) &= \sum_xp(x)\log\frac{p(x)}{q(x)}\\
          &= E_p \log \frac{p(X)}{q(X)}
\end{align}

where $E_p$ is the expectation with respect to $p$.

In information theory, information divergence is also known as relative entropy.  This makes sense as \ref{eq:13} could be interpreted as the entropy of $p$ relative to the entropy of $q$ as defined by the entropy measure in \ref{eq:5} and \ref{eq:7}.  Information divergence is also known as \emph{Kullback-Leibler distance} as well.  However, care must be taken when using this term as the information divergence measure is not symmetric in $p$ and $q$, and does not satisfy the triangular inequality.  i.e. $D(p||q) \neq D(q||p)$.  Whereas the word distance in many situations imply symmetry.


\cleardoublepage
\chapter{Table of notations}

\vspace{-4em}
\begin{table}[!h]
  \centering
  \begin{tabular}{c l}
    \hline
    Symbol & Definition\\
    \hline
    $DP(.)$  & Dirichlet process\\
    $G_0$    & base measure\\
    $\alpha$ & concentration parameter\\
    $\theta$ & Dirichlet distribution parameter space\\
    $\phi_{1:k}$ & Distinct values of $\theta$\\
    $m_k$    & number of parameter $\theta$ with $\phi_k$ value\\
    $m_{kt}'$ & prior weight of component $k$ at time $t$\\
    $\Delta$ & width of time decaying kernel\\
    $\lambda$& decay factor of a time decaying kernel\\
    $w$      & word in a document\\
    $H$      & base measure of the DP generating $G_0$\\
    $\gamma$ & concentration parameter for the DP generating $G_0$\\
    $\psi_{tdb}$   & topic $b$ for document $d$ at time $t$\\
    $n_{tdb}$      & number of words sampled from $\psi_{tdb}$\\
    $\psi_{tdb^{new}}$   & a new topic\\
    $B_d$    & number of topics in document $d$\\
    $\mathcal{N}(m,v_0)$ & a Gaussian distribution with mean $m$ and variance $v_0^2$\\
    $\beta_{i,k,w}$  & distribution of words over topic $k$ at time $i$ for word $w$\\
    $s_i$    & time stamp for time index $i$\\
    $\Delta_{s_j,s_i}$  & time duration between $s_i$ and $s_j$\\
    $Dir(.)$      & Dirichlet distribution\\
    $z_{t,n}$      & topic $n$ sampled at time $t$\\
    $I$      & identity matrix\\
    Mult(.)  & multinomial distribution\\
    $\pi(.)$ & mapping function\\
    \hline
  \end{tabular}

  \caption{Table of notations}
\end{table}


\end{document}